\documentclass[a4paper,11pt]{article}

\usepackage{classeJBArxive}

\usepackage{arydshln}
\usepackage{algorithm}
\usepackage[noend]{algpseudocode}
\usepackage{soul}

\newcommand*\unit[1]{\bigl[\, \mathsf{#1} \,\bigr]}
\newcommand*\unitt[1]{\mathsf{#1}}

\newcommand{\DF}{\textsc{D}u$\,$Fort--\textsc{F}rankel}

\newcommand{\cpu}{\mathrm{cpu}}
\newcommand{\ini}{\mathrm{ini}}
\newcommand{\obs}{\mathrm{obs}}
\newcommand{\apr}{\mathrm{apr}}

\newcommand{\Bi}{\mathrm{Bi}}
\newcommand{\Fo}{\mathrm{Fo}}

\newcommand{\degC}{^{\,\circ}C}

\newcommand*\prob[1]{\pi \,\bigl(\, #1 \,\bigr)}
\newcommand*\p{\boldsymbol{p}}

\title{
\vspace{-1.5cm}
Surface transfer coefficients estimation for heat conduction problem using the Bayesian framework \\	
\vspace{4pt}
}

\author{Julien Berger \textsuperscript{a}$^{\ast}$, Clemence Legros\textsuperscript{b,c}
\date{\today\vspace{-0.5cm}}}

\begin{document}

\maketitle

\begin{center}
\small
\textsuperscript{a} Laboratoire des Sciences de l’Ingénieur pour l’Environnement (LaSIE), UMR 7356 CNRS, La Rochelle Université, CNRS, 17000, La Rochelle, France\\
\textsuperscript{b} Univ. Savoie Mont Blanc, CNRS, LOCIE, 73000 Chambéry, France \\
\textsuperscript{c} Univ. Grenoble Alpes, CEA, LITEN, DTS, INES, 38000 Grenoble, France\\
$^{\ast}$corresponding author, e-mail address : julien.berger@univ-lr.fr\\
\end{center}

\begin{abstract}

\noindent This work deals with an inverse two-dimensional nonlinear heat conduction problem to determine the top and lateral surface transfer coefficients. For this, the \textsc{B}ayesian framework with the \textsc{M}arkov Chain \textsc{M}onte \textsc{C}arlo algorithm is used to determine the posterior distribution of unknown parameters. To handle the computational burden, a lumped one-dimensional model is proposed. \revision{The lumped model approximations are considered within the} \revision{parameter estimation procedure thanks to the Approximation Error Model.} The experiments are carried out for several configurations of chamber ventilator speed. Experimental observations are obtained through a complete measurement uncertainty propagation. By solving the inverse problem, accurate probability distributions are determined. Additional investigations are performed to demonstrate the reliability of the lumped model, in terms of accuracy and computational gains.~\\


\end{abstract}


\section{Introduction}

With the perspective of elaborating highly energy efficient buildings, several models have been proposed in the literature since the $70$'s. A recent review of the state-of-the art models has been proposed in \cite{mendes_numerical_2019}. Software such as \texttt{EnergyPlus} \cite{crawley_energyplus_2001} or \texttt{Domus} \cite{mendes_numerical_2019} are widely used to assess the building energy consumption.

Among all the physical phenomena, the wall conduction loads represent a non-negligible part of the energy consumption. The heat transfer through the building walls requires to be accurately represented. Several numerical models have been proposed to compute the temperature distribution \revision{through} the walls. Among others, a Spectral reduced order model has been proposed for fast and accurate computations in \cite{berger_comparison_2020}. In \cite{gasparin_innovative_2019}, an innovative approach has been proposed to compute the conduction loads using a numerical method with adaptive mesh. In \cite{berger_efficient_2021}, a two-dimensional model based on \DF ~numerical method is proposed to evaluate the wall efficiency considering the shading of the local urban environment. To validate the reliability of such models it is worth investigations to propose experimental observations. Several data-sets have been proposed in the literature as reported in the review proposed in \cite{busser_comparison_2019}. To benchmark the model predictions with experimental observations, the input parameter such as material properties are of capital importance. In those works, they are provided thanks to a precise determination according to the standards procedure. 

One important drawback of \revision{these} experimental data-sets is the lack of knowledge of the surface transfer coefficients. Indeed, the direct measurement of \revision{these} quantities using optical measurements or infra-red temperature mapping methods requires expensive equipment \cite{baughn_periodic_1998,naylor_recent_2003,bazargani_methodology_2015}. It has limitations depending on the temperature and flow gradients. Thus, the model predictions can only be evaluated considering first-type of \textsc{D}irichlet boundary conditions. The latter are provided by the measurement at the boundaries of the material. References \cite{berger_comparison_2020,berger_evaluation_2019,berger_new_2019}, can be consulted to illustrate this point. However, in real configurations, the inside and outside heat transfer coefficients of building walls play a crucial role on the energy efficiency \cite{berger_bayesian_2016}. In addition, the model reliability should be evaluated in realistic conditions. Thus, it is of major importance to propose experimental data-set with a good expertise of the surface transfer coefficients.

The objective of this article is to determine the surface transfer coefficients through an experimental campaign. The investigations aim at characterizing the heat transfer coefficient in a climatic chamber according to the ventilator speed velocity. Instead of measuring directly, an alternative method is investigated by solving an inverse two-dimensional non-linear heat conduction problem. Several methods have been proposed in the literature to solve such inverse problem as presented in \cite{orlande_thermal_2011} with primary examples for the estimation of heat transfer coefficients in \cite{farhanieh_inverse_2008,ryfa_determination_2021}. Here, the \textsc{B}ayesian framework statistics is selected for estimating the model parameters \cite{kaipio_bayesian_2011,mota_bayesian_2010}. It is based on stochastic simulation to generate samples and approximate the unknown parameters probability distribution. Thus, it provides a quantification of the uncertainties on the estimated parameter. It is a major advantage for our purpose of determining the surface transfer coefficient of a material placed in a climatic chamber. Nevertheless, the \textsc{B}ayesian estimation suffers from computational burden since it requires thousands of computations of the direct problem. To tackle this issue, a lumped one-dimensional model is proposed to decrease the computational cost. The lumped model approximations are considered within the parameter estimation procedure thanks to the Approximation Error Model.

The article is structured as follows. Next section presents the details of the mathematical model and the derivation of the lumped model. Section~\nameref{sec:pep} describes the parameter estimation within the \textsc{B}ayesian framework. Then, the experimental design with the climatic chamber is presented in Section~\nameref{sec:exp_design}. Last, the results of the inverse problem together with the validity of the proposed lumped model are discussed in Section~\nameref{sec:results}.

\section{Formulation of the mathematical model}
\label{sec:mathematical_model}

\subsection{\revision{Complete formulation}}
\label{sec:complete_formulation}

\begin{figure}[h!]
\centering
\includegraphics[width=.9\textwidth]{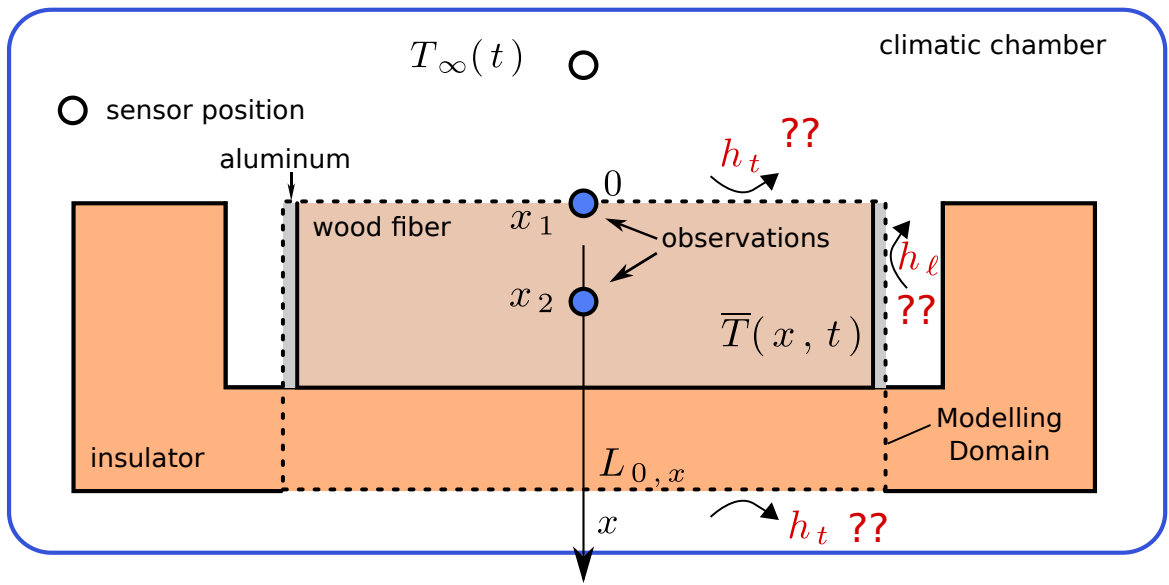}
\caption{\revision{Illustration of the domain.}}
\label{fig:domain}
\end{figure}

The heat transfer process occurs in materials placed in a climatic chamber as illustrated in Figure~\ref{fig:domain}. A wood fiber material is placed in a mold of insulation. The lateral surfaces of the wood fiber are covered by aluminum tape. There is a small gap between the side insulation and the aluminum. The materials are initially at $T_{\,\ini} \ \unit{K}$ and then submitted to variations of the climatic chamber temperature $T_{\,\infty}\,$. The time horizon of the investigations is $\Omega_{\,t} \, \eqdef \, \bigl[\, 0 \,,\, t_{\,0} \,\bigr]\,$, $t_{\,0} \ \unit{s}$ being the final time. The  materials have a volumetric heat capacity $c \ \unit{W\,.\,m^{\,-3}\,.\,K^{\,-1}}$ and a thermal conductivity $k \ \unit{W\,.\,m^{\,-1}\,.\,K^{\,-1}}$ that both depend on the temperature according to:
\begin{align*}
k_{\,i}(\,T\,) \egal k_{\,i\,,\,0} \plus k_{\,i\,,\,1} \cdot \frac{T}{T_{\,0}} \,, \qquad
c_{\,i}\revision{(\,T\,)} \egal c_{\,i\,,\,0} \plus c_{\,i\,,\,1} \cdot \frac{T}{T_{\,0}}  \,,
\end{align*}
where index $i \egal 1\,$, $i \egal 2$ and $i \egal 3$ stand for wood fiber, insulation and aluminum, respectively.
The temperature depends on the coordinates $(\,x\,,\,y\,,\,z\,) \, \in \, \Omega_{\,x}\, \times \, \Omega_{\,y}\, \times \, \Omega_{\,z}$ with:
\begin{align*}
\Omega_{\,x} \, \eqdef \, \bigl[\, 0 \,,\, L_{\,0\,,\,x}\,\bigr] \,, \qquad
\Omega_{\,y} \, \eqdef \, \bigl[\, 0 \,,\, L_{\,0\,,\,y}\,\bigr] \,, \qquad
\Omega_{\,z} \, \eqdef \, \bigl[\, 0 \,,\, L_{\,0\,,\,z}\,\bigr] \,,
\end{align*}
where $L_{\,0\,,\,x} \ \unit{m}\,$ $L_{\,0\,,\,y} \ \unit{m}$ and $L_{\,0\,,\,z}\ \unit{m}$ are the domain dimensions. The governing equation of heat transfer is written in three-dimension:
\begin{align}
\label{eq:heat_3D}
c_{\,i}\revision{(\,T\,)} \cdot \pd{T}{t} \egal \div \Bigl(\, k_{\,i}(\,T\,) \cdot \grad T \,\Bigr) \,,\, \qquad \forall i \,\in\,\bigl\{\,1 \,,\,2\,,\,3\,\bigr\} \,,
\end{align}
At the interface between the material and the ambient air of the climatic chamber, convective transfer occurs on the top and lateral sides, formulated as a \textsc{R}obin boundary condition. At the top surface, we have:
\begin{align}
\label{eq:BC_top}
k_{\,1}(\,T\,)  \cdot \pd{T}{x} \egal h_{\,t} \cdot \Bigl(\, T \moins T_{\,\infty}(\,t\,) \,\Bigr) \,, \qquad x \egal 0 \,, \qquad \forall (\,y\,,\,z\,) \, \in \Omega_{\,y} \, \times \, \Omega_{\,z}  \,,
\end{align}
with $h_{\,t} \ \unit{W\,.\,m^{\,-2}\,.\,K^{\,-1}}$ the top surface heat transfer coefficient and $T_{\,\infty}(\,t\,)$ is the time variation of the temperature imposed by the climatic chamber. For the lateral sides, the boundary conditions is assumed as:
\begin{subequations}
\label{eq:BC_sides}
\begin{align}
k_{\,3}(\,T\,)  \cdot \pd{T}{y} &\egal h_{\,\ell} \cdot \Bigl(\, T \moins T_{\,\infty}(\,t\,) \,\Bigr) \,, \qquad y \egal 0 \,, \qquad \forall (\,x\,,\,z\,) \, \in \, \Omega_{\,x} \, \times \, \Omega_{\,y} \,, \\[4pt]
k_{\,3}(\,T\,)  \cdot \pd{T}{y} &\egal - \, h_{\,\ell} \cdot \Bigl(\, T \moins T_{\,\infty}(\,t\,) \,\Bigr) \,, \qquad y \egal L_{\,0\,,\,y} \,, \qquad \forall (\,x\,,\,z\,)  \, \in \, \Omega_{\,x} \, \times \, \Omega_{\,y} \,, \\[4pt]
k_{\,3}(\,T\,)  \cdot \pd{T}{z} &\egal h_{\,\ell} \cdot \Bigl(\, T \moins T_{\,\infty}(\,t\,) \,\Bigr) \,, \qquad z \egal 0 \,, \qquad \forall (\,x\,,\,y\,) \, \in \, \Omega_{\,x} \, \times \, \Omega_{\,z}\,, \\[4pt]
k_{\,3}(\,T\,)  \cdot \pd{T}{z} &\egal - \, h_{\,\ell} \cdot \Bigl(\, T \moins T_{\,\infty}(\,t\,) \,\Bigr) \,, \qquad z \egal L_{\,0\,,\,z} \,, \qquad \forall (\,x\,,\,y\,)  \, \in \, \Omega_{\,x} \, \times \, \Omega_{\,z} \,.
\end{align}
\end{subequations}
With this description, it is assumed that there is one singular heat transfer coefficient for all lateral faces of the material denoted by $h_{\,\ell}\,$. In addition, the ambient temperature $T_{\,\infty}$ is equal at the top and lateral faces. At the bottom surface, the insulator is in contact with the ambient air according to:
\begin{align}
\label{eq:BC_bot}
k_{\,2}(\,T\,)  \cdot \pd{T}{x} \egal h_{\,t} \cdot \Bigl(\, T \moins T_{\,\infty}(\,t\,) \,\Bigr) \,, \qquad x \egal L_{\,0\,,\,x} \,, \qquad \forall (\,y\,,\,z\,) \, \in \Omega_{\,y} \, \times \, \Omega_{\,z}  \,,
\end{align}
assuming that the top and bottom surface transfer coefficients are equal. The two interface materials, insulator/wood fiber and aluminum/wood fiber, are denoted $\partial \Omega_{\,\mathrm{ins}}$ and $\partial \Omega_{\,\mathrm{alu}}\,$, respectively. For each, the continuity of the heat flux and of the temperature is assumed. At initial condition, the material is in equilibrium at a known temperature $T_{\,\ini}\,$:
\begin{align*}
T \egal T_{\,\ini} \,, \qquad \forall (\,x\,,\,y\,,\,z\,) \, \in \, \Omega_{\,x} \, \times \, \Omega_{\,y} \, \times \, \Omega_{\,z} \,, \qquad t \egal 0 \,.
\end{align*}

\subsection{\revision{Lumped formulation}}
\label{sec:lumped_formulation}

The physical model defined by Eqs.~\eqref{eq:heat_3D} to \eqref{eq:BC_bot} is a detailed formulation of the problem. To decrease its computational burden, particularly in the framework of \textsc{B}ayesian inferences, a lumped approach is proposed. \revision{Note that the error between the complete and lumped models} \revision{will be quantified as a \textsc{G}aussian variable and integrated in the parameter estimation problem.} The heat transfer is assumed mainly in the direction of $x\,$:
\begin{align*}
T(\,x\,,\,y\,,\,z\,,\,t\,) \, \approx \, \overline{T}(\,x\,,\,t\,) \,.
\end{align*}
In this way, only two materials are considered in the model, namely the wood fiber and the insulator. To obtain the differential equation for the wood fiber $i \egal 1$, Eq.~\eqref{eq:heat_3D} is integrated over $\Omega_{\,y}$ and $\Omega_{\,z}\,$:
\begin{align*}
\int_{\,\Omega_{\,y}} \, \int_{\,\Omega_{\,z}} \, c_{\,1}(\,\overline{T}\,) \, \pd{\overline{T}}{t} &  \, \mathrm{d}y \, \mathrm{d}z 
\egal 
\int_{\,\Omega_{\,y}} \, \int_{\,\Omega_{\,z}} \, \pd{}{x} \, \biggl(\, k_{\,1}(\,\overline{T}\,) \cdot \pd{\overline{T}}{x} \,\biggr) \, \mathrm{d}y \, \mathrm{d}z 
\plus \\[4pt]
& \int_{\,\Omega_{\,y}} \, \int_{\,\Omega_{\,z}} \, \pd{}{y} \, \biggl(\, k_{\,1}(\,\overline{T}\,) \cdot \pd{\overline{T}}{y} \,\biggr)\, \mathrm{d}y \, \mathrm{d}z 
\plus 
\int_{\,\Omega_{\,y}} \, \int_{\,\Omega_{\,z}} \, \pd{}{z} \ \biggl(\, k_{\,1}(\,\overline{T}\,) \cdot \pd{\overline{T}}{z}  \,\biggr)\, \mathrm{d}y \, \mathrm{d}z \,.
\end{align*}
It yields to:
\begin{align}
\label{eq:integrated_heat_equation}
c_{\,1}(\,\overline{T}\,) \cdot \pd{\overline{T}}{t} \egal 
\pd{}{x} \, \biggl(\, k(\,\overline{T}\,) \cdot \pd{\overline{T}}{x} \,\biggr)  \plus
\frac{2}{L_{\,0\,,\,y}} \, \biggl[\, k_{\,1}(\,\overline{T}\,) \, \pd{\overline{T}}{y}\,\biggr]_{\,\partial \Omega_{\,\mathrm{alu}}}  \plus
\frac{2}{L_{\,0\,,\,z}} \, \biggl[\, k_{\,1}(\,\overline{T}\,) \, \pd{\overline{T}}{z}\,\biggr]_{\,\partial \Omega_{\,\mathrm{alu}}}  \,.
\end{align}
A source term appears in the equation for the contribution of the lateral fluxes from the aluminum tape. It is assumed that $L_{\,0\,,\,y} \egal L_{\,0\,,\,z}$\,. In addition, the heat flux through the aluminum can be assessed through a thermal resistance $R_{\,\ell}$ related to $h_{\,\ell}$ and to be determined:
\begin{align*}
\biggl[\, k_{\,1}(\,\overline{T}\,) \, \pd{\overline{T}}{z}\,\biggr]_{\,\partial \Omega_{\,\mathrm{alu}}} \egal R_{\,\ell} \, \bigl(\, T_{\,\infty} \moins \overline{T} \,\bigr) \,.
\end{align*}
Such approximation can be justified by the fact that the heat transfer occurs faster in the aluminum tape than in the wood fiber. Thus, Eq.~\eqref{eq:integrated_heat_equation} becomes:
\begin{align}
\label{eq:heat_lumped_wf}
c_{\,1}(\,\overline{T}\,) \cdot \pd{\overline{T}}{t} \egal 
\pd{}{x} \, \biggl(\, k_{\,1}(\,\overline{T}\,) \cdot \pd{\overline{T}}{x} \,\biggr)  \plus
\frac{4 \cdot R_{\,\ell}}{L_{\,0\,,\,y}}\, \bigl(\, T_{\,\infty} \moins \overline{T} \,\bigr) \,.
\end{align}
For the second material $i \egal 2$, the insulation, the important width and depth dimensions enables to assume a one-dimensional transfer without source term arising from lateral contributions: 
\begin{align}
\label{eq:heat_lumped_ins}
c_{\,2}(\,\overline{T}\,) \cdot \pd{\overline{T}}{t} \egal 
\pd{}{x} \, \biggl(\, k_{\,2}(\,\overline{T}\,) \cdot \pd{\overline{T}}{x} \,\biggr) \,.
\end{align}
At the interface $\partial \Omega_{\,\mathrm{ins}}$ we have the following continuity conditions:
\begin{align*}
k_{\,1}(\,\overline{T}\,) \cdot \pd{\overline{T}}{x} & \egal k_{\,2}(\,\overline{T}\,) \cdot \pd{\overline{T}}{x} \,, \\[4pt]
\overline{T}(\,x \moins \epsilon \,,\,t\,) & \egal \overline{T}(\,x \plus \epsilon \,,\,t\,) \,,
\quad  \forall \, x \, \in \, \partial \Omega_{\,\mathrm{ins}} \,, 
\quad \epsilon \, \rightarrow \, 0 \,.
\end{align*}
The field $\overline{T}$  of the lumped model verifies the boundary condition:
\begin{align*}
k_{\,1}(\,\overline{T}\,) \cdot\pd{\overline{T}}{x} &\egal h_{\,t} \, \Bigl(\, \overline{T} \moins T_{\,\infty}(\,t\,) \,\Bigr) \,,  && x \egal 0 \,,  \\[4pt]
k_{\,2}(\,\overline{T}\,) \cdot\pd{\overline{T}}{x} &\egal - \, h_{\,t} \, \Bigl(\, \overline{T} \moins T_{\,\infty}(\,t\,) \,\Bigr) \,, && x \egal L_{\,0\,,\,x}  \,.
\end{align*}
and the initial condition: 
\begin{align*}
\overline{T} \egal T_{\,\ini} \,, \qquad t \egal 0 \,.
\end{align*}

\subsection{\revision{Dimensionless formulation}}

With the issue of solving an inverse problem, a dimensionless formulation of the lumped mathematical model is defined for each of the two materials $i \, \in \, \bigl\{\, 1\,,\,2\,\bigr\}\,$. For this, the following transformation is defined: 
\begin{align*}
u \, \eqdef \, \frac{\overline{T}}{T_{\,0}} \qquad
\tau \, \eqdef \, \frac{t}{t_{\,0}}\,, \qquad
\chi \, \eqdef \, \frac{x}{L_{\,0\,,x}} \,.
\end{align*}
Thus, the dimensionless formulation of Eq.~\eqref{eq:heat_lumped_wf} is for the wood fiber $i \egal 1\,$:
\begin{align}
\label{eq:heat_lumped_dimless_wf}
\zeta_{\,1}\revision{(\,u\,)} \pd{u}{\tau} \egal 
\Fo_{\,1} \ \pd{}{\chi} \, \biggl(\, \kappa_{\,1}(\,u\,) \cdot \pd{u}{\chi} \,\biggr) 
\plus \Bi_{\,\ell} \ r \ \Fo_{\,1}  \  \bigl(\, u_{\,\infty} \moins u \,\bigr)  \,,
\end{align}
and for the insulation $i \egal 2\,$:
\begin{align}
\label{eq:heat_lumped_dimless_ins}
\zeta_{\,2}\revision{(\,u\,) }\pd{u}{\tau} \egal 
\Fo_{\,2} \ \pd{}{\chi} \, \biggl(\, \kappa_{\,2}(\,u\,) \cdot \pd{u}{\chi} \,\biggr)  \,.
\end{align}
The \textsc{F}ourier number, the lateral \textsc{B}iot numbers and the dimensionless length ratio are defined by:
\begin{subequations}
\label{eq:def_Fourier}
\begin{align}
\Fo_{\,i} & \eqdef \, \frac{k_{\,i\,,\,0} \ t_{\,0}}{c_{\,i\,,\,0} \ L_{\,0\,,\,x}^{\,2}} \,,
\qquad \forall i \, \in \, \bigl\{\, 1\,,\,2\,\bigr\} \,, \\[4pt]
\Bi_{\,\ell} & \eqdef \, \frac{4 \ R_{\,\ell} \ L_{\,0\,,\,x} }{k_{\,1\,,\,0}} \,, \\[4pt]
r & \eqdef \, \frac{L_{\,0\,,\,x}}{L_{\,0\,,\,y}} \,,
\end{align}
\end{subequations}
and the dimensionless thermal conductivity and heat capacity are:
\begin{align*}
\zeta_{\,i}(\,u\,) & \egal 1 \plus \zeta_{\,i\,,1} \, u \,, \qquad \zeta_{\,i\,,1} \, \eqdef \, \frac{c_{\,i\,,1}}{c_{\,i\,,0}} \,, \\[4pt]
\kappa_{\,i}(\,u\,) & \egal 1 \plus \kappa_{\,i\,,1} \, u \,, \qquad \kappa_{\,i\,,1} \, \eqdef \, \frac{k_{\,i\,,1}}{k_{\,i\,,0}} \,, \qquad \forall i \, \in \, \bigl\{\, 1\,,\,2\,\bigr\} \,.
\end{align*}
The boundary conditions of the dimensionless problem are: 
\begin{subequations}
\label{eq:bc_lumped_dimless}
\begin{align}
\kappa_{\,1}(\,u\,) \cdot \pd{u}{\chi} & \egal \Bi_{\,t} \ \bigl(\, u \moins u_{\,\infty} \,\bigr) \,, \qquad \chi \egal 0 \,, \qquad \tau \, \geqslant \, 0 \,, \\[4pt]
\kappa_{\,21} \cdot \kappa_{\,2}(\,u\,) \cdot \pd{u}{\chi} & \egal - \, \Bi_{\,t} \ \bigl(\, u \moins u_{\,\infty} \,\bigr)  \,, \qquad \chi \egal 1 \,, \qquad \tau \, \geqslant \, 0 \,,
\end{align}
\end{subequations}
with the top \textsc{B}iot number and the dimensionless thermal conductivity are given by:
\begin{align}
\label{eq:def_Biot}
\Bi_{\,t} \, \eqdef \, \frac{h_{\,t} \ L_{\,0\,,\,x}}{k_{\,1\,,\,0}} \quad \,, \quad
\kappa_{\,21} \egal \frac{k_{\,2\,,\,0}}{k_{\,1\,,\,0}} \,.
\end{align}
The dimensionless interface $\partial \Omega_{\,\mathrm{ins}}$ conditions are:
\begin{align*}
\kappa_{\,1}(\,u\,) \cdot \pd{u}{\chi} & \egal \kappa_{\,21} \cdot \kappa_{\,2}(\,u\,) \cdot \pd{u}{\chi} \,, \\[4pt]
u(\,\chi \moins \epsilon \,,\,t\,) & \egal u(\,\chi \plus \epsilon \,,\,t\,) \,, \quad  \chi \, \in \, \partial \Omega_{\,\mathrm{ins}}
\quad \epsilon \, \rightarrow \, 0\,.
\end{align*}
Last, the initial condition is:
\begin{align}
\label{eq:ic_lumped_dimless}
u \egal u_{\,\ini} \,, \qquad \forall \chi \, \in \, \bigl[\, 0 \,,\, 1 \,\bigr] \,, \qquad \tau \egal 0 \,.
\end{align}
The governing equation is solved using a numerical method based on a central second-order finite-differences with a variable-step, variable-order solver for the time integration \cite{shampine_matlab_1997}. All the implementation is realized in the \texttt{Matlab}\texttrademark ~environment.

\section{Parameter estimation within the \textsc{B}ayesian framework}
\label{sec:pep}

The unknown parameters are the top $h_{\,t}$ and side $R_{\,\ell}$ heat transfer coefficients, corresponding to the dimensionless \textsc{B}iot numbers $\Bi_{\,t}$ and $\Bi_{\,\ell}\,$, respectively. They are denoted by:
\begin{align*}
\p \, \eqdef \, \Bigl(\, p_{\,1} \,,\, p_{\,2} \,\Bigr) \, \equiv \, \Bigl(\, \Bi_{\,t} \,,\, \Bi_{\,\ell} \,\Bigr) \,.
\end{align*}
The set of parameters is denoted by $\Omega_{\,p}\,$. The \emph{a priori} parameters are denoted with the superscript $\apr\,$. It supposed that a total of $N_{\,x} \, \times \, N_{\,t}$ transient measurement of the temperature field $\hat{u}_{\,j}(\,t^{\,n}\,)$ are available at the (dimensionless) time $\tau^{\,n}\,, \ n \,\in\, \bigl\{\, 1 \,,\,\ldots \,,\, N_{\,t}\,\bigr\}$ and at the (dimensionless) sensor positions $\chi_{\,j}\,, \ j \,\in\, \bigl\{\, 1 \,,\,\ldots \,,\, N_{\,x}\,\bigr\}\,$. The measurement error are supposed as \textsc{G}aussian random variables, with zero means and known variance $\sigma(\,\chi_{\,j}\,,\,\tau^{\,n}\,)\,$. The measurement errors are additive and independent of the unknown parameter $\p\,$. 

\subsection{Structural identifiability}

Before performing the estimation, the theoretical, or structural, identifiability of the parameter needs to be demonstrated. For this the definition provided by \textsc{W}alter and co-workers in \cite{walter_global_1982,walter_qualitative_1990} is recalled. A parameter $\p \, \in \, \Omega_{\,p}$ is Structurally Globally Identifiable (SGI) in the model $f$ if the following condition is verified:
\begin{align*}
\forall \, (\,x\,,\,t\,) \,, \qquad f\,\bigl(\,x\,,\,t\,,\,\p \,\bigr) \, \equiv \, f\, \bigl(\,x\,,\,t\,,\,\widetilde{\p} \,\bigr) \, \Longrightarrow \, \p \egal \widetilde{\p} \,.
\end{align*}
From a mathematical point of view, this property is equivalent to the injection of the model $f$ according to the unknown parameter. It must be evaluated for each parameter $\Bi_{\,t}$ and $\Bi_{\,\ell}$ of the model.

\subsection{\revision{Sensitivity equations}}

To evaluate the practical identifiability of the unknown parameters before the estimation, the following sensitivity functions are defined: 
\begin{align}
\label{eq:sensitivity_functions}
\theta \, \eqdef \, \pd{u}{\Bi_{\,t}} \,, \qquad \psi \, \eqdef \, \pd{u}{\Bi_{\,\ell}} \,.
\end{align}
Their equations are obtained by differentiating Eqs.~\eqref{eq:heat_lumped_dimless_wf} and \eqref{eq:heat_lumped_dimless_ins} according to the parameters. Namely, the function $\theta$ verifies for the wood fiber: 
\begin{align*}
\zeta_{\,1}\revision{(\,u\,)} \cdot \pd{\theta}{\tau} \egal 
\Fo_{\,1} \ \pd{}{\chi} \, \biggl(\, \kappa_{\,1}(\,u\,) \cdot \pd{\theta}{\chi} \,\biggr) 
\moins \Bi_{\,\ell} \ r \ \Fo_{\,1} \  \theta 
\plus \Fo_{\,1} \ \pd{}{\chi} \, \biggl(\, \kappa_{\,1\,,\,1} \ \theta \ \pd{u}{\chi} \,\biggr) 
\moins \zeta_{\,1\,,\,1} \ \theta \ \pd{u}{\tau}  \,,
\end{align*}
and for the insulation:
\begin{align*}
\zeta_{\,2}\revision{(\,u\,)} \cdot \pd{\theta}{\tau} \egal 
\Fo_{\,2} \ \pd{}{\chi} \, \biggl(\, \kappa_{\,2}(\,u\,) \cdot \pd{\theta}{\chi} \,\biggr) 
\plus \Fo_{\,2} \ \pd{}{\chi} \, \biggl(\, \kappa_{\,2\,,\,1} \ \theta \ \pd{u}{\chi} \,\biggr) 
\moins \zeta_{\,2\,,\,1} \ \theta \ \pd{u}{\tau}  \,,
\end{align*}
with the corresponding boundary, interface and initial conditions.
Then, the following differential equations is verified by the function $\psi\,$ in the wood fiber material:
\begin{align*}
\zeta_{\,1}\revision{(\,u\,)} \cdot \pd{\psi}{\tau} \egal &
\Fo_{\,1} \ \pd{}{\chi} \, \biggl(\, \kappa_{\,1}(\,u\,) \cdot \pd{\psi}{\chi} \,\biggr) 
\moins \Bi_{\,\ell} \ r \ \Fo_{\,1} \  \psi 
\plus r \ \Fo_{\,1} \  \bigl(\, u_{\,\infty} \moins u_{\,1} \,\bigr) \\[4pt]
& \plus \Fo_{\,1} \ \pd{}{\chi} \, \biggl(\, \kappa_{\,1\,,\,1} \ \psi \ \pd{u}{\chi} \,\biggr) 
\moins \zeta_{\,1\,,\,1} \ \theta \ \pd{u}{\tau}  \,,
\end{align*}
and for the insulation one:
\begin{align*}
\zeta_{\,2}\revision{(\,u\,)} \cdot \pd{\psi}{\tau} \egal
\Fo_{\,2} \ \pd{}{\chi} \, \biggl(\, \kappa_{\,2}(\,u\,) \cdot \pd{\psi}{\chi} \,\biggr) 
\plus \Fo_{\,2} \ \pd{}{\chi} \, \biggl(\, \kappa_{\,2\,,\,1} \ \psi \ \pd{u}{\chi} \,\biggr) 
\moins \zeta_{\,2\,,\,1} \ \theta \ \pd{u}{\tau}\,,
\end{align*}
completed with the corresponding interface, initial and boundary conditions. Both sensitivity equations are solved using the same numerical methods as for the governing equation of $u\,$. With the sensitivity function, an error estimator can be defined based on the \textsc{F}isher information matrix $\mathcal{F} \, \in \mathcal{M}at\,\bigl(\, \mathbb{R}_{\,>\,0} \,,\, 2 \,,\, 2 \,\bigr)$ :
\begin{align*}
\displaystyle \mathcal{F}_{\,11} \, \eqdef \, \bigl[\, \mathcal{F}_{\,i\,j} \,\bigr] \,, & \qquad
\bigl(\,i\,,\,j\,\bigr) \, \in \, \bigl\{\, 1 \,,\, 2 \,\bigr\}^{\,2} \,,
 \\[4pt]
\displaystyle \mathcal{F}_{\,11} \, \eqdef \, \sum_{j \egal 1}^{\,N_{\,x}} \, \int_{\,\Omega_{\,t}} 
\frac{\theta(\,\chi_{\,j} \,,\,t\,)^{\,2}}{\sigma(\,\chi_{\,j} \,,\,t\,)^{\,2}}  
\ \mathrm{d}t \,, & \qquad 
\mathcal{F}_{\,22} \, \eqdef \, \sum_{j \egal 1}^{\,N_{\,x}} \, \int_{\,\Omega_{\,t}} 
\frac{\psi(\,\chi_{\,j} \,,\,t\,)^{\,2}}{\sigma(\,\chi_{\,j} \,,\,t\,)^{\,2}} 
\ \mathrm{d}t \,, \\[4pt]
\mathcal{F}_{\,12} \egal \mathcal{F}_{\,21} \, \eqdef \,  \sum_{j \egal 1}^{\,N_{\,x}} \, \int_{\,\Omega_{\,t}} &
\frac{\theta(\,\chi_{\,j} \,,\,t\,) \cdot \psi(\,\chi_{\,j} \,,\,t\,) }{\sigma(\,\chi_{\,j} \,,\,t\,)^{\,2}}  
\ \mathrm{d}t \,.
\end{align*}
Then, the error indicators are given by
\begin{align*}
\eta_{\,t} \, \eqdef \, \sqrt{\bigl(\, \mathcal{F}_{\,11} \,\bigr)^{\,-1}} \,, \qquad 
\eta_{\,\ell} \, \eqdef \, \sqrt{\bigl(\, \mathcal{F}_{\,22} \,\bigr)^{\,-1}} \,.
\end{align*}
for $\Bi_{\,t}$ and $\Bi_{\,\ell}\,$, respectively.

\subsection{\textsc{B}ayesian estimation}

The \textsc{M}arkov Chain \textsc{M}onte \textsc{C}arlo (MCMC) \cite{christen_markov_2005,gamerman_markov_2006} is employed to estimate the so-called posterior distribution $\prob{\p\,\bigl|\,\hat{u}} $ of the parameter $\p$ within the \textsc{B}ayesian framework \cite{kaipio_statistical_2005,kaipio_bayesian_2011,lee_bayesian_2012}. According to \textsc{B}aye's theorem, we have:
\begin{align*}
\prob{\p\,\bigl|\,\hat{u}} \egal \frac{\prob{\p} \cdot \prob{\hat{u}\,\bigl|\, \p}}{\prob{\hat{u}}} \,,
\end{align*}
where $\prob{\p}$ is the prior probability density of the parameters $\p\,$,  $\prob{\hat{u}\,\bigl|\, \p}$ is the likelihood function and $\prob{\hat{u}}$ is the marginal probability density of measurements. The latter is generally difficult to estimate and not required from a practical point of view to determine the unknown parameters. Given that, the theorem is simplified to:
\begin{align*}
\prob{\p\,\bigl|\,\hat{u}} \, \propto \, \prob{\p} \cdot \prob{\hat{u}\,\bigl|\, \p} \,.
\end{align*}
With the hypothesis \revision{assuming that measurement errors are \textsc{G}aussian random variables,} \revision{with zero means, known covariance matrix and independent of the unknown parameters}, the likelihood is expressed as: 
\begin{align*}
\prob{\hat{u}\,\bigl|\, \p} \, \propto \, \exp 
\Biggl(\, 
\sum_{j \egal 1}^{N_{\,x}} \, \sum_{n \egal 1}^{N_{\,t}} \ \frac{1}{\sigma(\,\chi_{\,j} \,,\,\tau_{\,n}\,)^{\,2}} \, 
\Bigl(\,\hat{u}_{\,j}(\,t^{\,n}\,) \moins u(\,\p \,,\, \chi_{\,j} \,,\, \tau^{\,n} \,) \,\Bigr)^{\,2}
\,\Biggr) \,,
\end{align*}
where $u(\,\p \,,\, \chi_{\,j} \,,\, \tau^{\,n} \,)$ is the solution of the direct problem, Eqs.~\eqref{eq:heat_lumped_wf} and \eqref{eq:heat_lumped_ins}, with boundary condition, Eq.~\eqref{eq:bc_lumped_dimless}, and initial conditions, Eq.~\eqref{eq:ic_lumped_dimless}, obtained at the sensor position $\chi_{\,j}\,$, the time $\tau^{\,n}$ and given parameters $\p\,$. \revision{Here, the prior probability density is assumed as a \textsc{G}aussian distribution in the domain $\Omega_{\,p}\,$:}
\begin{align*}
\prob{\p} \egal \mathcal{N}\bigl(\,\Omega_{\,p} \,\bigr)\,.
\end{align*}
The \textsc{M}etropolis--\textsc{H}astings algorithm is used to explore the posterior distribution \cite{metropolis_equation_1953,hastings_monte_1970}. It runs over the states $k$ of the \textsc{M}arkov chain as synthesized in the Table~\ref{alg:MCMC}. It starts by sampling a new candidate $\p^{\,\star}$ according to a proposal distribution $\mathcal{P}$ and the given state $\p^{\,k-1}$ of the \textsc{M}arkov chain (Step~\ref{alg:step_proposal_parameter}). Here a uniform proposal is used: 
\begin{align*}
\mathcal{P}(\,\p^{\,\star}\,\bigl|\,\p^{\,k-1}\,) \egal \p^{\,k-1} \plus \boldsymbol{w} \cdot \mathcal{U}\bigl(\,\bigl[\,-1\,,\,1\,\bigr] \,\bigr)  \,,
\end{align*}
where $ \boldsymbol{w}$ is the parameters walk. Then, the solution of the direct problem,  Eqs.~\eqref{eq:heat_lumped_dimless_wf} and \eqref{eq:heat_lumped_dimless_ins}, is computed given the sampled parameter $\p^{\,\star}$ (Step~\ref{alg:step_direct_problem}). After that, the posterior distribution $\prob{\p^{\,\star}\,\bigl|\,\hat{u}}$ is evaluated together with the acceptance factor defined in the case of a random walk by: 
\begin{align}
\label{eq:acceptance_factor}
\beta \, \eqdef \, \min \, \Biggl(\, 1 \,,\,
\frac{\prob{\p^{\,\star}\,\bigl|\,\hat{u}}}{\prob{\p^{\,k-1}\,\bigl|\,\hat{u}}}
\,\Biggr) \,.
\end{align}
A random value $U$ uniformly distributed on $\bigl]\,0\,,\,1\,\bigr[$ is then generated (Step~\ref{alg:step_sample_random_value}). At step~\ref{alg:step_test_parameter}, If $U \, \leqslant \beta$ then the candidate parameter is retained $\p^{\,k} \egal \p^{\,\star}\,$. Otherwise, it is stated that $\p^{\,k} \egal \p^{\,k-1}\,$. At the end of the total states number $N_{\,s}\,$, a set of parameter is obtained $\bigl\{\, \p^{\,1} \,,\, \ldots \,,\, \p^{\,N_{\,s}}\,\bigr\}\,$. Values while the chain has not converged to the equilibrium (burn-in period $N_{\,b}$) should be ignored and the posterior distribution can be represented. Furthermore, the mean and standard deviation of the estimated parameters can be evaluated for $i \, \in \, \bigl\{\, 1 \,,\,2 \,\bigr\}\,$: 
\begin{align*}
\mu\bigl(\, p_{\,i}\,\bigr) \egal \frac{1}{N_{\,s} \moins N_{\,b}} \ \sum_{k \egal N_{\,b}}^{\,N_{\,s}} \ p_{\,i}^{\,k} \,, \qquad 
\upsilon (\, p_{\,i}\,) \egal \sqrt{ \frac{1}{N_{\,s} \moins N_{\,b}} \ \sum_{k \egal N_{\,b}}^{\,N_{\,s}} \ \Bigl(\, p_{\,i}^{\,k} \moins \mu \, \circ p_{\,i} \,\Bigr)^{\,2} } \,.
\end{align*}
Interested readers are invited to consult \cite{orlande_inverse_2012,mota_bayesian_2010} for complementary examples of \textsc{B}ayesian estimation for inverse heat conduction problems.

\begin{algorithm}
\caption{MCMC Algorithm.}\label{alg:MCMC}
\begin{algorithmic}[1]
\State Set state indicator $k \egal 1\,$
\While{$k\,\leqslant\,N_{\,s}$}
\State Sample the parameter $\p^{\,\star}  \, \in \, \Omega_{\,p}$ from proposal distribution $\mathcal{P}(\,\p^{\,\star} \,\bigl|\,\p^{\,k-1}\,)$ \label{alg:step_proposal_parameter}
\State Compute direct problem $u$ from Eqs.~\eqref{eq:heat_lumped_dimless_wf},\eqref{eq:heat_lumped_dimless_ins}  \label{alg:step_direct_problem}
\State Compute acceptance factor $\beta$ with Eq.~\eqref{eq:acceptance_factor}
\State Sample random value $U \egal \mathcal{U}\bigl(\,\bigl]\,0\,,\,1\,\bigr[\,\bigl)$ \label{alg:step_sample_random_value}
\If{$U \, \leqslant \, \beta$}  \label{alg:step_test_parameter}
\State $\p^{\,k}\egal \p^{\,\star}$
\Else{ }
\State $\p^{\,k}\egal \p^{\,k-1}$
\EndIf 
\State \textbf{end}
\State $k \egal k \plus 1$ 
\EndWhile 
\State \textbf{end}
\end{algorithmic}
\end{algorithm}

\subsection{\revision{Approximation error model}}

A lumped model is proposed for the direct problem computation to cut the computational cost of the inverse problem. However, an error in the formulation of the physical problem is induced with this strategy. A possibility to take into account those approximations, is to use the Approximation Error Model (AEM) introduced by \cite{kaipio_statistical_2005,nissinen_bayesian_2007,orlande_accelerated_2014}. The likelihood function is modified by adding a noise in the measurement model:
\begin{align*}
\prob{\hat{u}\,\bigl|\, \p} \, \propto \, \exp 
\Biggl(\, 
\sum_{j \egal 1}^{N_{\,x}} \, \sum_{n \egal 1}^{N_{\,t}} \ \frac{1}{\tilde{\sigma}(\,\chi_{\,j} \,,\,\tau_{\,n}\,)^{\,2}} \, 
\Bigl(\,\hat{u}_{\,j}(\,t^{\,n}\,) \moins u(\,\p \,,\, \chi_{\,j} \,,\, \tau^{\,n} \,)
\moins e(\,\chi_{\,j} \,,\, \tau^{\,n} \,) \,\Bigr)^{\,2}
\,\Biggr) \,,
\end{align*}
where $\tilde{\sigma}$ is the approximated variance as described in \cite{kaipio_statistical_2005,nissinen_bayesian_2007,orlande_accelerated_2014} and $e$ is the mean of the error between the lumped model and the complete model, described in Section~\nameref{sec:complete_formulation}:
\begin{align*}
e \egal \overline{T} \moins T \,.
\end{align*}
Such error is computed before solving the inverse problem considering a sampling according to the prior distribution of the unknown parameters.

\section{Experimental design}
\label{sec:exp_design}

\subsection{Material and methods}

\begin{figure}[h!]
\centering 
\subfigure[\label{fig:insulator}]{\includegraphics[width=.45\textwidth]{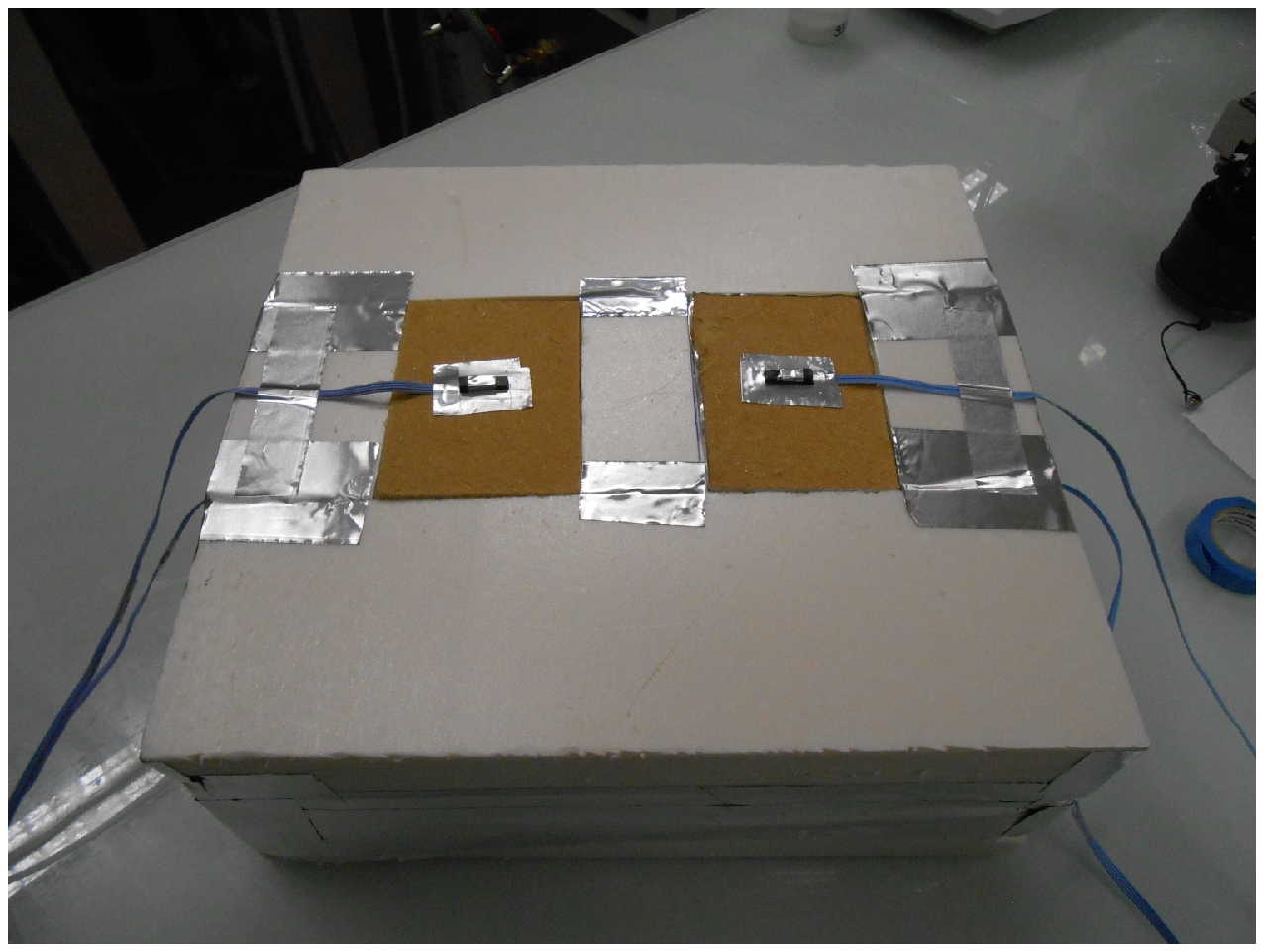}}\hspace{0.2cm}
\subfigure[\label{fig:climatic_chamber}]{\includegraphics[width=.45\textwidth]{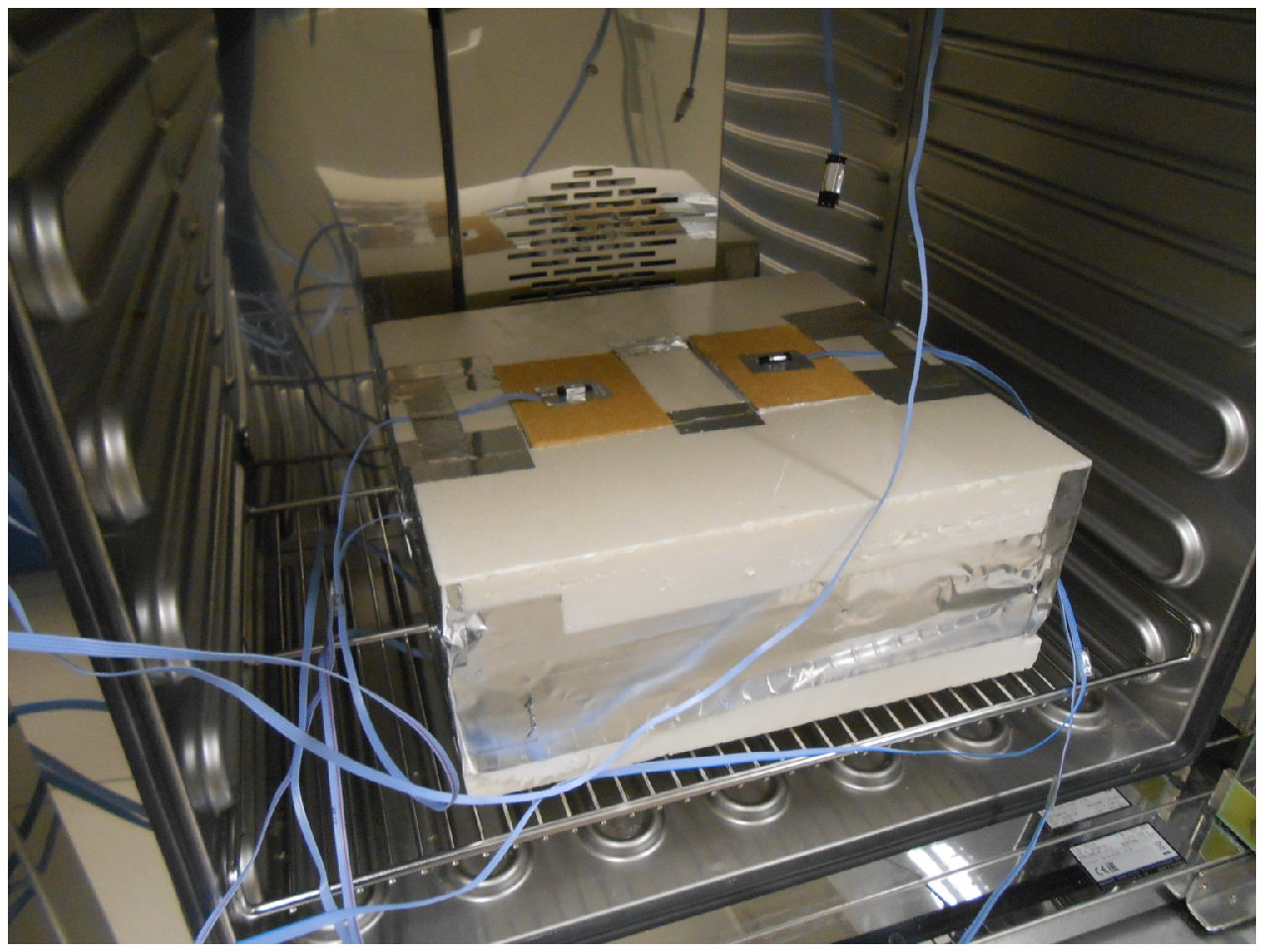}}\\
\subfigure[\label{fig:inside_sensor}]{\includegraphics[width=.45\textwidth]{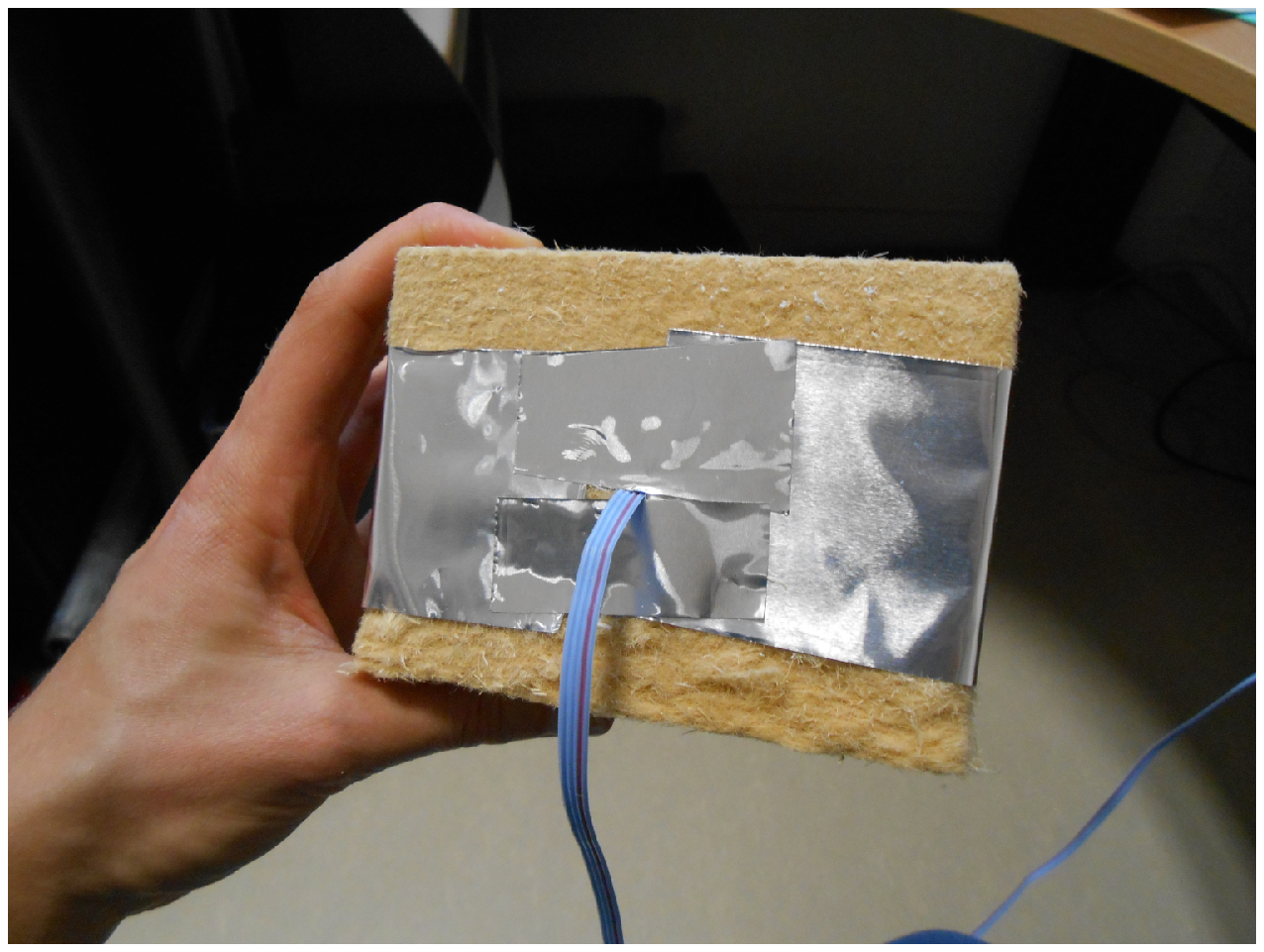}} \hspace{0.2cm}
\subfigure[\label{fig:Ta_ft}]{\includegraphics[width=.45\textwidth]{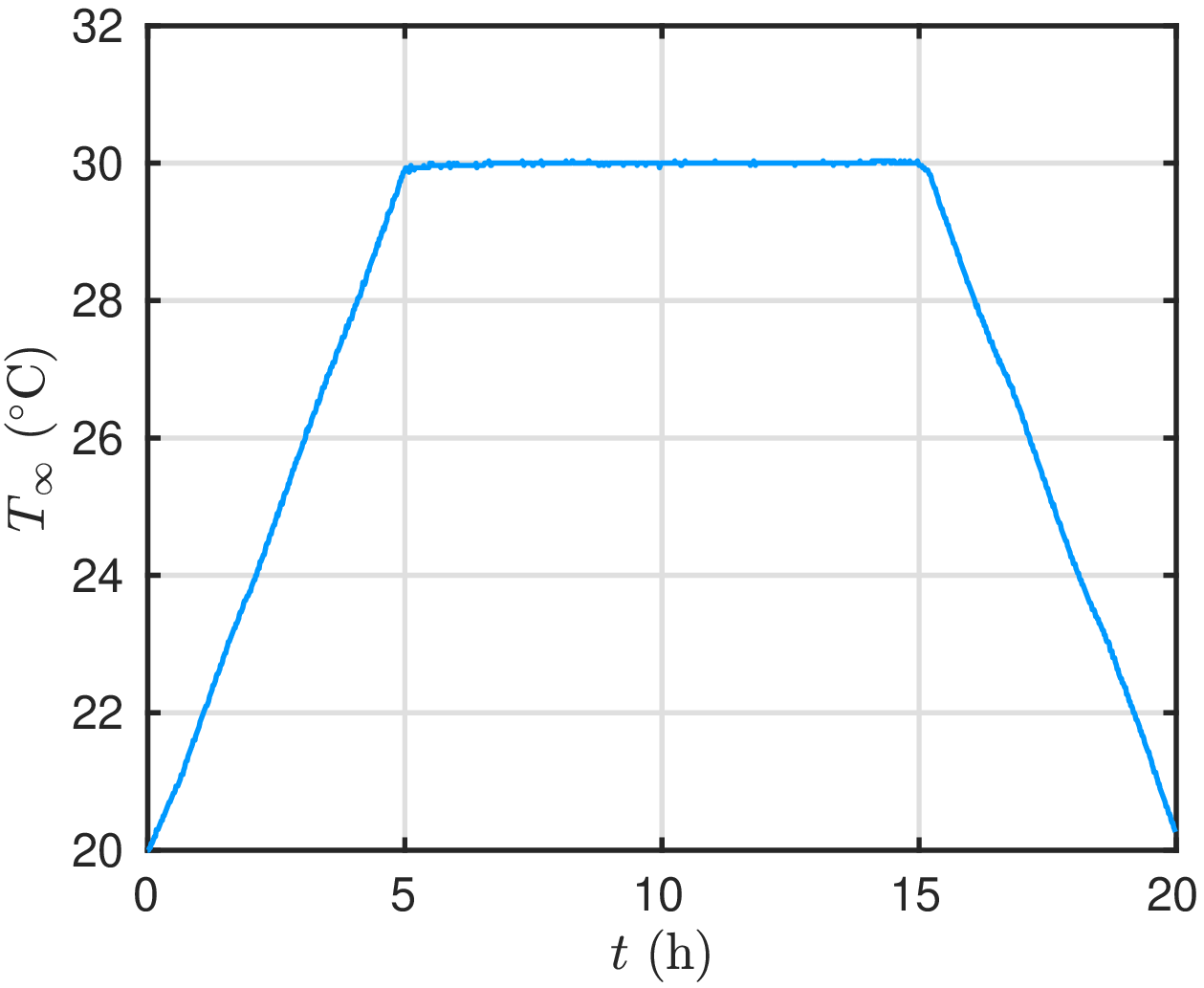}}
\caption{Illustration of the samples inserted in the insulator \emph{(a)}, in the climatic chamber \emph{(b)} and of the insertion of the sensor at $x \egal L_{\,0}\,$ \emph{(c)}. Time variation of the set-up temperature in the climatic chamber \emph{(d)}.}
\end{figure}

To solve the inverse problem, an experimental campaign is carried out. Samples of wood fiber are placed inside a climatic chamber of type \texttt{Memmert} ICP $110$. The surrounding of the wood fiber are put inside a polyurethane insulator mold as illustrated in Figures~\ref{fig:domain} and \ref{fig:insulator}. The contact between the wood fiber and the insulator is not perfect. A gap, measured between $1$ and $3 \ \mathsf{mm}\,$, exists so that convection exchanges occur at the interface with the climatic chamber ambient air. Given the ventilation rate and the mixing of the air, it is assumed that the ambient temperature is equal at top and lateral faces. Differences are very negligible compared to the evaluated measurement uncertainties. Figure~\ref{fig:climatic_chamber} shows a picture of the samples inside the chamber. It is assumed that no phenomenon of conduction in a closed domain occurs. The exposed face of the samples are placed parallel to the ventilator airflow. Furthermore, the laminar boundary layer is assumed of a few centimeters above the samples. The material properties have been determined in \cite{busser_dynamic_2018}. The thermal conductivity are $k_{\,1\,,\,0} \egal 8.5 \e{-3} \ \mathsf{W\,.\,m^{\,-1}\,.\,K^{\,-1}}$ and $k_{\,1\,,\,1} \egal 3.17 \e{-2} \ \mathsf{W\,.\,m^{\,-1}\,.\,K^{\,-1}}\,$. The volumetric heat capacity coefficients are $c_{\,1\,,\,0} \egal -2.77 \e{5}\ \mathsf{W\,.\,m^{\,-3}\,.\,K^{\,-1}}\,$ $c_{\,1\,,\,1} \egal 4.7 \e{5}\ \mathsf{W\,.\,m^{\,-3}\,.\,K^{\,-1}}\,$, valid for building temperature range. The temperature constant is $T_{\,0} \egal 20 \ \mathsf{\degC}\,$. The dimensions of the wood fiber material is $8 \ \mathsf{cm}\,$. 
The insulator and aluminum have temperature invariant material properties: $k_{\,2\,,\,0} \egal 4 \e{-3} \ \mathsf{W\,.\,m^{\,-1}\,.\,K^{\,-1}}\,$, $k_{\,3\,,\,0} \egal 240 \ \mathsf{W\,.\,m^{\,-1}\,.\,K^{\,-1}}\,$, $c_{\,2\,,\,0} \egal 7 \e{4}\ \mathsf{W\,.\,m^{\,-3}\,.\,K^{\,-1}}\,$, and $c_{\,3\,,\,0} \egal 2.37 \e{6}\ \mathsf{W\,.\,m^{\,-3}\,.\,K^{\,-1}}\,$. Their respective lengths are $8 \ \mathsf{cm}$ and $0.1 \ \mathsf{mm}\,$. The dimension of the whole domain are $L_{\,0\,,\,x} \egal 16 \ \mathsf{cm}$ and $L_{\,0\,,\,y} \egal L_{\,0\,,\,z} \egal 8  \ \mathsf{cm}\,$.

The chamber is programmed as follows. A first period of $24 \ \mathsf{h}$ occurs to ensure a homogeneous initial temperature at  $T_{\,0} \egal 20 \ \mathsf{\degC}\,$. Then, during the first phase, the temperature increases slowly from $20$ to $30 \ \mathsf{\degC}$ at a rate of $2 \ \mathsf{\degC\,.\,h^{\,-1}}$ during the time interval $\bigl[\, 0 \,,\, 5 \,\bigr] \ \mathsf{h}\,$. The temperature is then maintained at $30 \ \mathsf{\degC}$ within the time interval $\bigl[\, 5 \,,\, 15\,\bigr]\ \mathsf{h}\,$. During the last phase $\bigl[\, 15 \,,\, 20 \,\bigr]\ \mathsf{h}\,$, the temperature decreases to $20 \ \mathsf{\degC}$ at a rate $-\,2 \ \mathsf{\degC\,.\,h^{\,-1}}\,$. The imposed boundary conditions is illustrated in Figure~\ref{fig:Ta_ft}. With the issue of characterizing the climatic chamber, four configurations are performed according to different values of the maximal rate of the chamber ventilator $\bigl\{\,0.1 \,,\, 0.4 \,,\, 0.7 \,,\, 1 \,\bigr\} \cdot v_{\,\max}\,$. 

To obtain the experimental observation, two \texttt{Sensirion}\texttrademark ~SHT sensors are used ($N_{\,x} \egal 2$). One is placed at the interface between the material and the ambient air at $x_{\,1} \egal 0\,$. To ensure a perfect contact, a thermal paste is used. A complementary sensor is set in the middle of the material at $x_{\,2} \egal 0.5 \ L_{\,0\,,\,x}\,$. For this, the material is cut in two parts to set the sensor. Then, the material is sealed using an aluminum tape as shown in Figure~\ref{fig:inside_sensor}. Note that the dimensionless sensor positions are denoted by $\chi_{\,1}$ and $\chi_{\,2}\,$, respectively. The ambient air temperature is recorded twice, by the chamber sensor and by an additional SHT sensor. All sensors have been calibrated previous to the experimental campaign. The time step between two measurements is $1 \ \mathsf{min}\,$.

\begin{figure}[h!]
\centering 
\subfigure[\label{fig:phi_ft}]{\includegraphics[width=.45\textwidth]{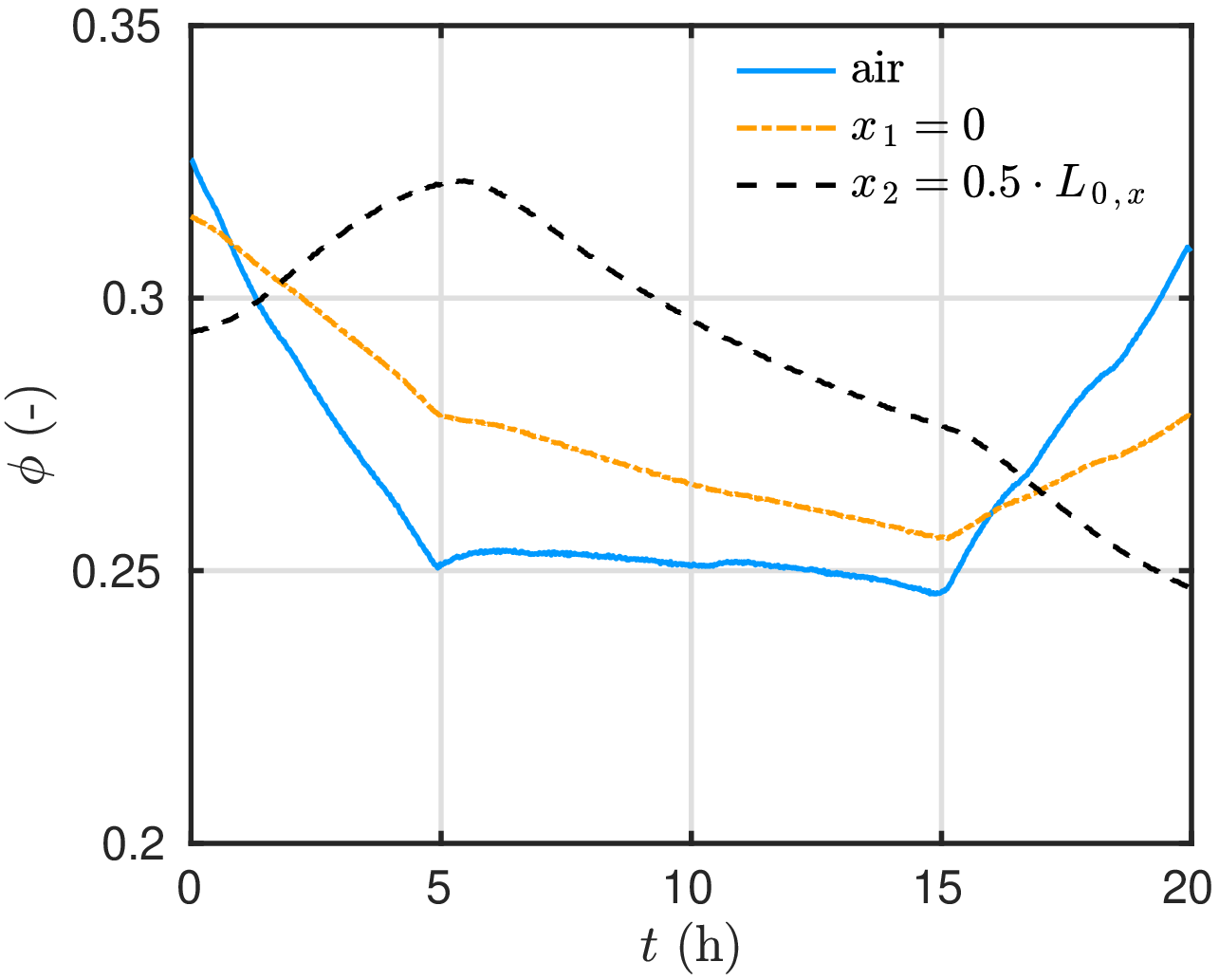}} \hspace{0.2cm}
\subfigure[\label{fig:w_ft}]{\includegraphics[width=.45\textwidth]{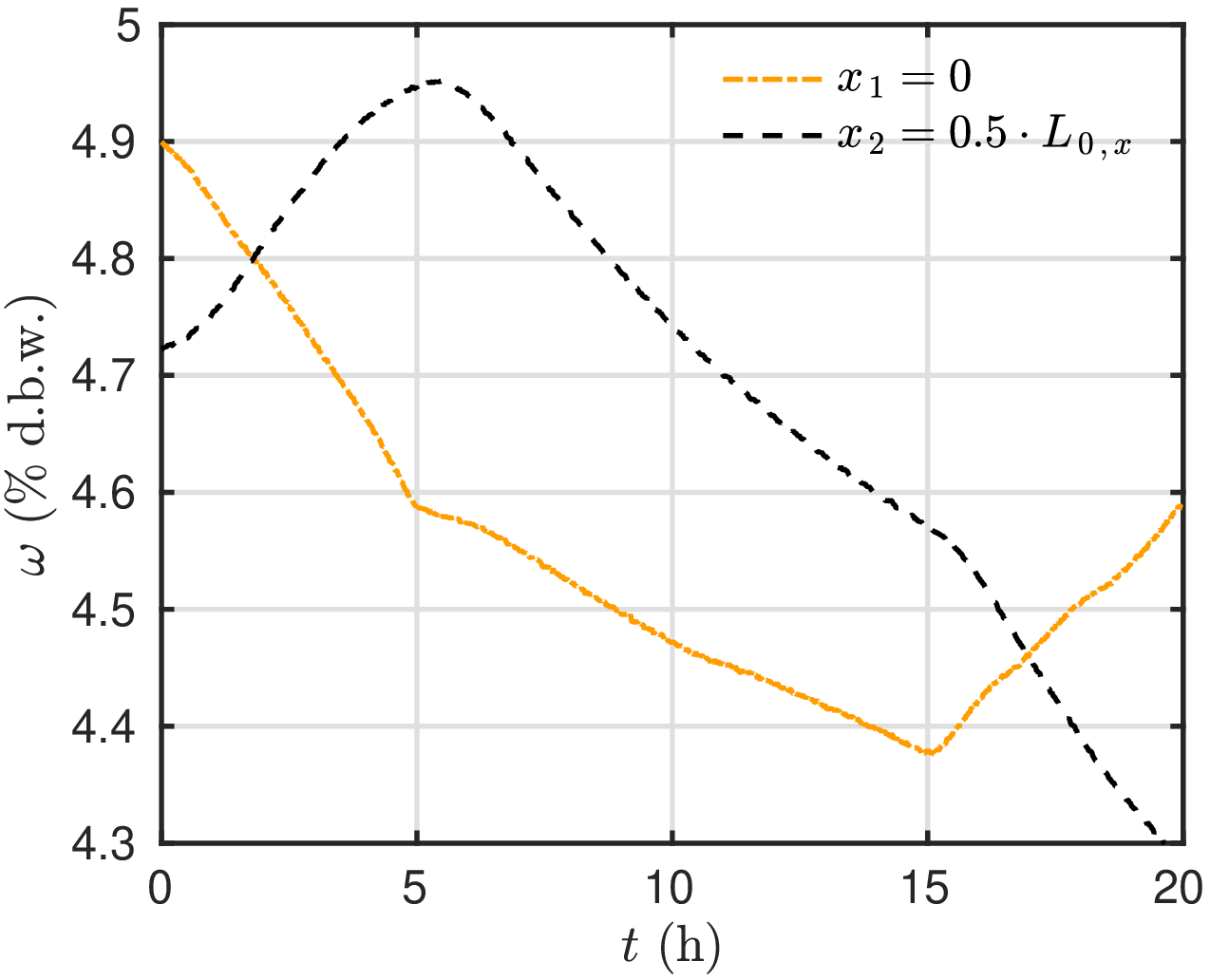}} 
\caption{Time variation of the relative humidity \emph{(a)} and moisture content \emph{(b)} in the climatic chamber and at the two sensor positions.}
\end{figure}

Last, to verify that the heat transfer is not influenced by the latent effects, the relative humidity $\phi \ \unit{-}$ is monitored in the climatic chamber and in the material. The moisture content $\omega \ \unit{-}$ can be evaluated using the adsorption curve from \cite{busser_dynamic_2018}. Figures~\ref{fig:phi_ft} and \ref{fig:w_ft} present the time variation of the relative humidity and the dry basis weight (d.b.w.) moisture content, respectively. The relative humidity remains stable around $0.3$ during the whole experiment. The moisture content in the material is lower than $5 \, \%$ and has a relative variation lower than $10\, \%\,$. From this analysis, it can be assumed that the latent heat transfer is negligible in the material.

\subsection{Measurement uncertainties}

\begin{figure}[h!]
\centering 
\includegraphics[width=.6\textwidth]{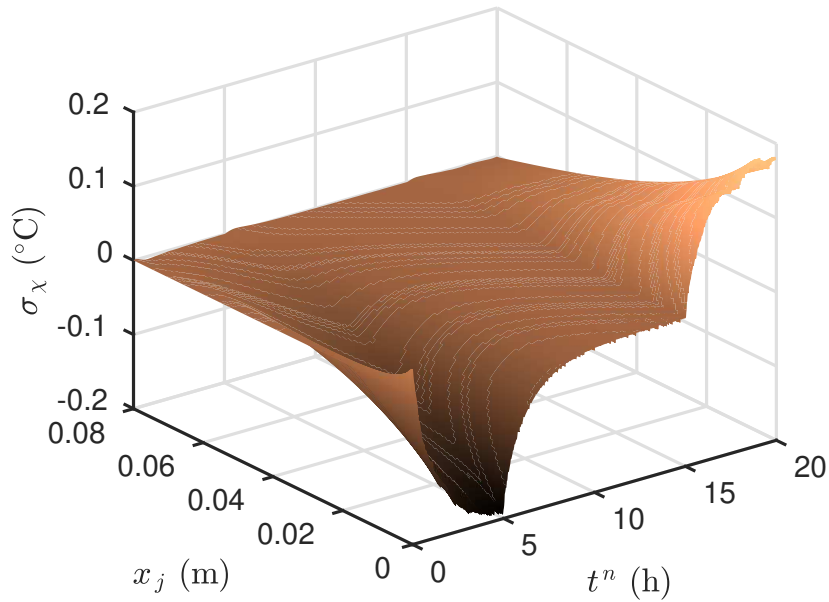}
\caption{\revision{Variation of the measurement uncertainty due to sensor position according to space and time.}}
\label{fig:sigT_fxt}
\end{figure}

For each configuration, the campaign is repeated $N_{\,e} \egal 3$ times to decrease the random aspect of the measurement uncertainties. Thus, the best estimates of the temperature measurement in the material is given by \cite{taylor_introduction_2008}:
\begin{align*}
\hat{T}_{\,j}(\,t^{\,n}\,) \, \eqdef \, \frac{1}{N_{\,e}} \, \sum_{i \egal 1}^{N_{\,e}} \, \hat{T}_{\,\obs\,,\,i}(\,x_{\,j}\,,\,t^{\,n}\,) \,, \qquad j \, \in \, \bigl\{\, 1 \,,\, \ldots \,,\, N_{\,x} \,\bigr\} \,, \qquad n \,\in\, \bigl\{\, 1 \,,\,\ldots \,,\, N_{\,t}\,\bigr\} \,,
\end{align*}
where $\hat{T}_{\,\obs\,,\,i}$ is the temperature measurement obtained from the sensor at position $x_{\,j}\,$. The total uncertainty $\sigma \ \unit{K}$ of the temperature measurement is evaluated through a complete propagation of each uncertainties:
\begin{align*}
\sigma(\,x_{\,j}\,,\,t^{\,n}\,) \, \eqdef \,  \sqrt{\sigma_{\,s}^{\,2} \plus \sigma_{\,\sim}^{\,2}(\,x_{\,j}\,,\,t^{\,n}\,) \plus  \sigma_{\,\chi}^{\,2}(\,x_{\,j}\,,\,t^{\,n}\,) } \,,
\end{align*}
where $\sigma_{\,s} \egal 0.3 \ \mathsf{\degC}$ is the sensor measurement uncertainty. The random part $\sigma_{\,\sim}$ of the uncertainty:
\begin{align*}
\sigma_{\,\sim}(\,x_{\,j}\,,\,t^{\,n}\,) \, \eqdef \, \frac{1}{\sqrt{N_{\,e}}} \ 
\sqrt{ 
\frac{1}{N_{\,e}} \, \sum_{i \egal 1}^{N_{\,e}} \, \Bigl(\, \hat{T}_{\,\obs\,,\,i}(\,x_{\,j}\,,\,t^{\,n}\,) - \hat{T}_{\,j}(\,t^{\,n}\,)  \,\Bigr)^{\,2}
 } \,,
\end{align*}
The uncertainty on the sensor position, denoted by $\sigma_{\,\chi}\,$, is evaluated by:
\begin{align*}
\sigma_{\,\chi}(\,x_{\,j}\,,\,t^{\,n}\,) \, \eqdef \, \pd{T}{x}\,\biggl|_{\,x\egal x_{\,j}\,,\,t\egal t^{\,n}} \cdot \, \delta_{\,\chi} \,,
\end{align*}
where $\delta_{\,\chi} \egal 5 \ \mathsf{mm}$ is the position uncertainty and $\displaystyle \pd{T}{x}$ is evaluated using the numerical solution of the governing equation. Figure~\ref{fig:sigT_fxt} shows the variation of the measurement uncertainty due to the sensor position according to space and time. It can be remarked that the uncertainty is higher near the top surface $x \egal 0\,$. It varies according to the boundary condition variation with an absolute maximum of $0.2 \ \mathsf{\degC}\,$.

\section{Results} 
\label{sec:results}

\subsection{\revision{Approximation error model of the lumped formulation}}

\begin{figure}[h!]
\centering 
\subfigure[\label{fig:aem_prior}]{\includegraphics[width=.45\textwidth]{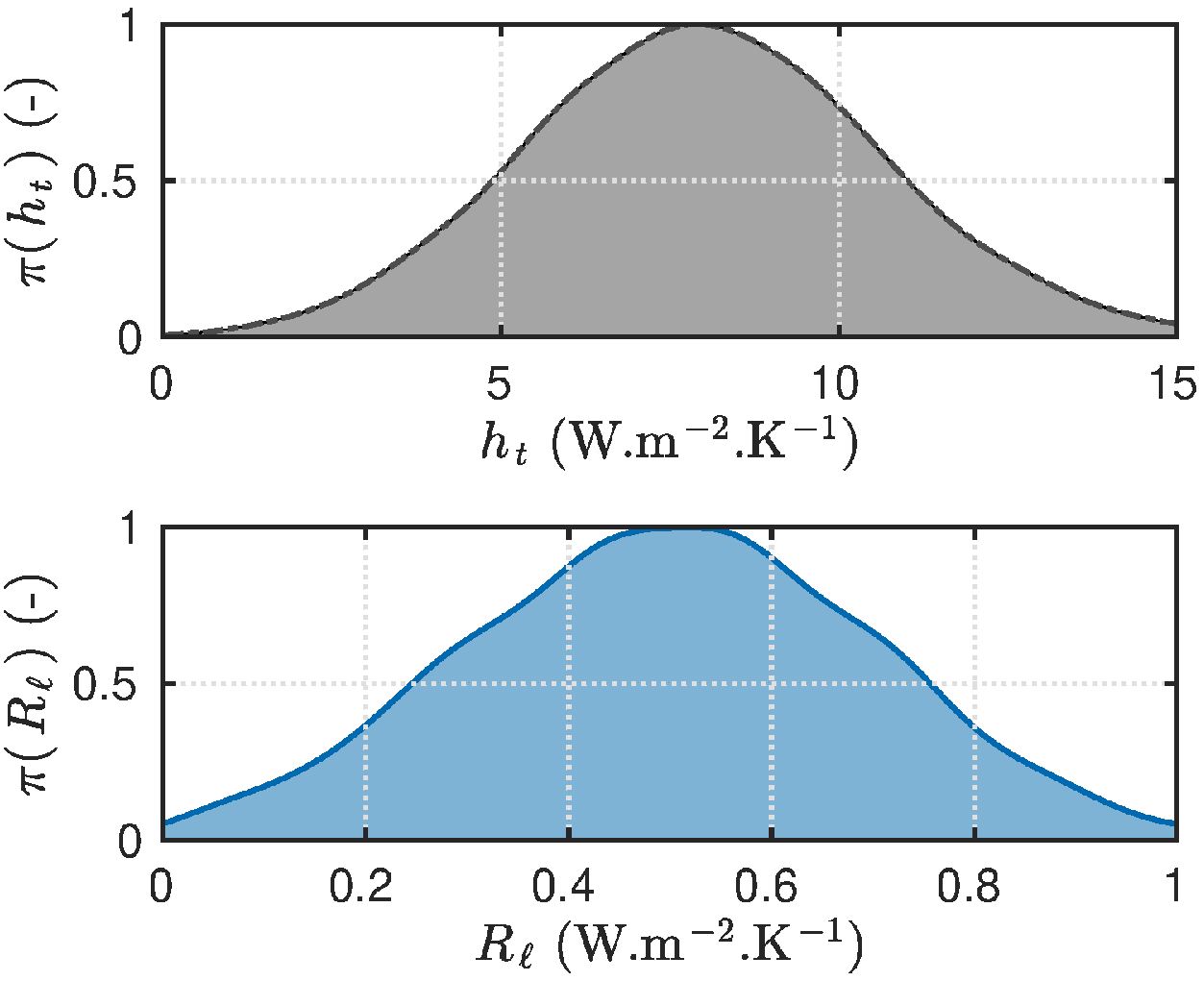}}
\hspace{0.2cm}
\subfigure[\label{fig:aem_error}]{\includegraphics[width=.45\textwidth]{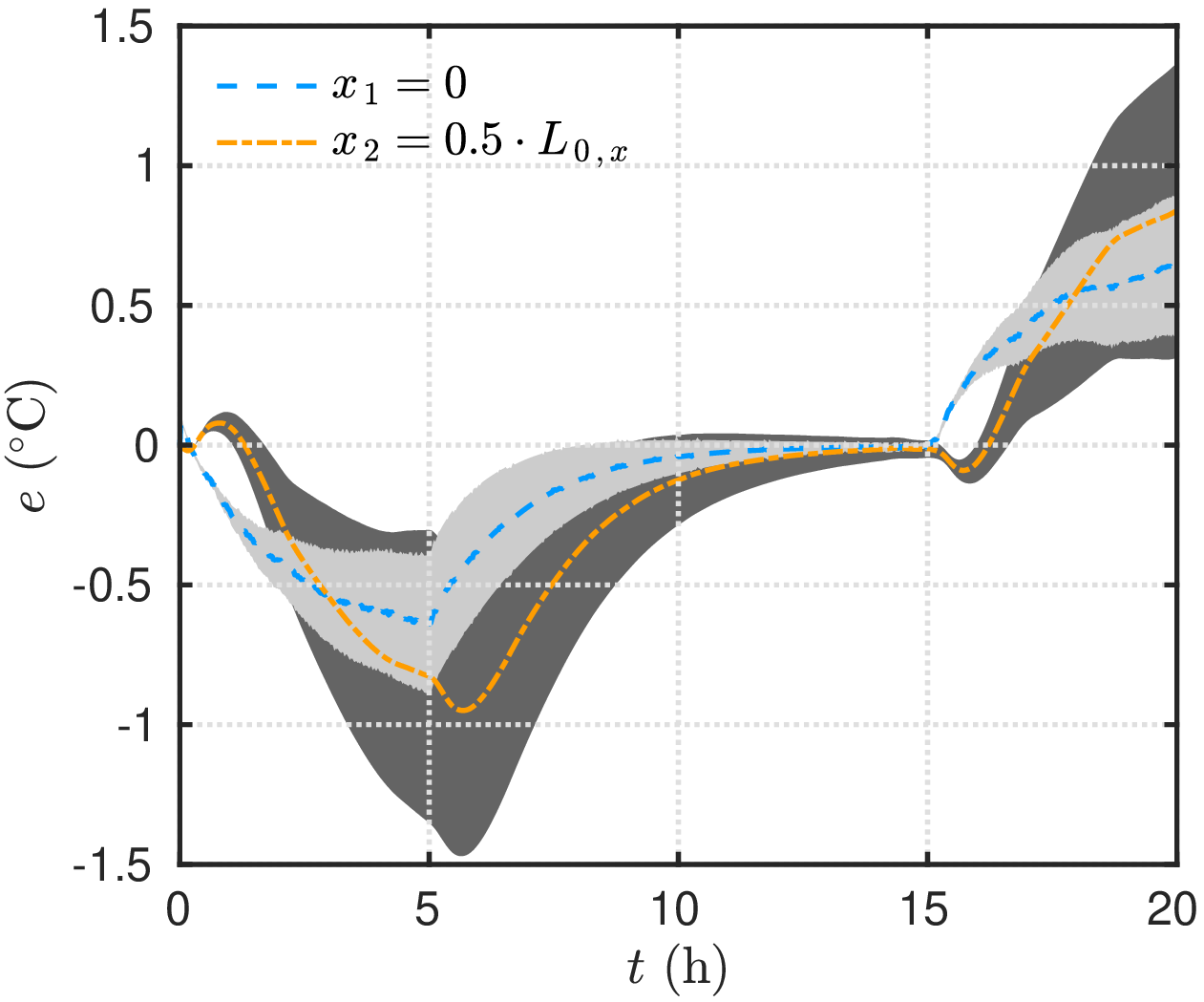}}
\caption{\revision{Prior distribution of the unknown parameters $h_{\,t}$ and $R_{\,\ell}$ \emph{(a)}. Time variation of the mean error} \revision{ and its standard deviation between the lumped and complete model \emph{(b)}.}}
\end{figure}

Before solving the parameter estimation problem, the AEM is first computed. For this, $1000$ samples of the unknown parameters $h_{\,t}$ and $R_{\,\ell}$  are sampled according to the \textsc{G}aussian prior distributions $\mathcal{N}(\,0.5\,,\,0.2\,)$ and $\mathcal{N}(\,8\,,\,2.5\,)\,$, respectively. The distributions are illustrated in Figure~\ref{fig:aem_prior}. The time evolution of the mean error for both sensor position is shown in Figure~\ref{fig:aem_error}. The error is higher for the sensor placed inside the material. It scales with $1 \mathsf{\degC}$ at its maximum, corresponding with the changes in the boundary conditions. During the first phase $t \, \leqslant \, 5 \ \mathsf{h}\,$, the error is negative indicating that the lumped model underestimates the predictions of the complete one. For the last phase $t \, \geqslant \, 15 \ \mathsf{h}\,$, it inverses. Those preliminary computations enable to build a lumped model, with a lower computational cost and with an error model compared to the complete formulation.

\subsection{Structurally Globally Identifiability}

First, the theoretical identifiability of the parameter is demonstrated using the dimensionless representation of the problem. It is assumed that two sets of temperature observations are obtained, denoted $\hat{u}$ and $\widehat{\widetilde{u}}$, corresponding to two sets of parameters $\p \egal \Bigl(\, \Bi_{\,t} \,,\, \Bi_{\,\ell} \,\Bigr)$ and $\widetilde{\p} \egal \Bigl(\, \widetilde{\Bi}_{\,t} \,,\, \widetilde{\Bi}_{\,\ell} \,\Bigr)\,$, respectively. The two observations verifies the governing equation~\eqref{eq:heat_lumped_dimless_wf}. Thus, for the first observation at the sensor $\chi_{\,2} \,$, we have: 
\begin{align}
\label{eq:sgi_gov_eq}
\revision{ (\,1 \plus }\zeta_{\,1\,,\,1} \, \hat{u}\,)  \cdot \pd{\hat{u}}{\tau} \egal 
\Fo_{\,1} \ \pd{}{\chi} \, \biggl(\, (\,1 \plus \kappa_{\,1\,,\,1} \, \hat{u}\,) \cdot \pd{\hat{u}}{\chi} \,\biggr) 
\plus \Bi_{\,\ell} \ r \ \Fo_{\,1}  \  \bigl(\, u_{\,\infty} \moins \hat{u} \,\bigr) \,,
\end{align}
and for the second:
\begin{align}
\label{eq:sgi_gov_eq_tilde}
\revision{ (\,1 \plus} \zeta_{\,1\,,\,1} \, \widehat{\widetilde{u}}\,) \cdot \pd{\widehat{\widetilde{u}}}{\tau} \egal 
\Fo_{\,1} \ \pd{}{\chi} \, \biggl(\, (\,1 \plus \kappa_{\,1\,,\,1} \, \widehat{\widetilde{u}}\,) \cdot \pd{\widehat{\widetilde{u}}}{\chi} \,\biggr) 
\plus \widetilde{\Bi}_{\,\ell} \ r \ \Fo_{\,1} \  \bigl(\, u_{\,\infty} \moins \widehat{\widetilde{u}} \,\bigr)  \,.
\end{align}
Now, it is assumed that $\hat{u} \, \equiv \, \widehat{\widetilde{u}}$ and so $ \displaystyle \pd{\hat{u}}{\tau} \, \equiv \, \pd{\widehat{\widetilde{u}}}{\tau}$ and $ \displaystyle  \pd{\hat{u}}{\chi} \, \equiv \, \pd{\widehat{\widetilde{u}}}{\chi}$ since they are  supposed independent. In addition, the operation Eq.~\eqref{eq:sgi_gov_eq} minus Eq.~\eqref{eq:sgi_gov_eq_tilde} is carried to obtain:
\begin{align}
\label{eq:sgi_bi_ell}
0 \egal 
- \, \bigl(\, \Bi_{\,\ell} \moins \widetilde{\Bi}_{\,\ell} \,\bigr) \ r \ \Fo \   \hat{u} \,.
\end{align}
The unique solution to Eq.~\eqref{eq:sgi_bi_ell} is $\Bi_{\,\ell} \, \equiv \, \widetilde{\Bi}_{\,\ell}$ and the parameter is SGI. For the second parameter, the second measurement at the sensor position $\chi_{\,1}$ is used. Using the dimensionless formulation of the boundary condition Eq.~\eqref{eq:bc_lumped_dimless} at $\chi \egal 0\,$, we have: 
\begin{align}
\label{eq:sgi_bc}
\kappa_{\,1}(\,\hat{u}\,) \cdot \pd{\hat{u}}{\chi} & \egal \Bi_{\,t} \ \bigl(\, \hat{u} \moins u_{\,\infty} \,\bigr)  \,,
\end{align}
and
\begin{align}
\label{eq:sgi_bc_tilde}
\kappa_{\,1}(\,\widehat{\widetilde{u}}\,) \cdot \pd{\widehat{\widetilde{u}}}{\chi} & \egal \widetilde{\Bi}_{\,t} \ \bigl(\, \widehat{\widetilde{u}} \moins u_{\,\infty} \,\bigr) \,.
\end{align}
Similarly, by assuming $\hat{u} \, \equiv \, \widehat{\widetilde{u}}$ and $ \displaystyle  \pd{\hat{u}}{\chi} \, \equiv \, \pd{\widehat{\widetilde{u}}}{\chi}$, performing the operation Eq.~\eqref{eq:sgi_bc} minus Eq.~\eqref{eq:sgi_bc_tilde} yields to:
\begin{align*}
0 \egal \bigl(\, \Bi_{\,t} \moins \widetilde{\Bi}_{\,t} \, \bigr) \ \hat{u} \,.
\end{align*}
Therefore, $\Bi_{\,t} \, \equiv \, \widetilde{\Bi}_{\,t}$ and the parameter is SGI.

\subsection{\revision{Practical identifiability}}
\label{sec:practical_identifiability}

\begin{figure}[h!]
\centering 
\subfigure[\label{fig:TetaS_ft}]{\includegraphics[width=.45\textwidth]{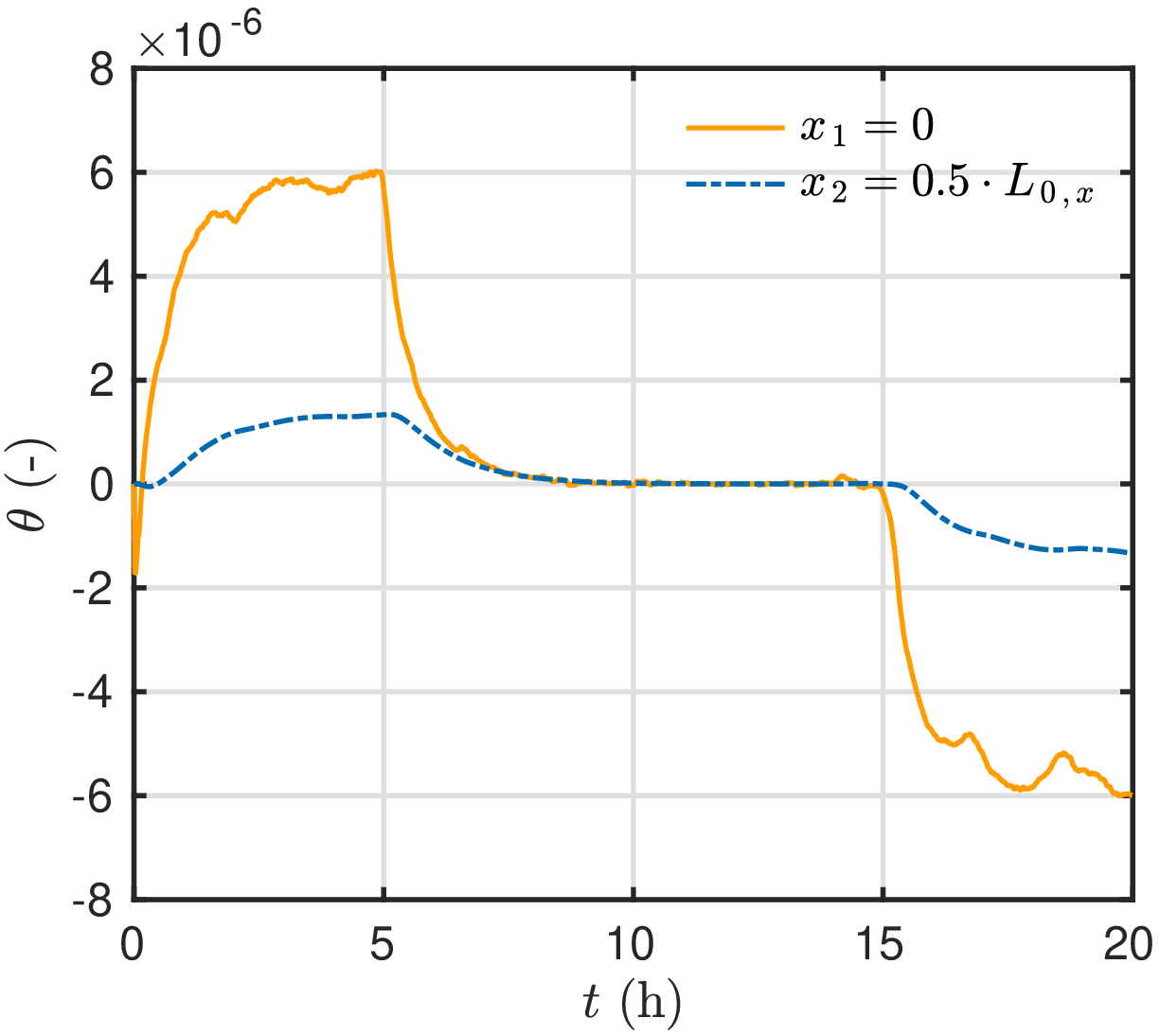}}
\hspace{0.2cm}
\subfigure[\label{fig:PsiS_ft}]{\includegraphics[width=.45\textwidth]{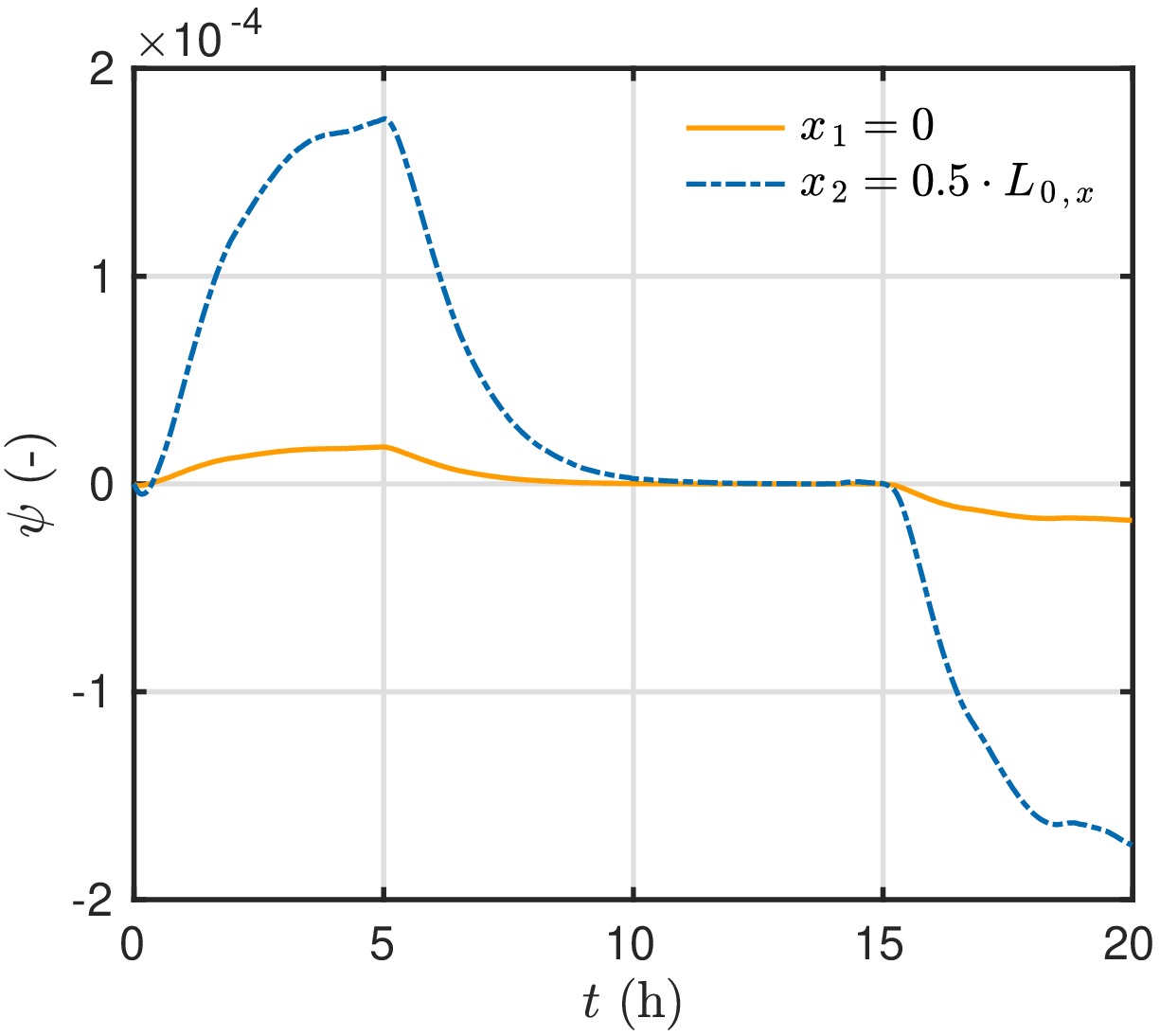}} 
\caption{\revision{Time variation of the sensitivity functions of $h_{\,t}$ \emph{(a)} and $R_{\,\ell}$ \emph{(b)} computed with the \emph{a priori} coefficients.}}
\end{figure}

Before estimating the unknown parameters $h_{\,t}$ and $R_{\,\ell}\,$, their practical identifiability is evaluated by computing the two sensitivity functions. The following \emph{a priori} parameters are used $h_{\,t}^{\,\apr} \egal 8 \ \mathsf{W\,.\,m^{\,-2}\,.\,K^{\,-1}} $ and $R_{\,\ell}^{\,\apr} \egal 0.5 \ \mathsf{W\,.\,m^{\,-2}\,.\,K^{\,-1}} \,$, corresponding to the mean of the prior distribution Eq.~\ref{fig:aem_prior}. The results are shown in Figures~\ref{fig:TetaS_ft} and \ref{fig:PsiS_ft} for both sensor positions $x_{\,1} \egal 0 \ \mathsf{cm}$ and  $x_{\,2}\egal 4 \ \mathsf{cm}\,$. The magnitude of the sensitivity function $\theta$ is higher at the position $x_{\,1}$ corresponding to the interface of the material with the ambient air. This is consistent with the fact that $\theta$ is the sensitivity function of $h_{\,t}\,$, which is involved in the physical model at the boundary $x \egal 0\,$. Conversely, the parameter $R_{\,\ell}$ appears in the source term of the governing equation. Thus, the order of magnitude is higher about $1$ order at the position $x_{\,2}$ than at $x_{\,1}\,$. 

The correlation coefficient of the sensitivity functions are computed and results are reported in Table~\ref{tab:corr_coeff}. If one deals with a single response of the sensor, at $x_{\,1}$ or $x_{\,2}\,$, for the parameter estimation, then the sensitivity functions are almost linearly \revision{dependent}. The parameter estimation might suffer from accuracy and difficulties of convergence. However, by considering multiple sensors, the correlation coefficient decreases strongly. Therefore, the two unknown parameters are identifiable from a practical point of view.

Note that the uncertainties on the thermophysical properties of the wood fiber are not integrated in the prior density distributions. First, previous works \cite{busser_dynamic_2018,vololonirina_characterization_2014} provide a good \emph{a priori} knowledge on these properties. In addition, the sensitivity coefficients have been computed for the two surface coefficients and for the \textsc{F}ourier number, representing the thermophysical properties of the wood fiber material. So a total of three sensitivity coefficients are computed. Two are given by Eqs.~\eqref{eq:sensitivity_functions}. And one additional for the \textsc{F}ourier number is denoted by $\displaystyle \varphi \egal \pd{u}{\Fo_{\,1}}\,$. As presented in Table~\ref{tab:corr_coeff}, the correlation coefficients between $\psi$ and $\varphi$ is very high. 

\begin{figure}[h!]
\centering 
\subfigure[\label{fig:eta_ht}]{\includegraphics[width=.45\textwidth]{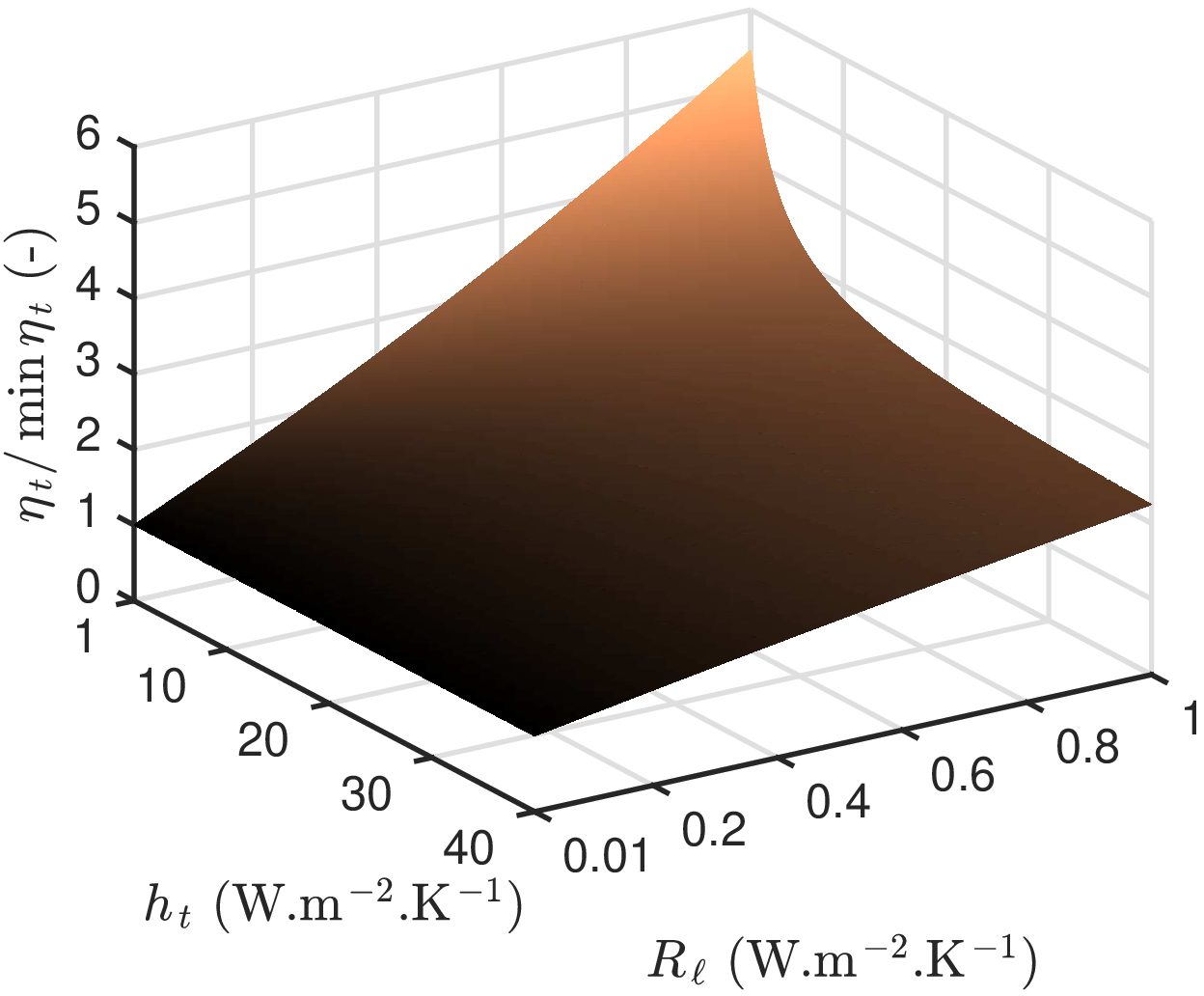}}
\hspace{0.2cm}
\subfigure[\label{fig:eta_hl}]{\includegraphics[width=.45\textwidth]{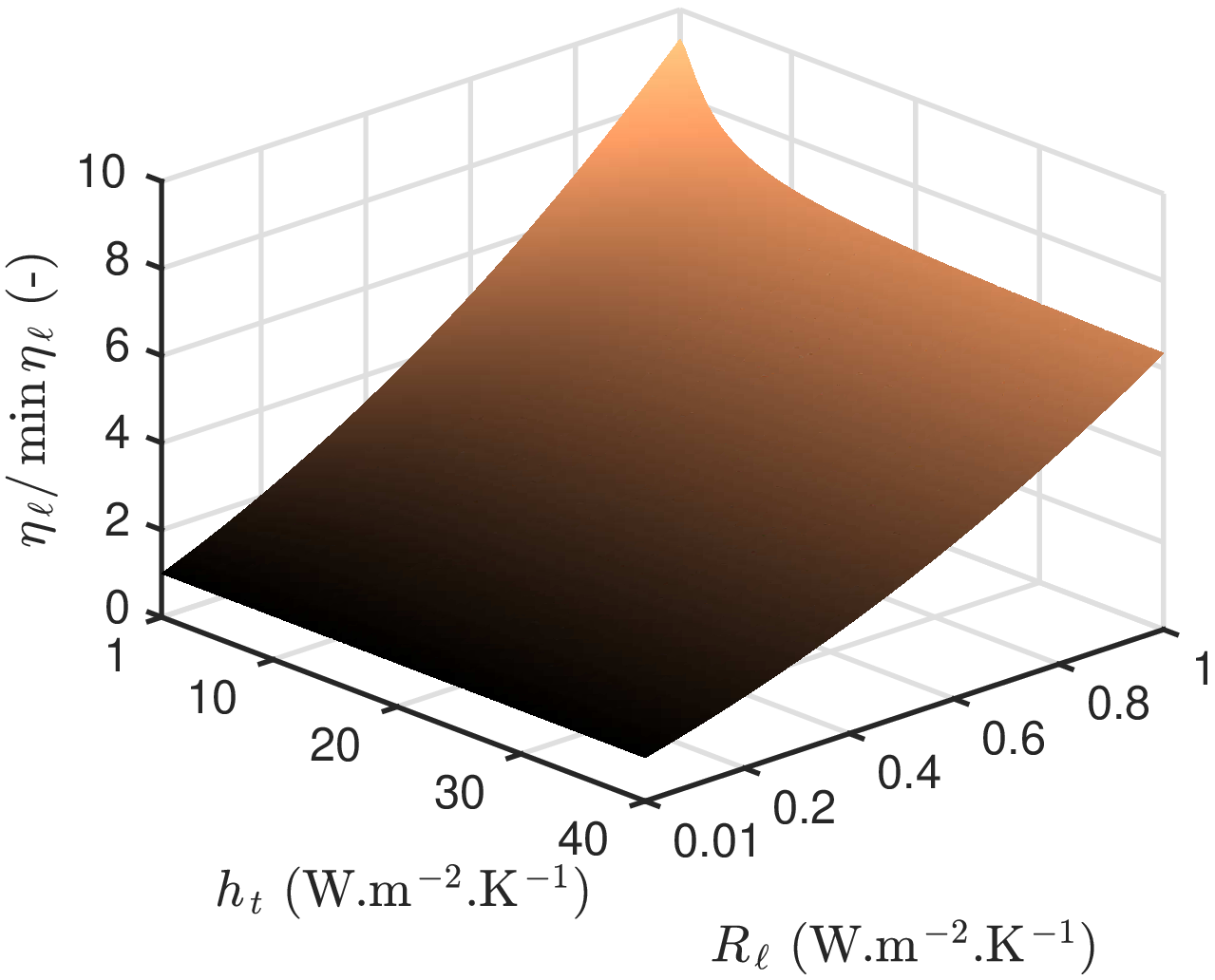}}
\caption{\revision{Variation of the error estimator of parameters $h_{\,t}$ \emph{(a)} and $R_{\,\ell}$ \emph{(b)} according to their domain.}}
\end{figure}

The two error estimators are computed for the domain of variation of the two parameters $\Omega_{\,p} \egal \bigl[\, 1 \,,\, 40 \,\bigr] \, \times \, \bigl[\, 0.01 \,,\, 0.1 \,\bigr] \, \bigl(\,\mathsf{W\,.\,m^{\,-2}\,.\,K^{\,-1}}\bigl)^{\,2}\,$. Results are shown in Figures~\ref{fig:eta_hl} and \ref{fig:eta_ht}. For both coefficients, the error estimators almost not vary with $h_{\,t}$ except for extreme values. A higher variation is observed according to $R_{\,\ell}\,$. These results validate the practical identifiability if the estimated parameters are not very different from the \emph{a priori} parameters.

\begin{table}[h!]
\centering
\caption{Correlation coefficients of the sensitivity functions, computed with the \emph{a priori} coefficients.}
\label{tab:corr_coeff}
\setlength{\extrarowheight}{.5em}
\begin{tabular}[l]{@{} c c}
\hline
\hline
\textit{Sensitivity functions}
& \textit{Correlation coefficient} \\
$\Bigl(\,\theta(\,x_{\,1}\,,\,t\,) \,,\, \psi(\,x_{\,1}\,,\,t\,) \, \Bigr)$
& $0.96$\\
$\Bigl(\,\theta(\,x_{\,2}\,,\,t\,) \,,\, \psi(\,x_{\,2}\,,\,t\,) \, \Bigr)$
& $0.99$ \\
$\Bigl(\,\theta(\,\bigl\{\,x_{\,1}\,,\,x_{\,2} \,\bigr\}\,,\,t\,) \,,\, \psi(\, \bigl\{\,x_{\,1}\,,\,x_{\,2} \,\bigr\}\,,\,t\,) \, \Bigr)$ 
& $0.29$\\
$\Bigl(\,\theta(\,\bigl\{\,x_{\,1}\,,\,x_{\,2} \,\bigr\}\,,\,t\,) \,,\, \varphi(\, \bigl\{\,x_{\,1}\,,\,x_{\,2} \,\bigr\}\,,\,t\,) \, \Bigr)$  & $0.69$ \\
$\Bigl(\,\psi(\,\bigl\{\,x_{\,1}\,,\,x_{\,2} \,\bigr\}\,,\,t\,) \,,\, \varphi(\, \bigl\{\,x_{\,1}\,,\,x_{\,2} \,\bigr\}\,,\,t\,) \, \Bigr)$  & $0.87$ \\
\hline
\hline
\end{tabular}
\end{table}

\subsection{\revision{Parameter estimation results}}

The \textsc{B}ayesian estimation is performed for the four experiments. The total number of state is $N_{\,s} \egal 10^{\,5}\,$. The prior distribution of the unknown parameters is set as a uniform considering the \emph{a priori} bounds given in previous section. The proposal distribution uses a parameters walk $ \boldsymbol{w} \egal \bigl(\, 5 \e{-4} \,,\,  5 \e{-4}\,\bigr)\,$. The maximum and relative tolerances of the time integration solvers are set to $10^{\,-3}\,$. Computations of the direct and sensitivity problems are done considering a uniform mesh of $101$ points.

\subsubsection{Case $10\%\,$, $40\%$ and $70\%$ of $v_{\,\max}\,$}

\begin{figure}[h!]
\centering 
\subfigure[\label{fig:likelihood_fn}]{\includegraphics[width=0.45\textwidth]{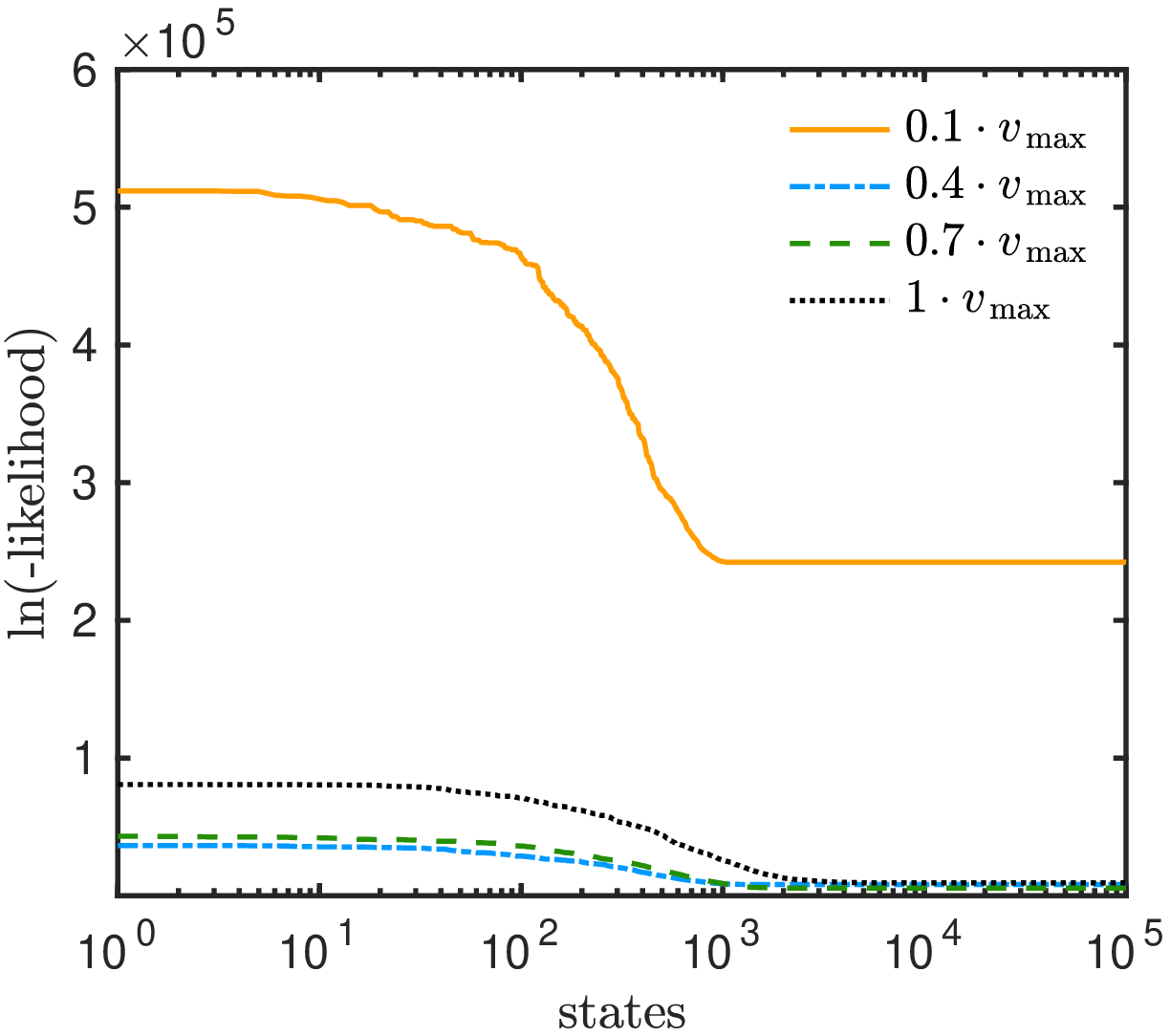}}
\subfigure[\label{fig:accpt_state}]{\includegraphics[width=0.45\textwidth]{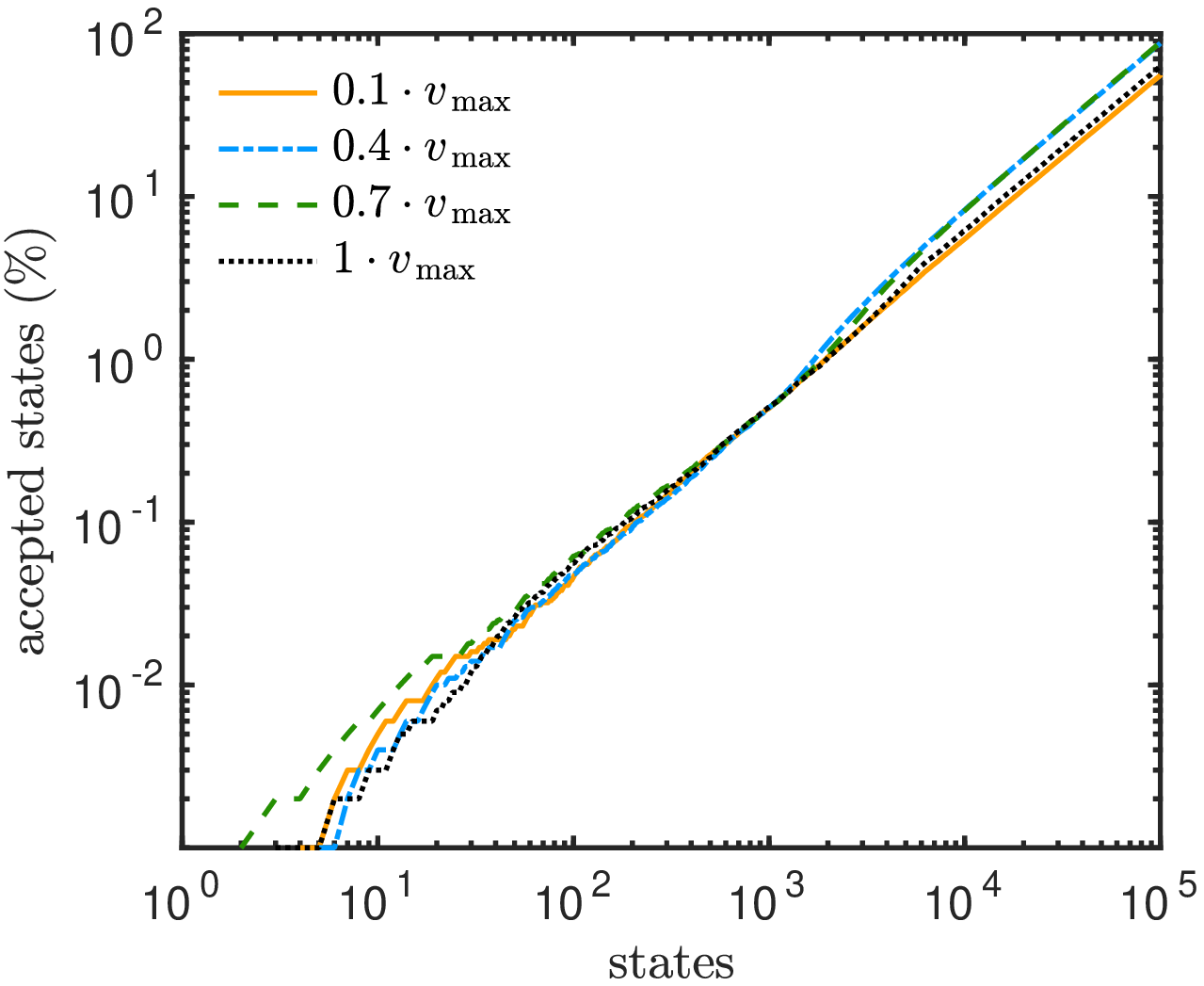}} 
\subfigure[\label{fig:htop_fn}]{\includegraphics[width=0.45\textwidth]{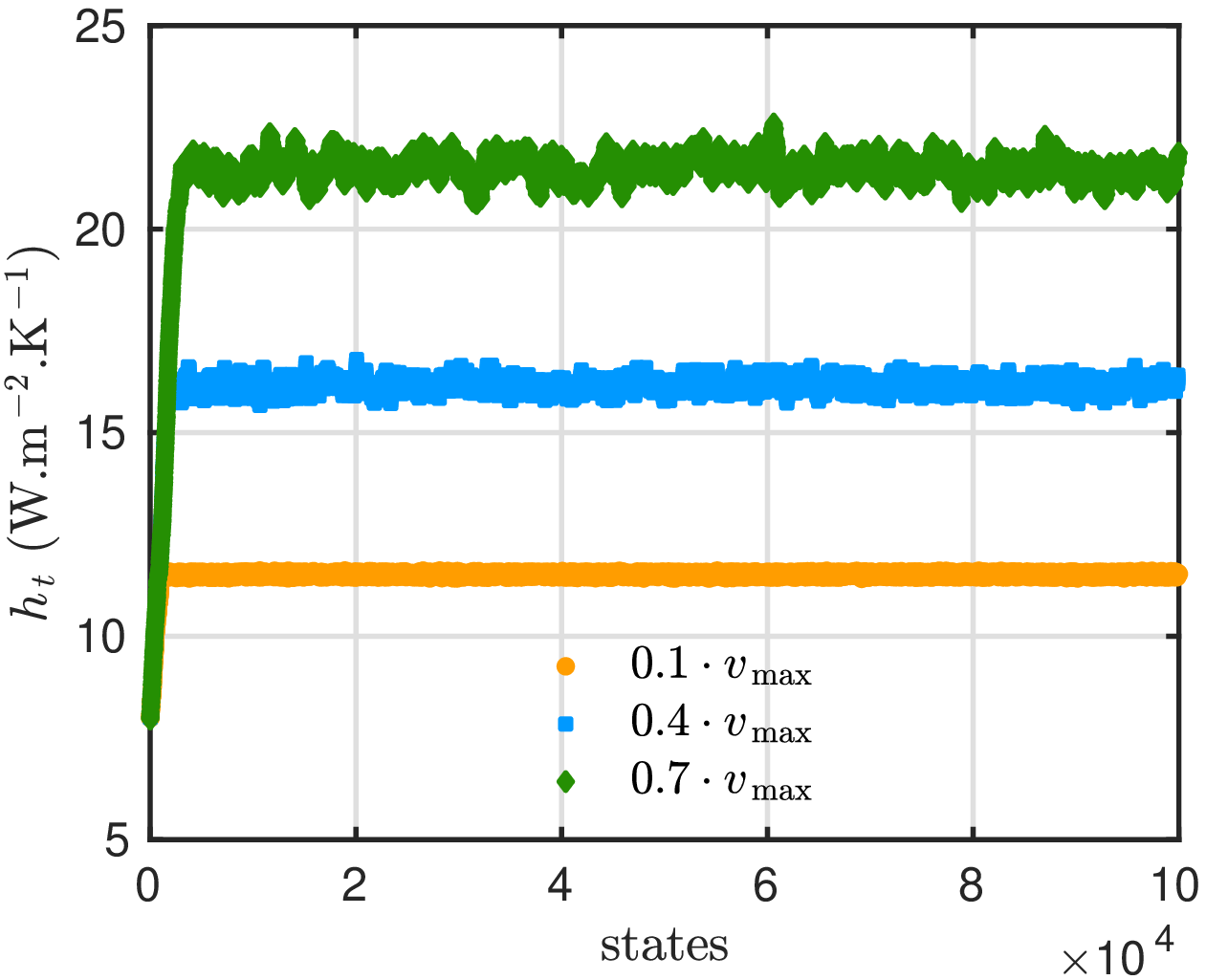}}
\subfigure[\label{fig:hlat_fn}]{\includegraphics[width=0.45\textwidth]{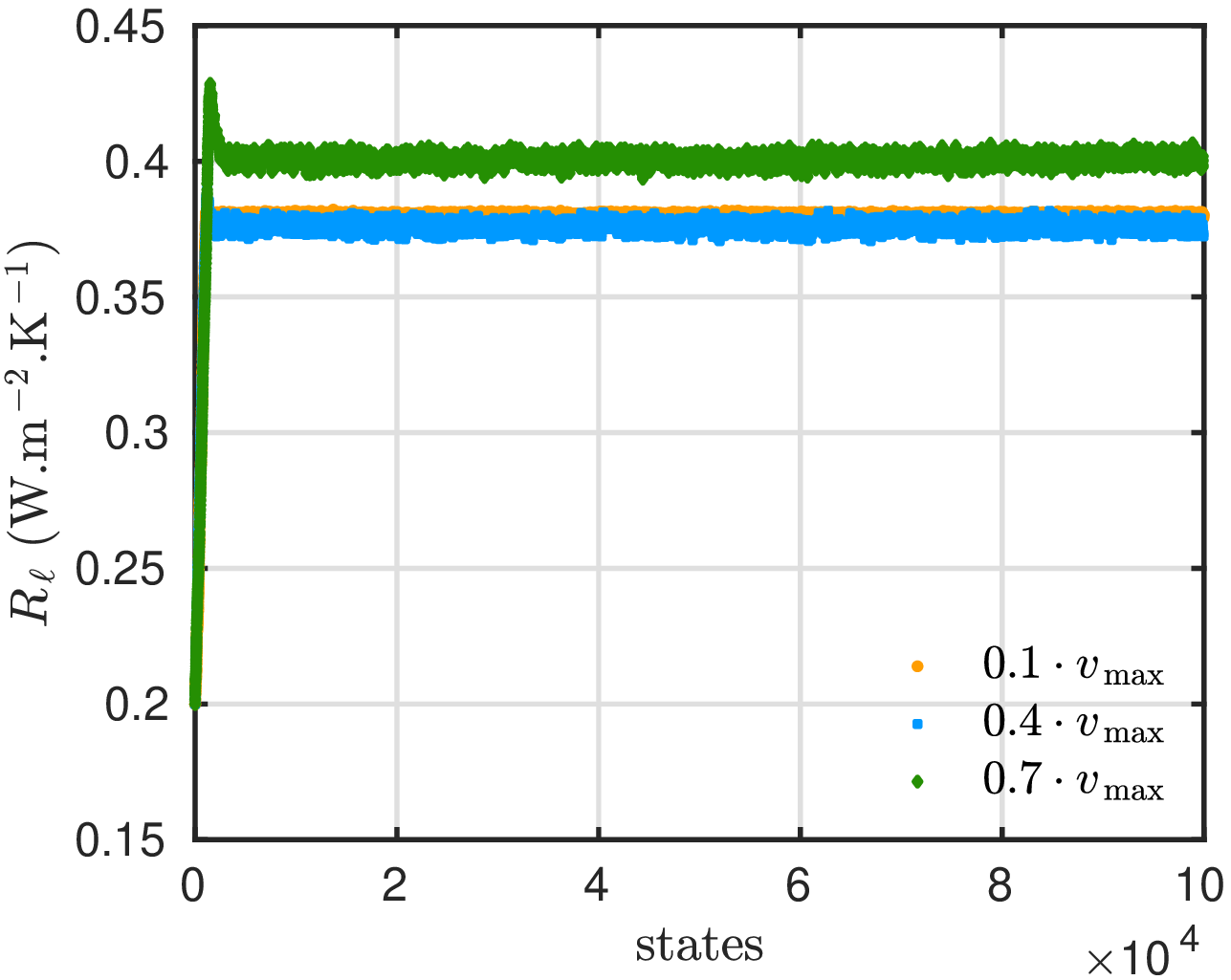}} 
\caption{\revision{Variation of the likelihood \emph{(a)}, of the acceptance rate \emph{(b)} and of the two estimated surface} \revision{ transfer coefficients $h_{\,t}$ \emph{(c)} and $R_{\,\ell}$ \emph{(d)} according to the number of states for the different cases.}}
\end{figure}

First, the \textsc{B}ayesian estimation results are presented for the three experiments $\bigl\{\,0.1 \,,\, 0.4 \,,\, 0.7 \,\bigr\} \cdot v_{\,\max}$. The variation of the likelihood according to the number of states, considering the model with AEM, is shown in Figure~\ref{fig:likelihood_fn}. The burn-in period scales with $N_{\,b} \egal 10^{\,3}$ states. Afterwards, the chains reach an equilibrium. At the end, the acceptance rate, given in Figure~\ref{fig:accpt_state}, is of an order of $80\%\,$. \revision{The variation of the estimated parameters according to the \textsc{M}arkov} Chain is presented in Figures~\ref{fig:htop_fn} and \ref{fig:hlat_fn}. The burn-in period is confirmed. It can also be remarked that the initial state is different from the estimations.

\begin{figure}[h!]
\centering 
\subfigure[\label{fig:pdf_htop_aem}]{\includegraphics[width=.45\textwidth]{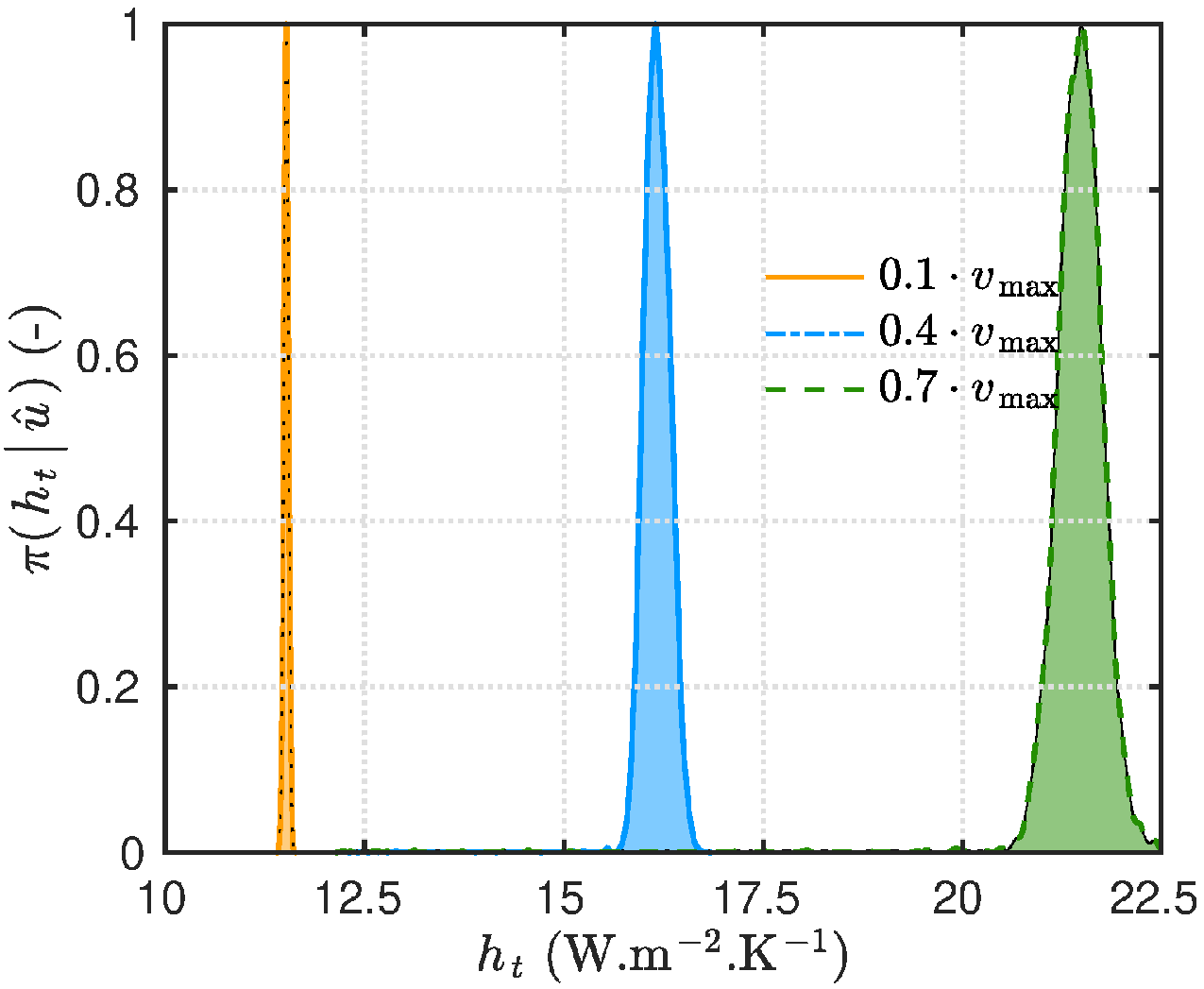}}
\hspace{0.2cm}
\subfigure[\label{fig:pdf_hlat_aem}]{\includegraphics[width=.45\textwidth]{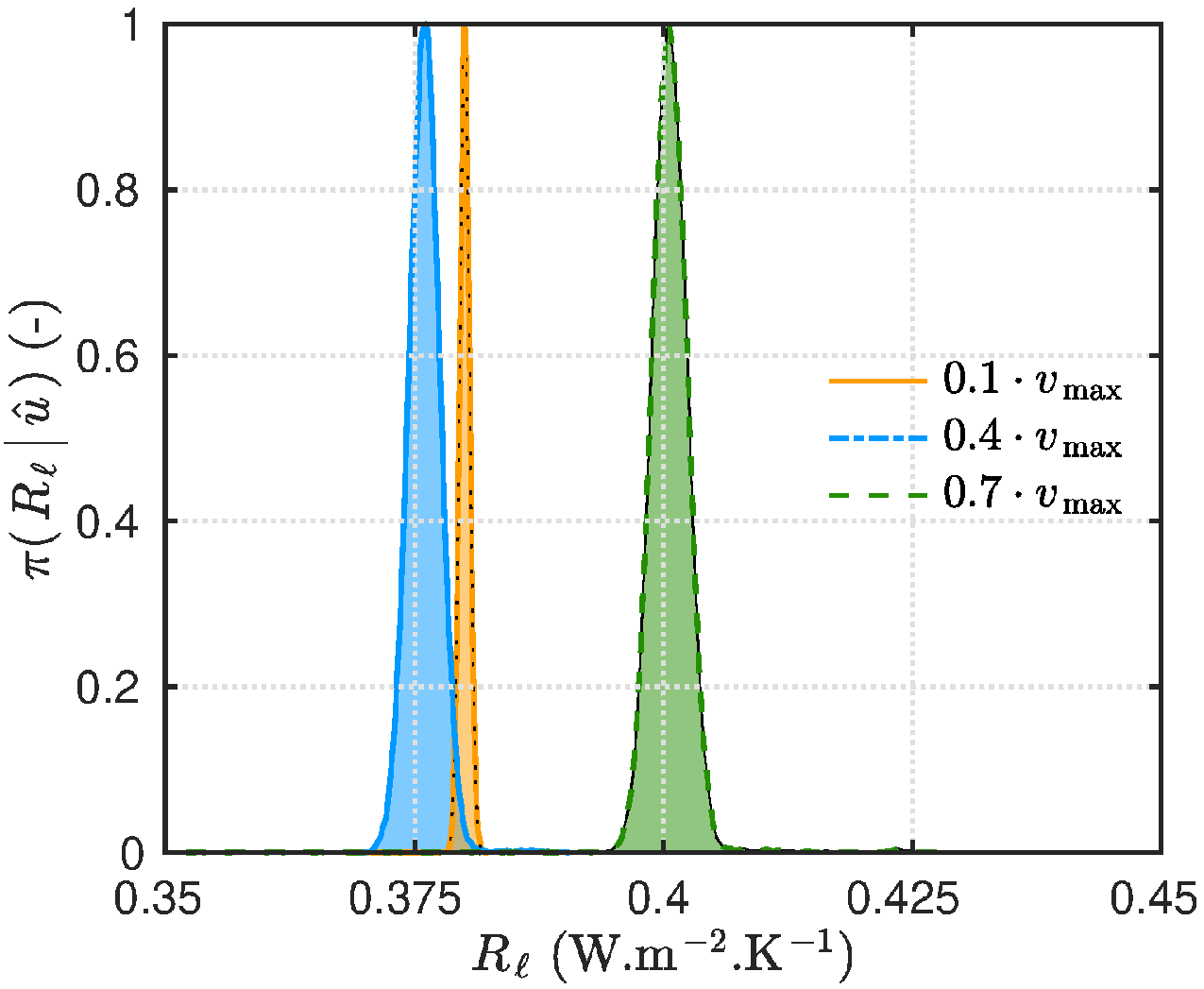}} \\
\subfigure[\label{fig:pdf_htop}]{\includegraphics[width=.45\textwidth]{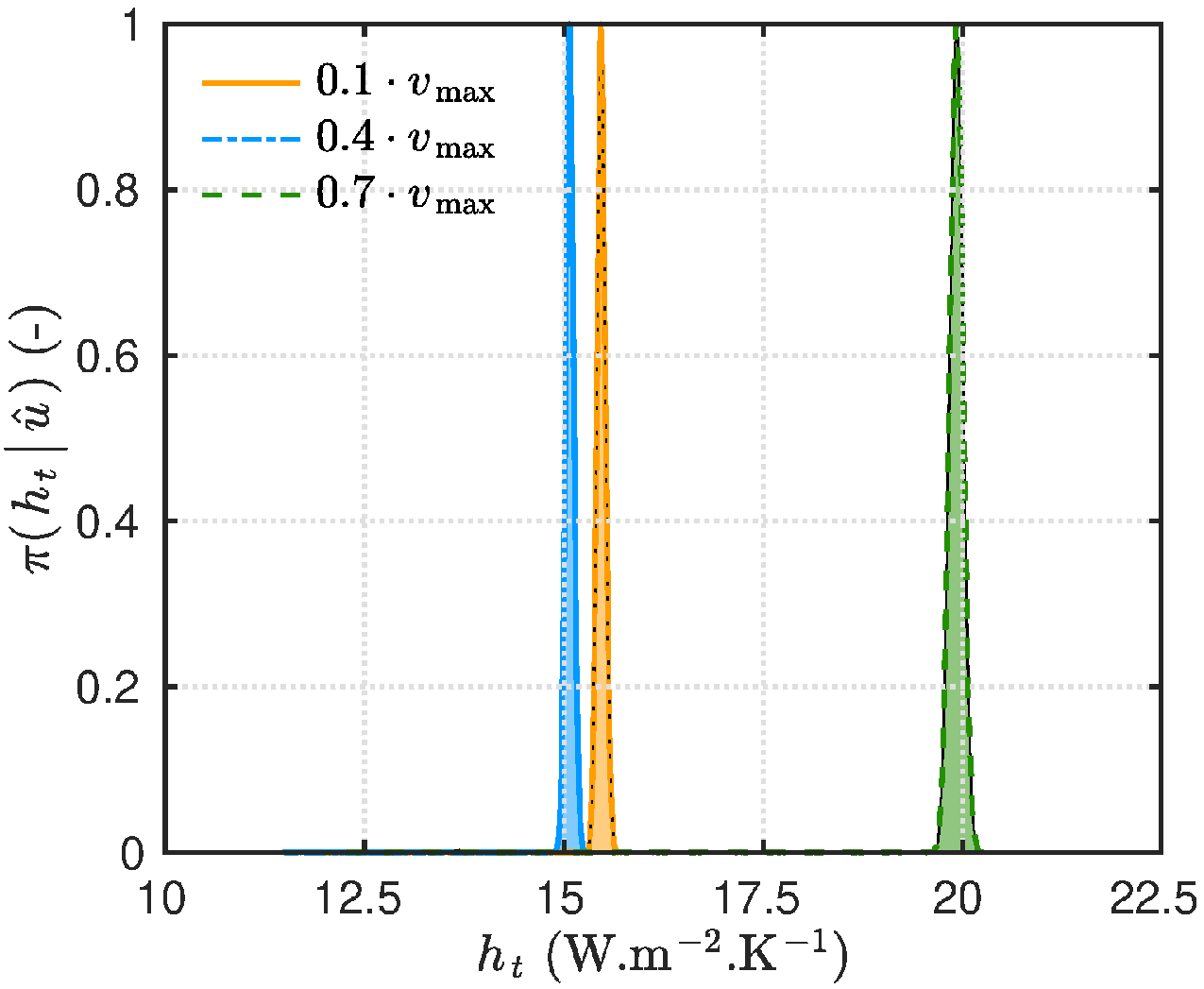}}
\hspace{0.2cm}
\subfigure[\label{fig:pdf_hlat}]{\includegraphics[width=.45\textwidth]{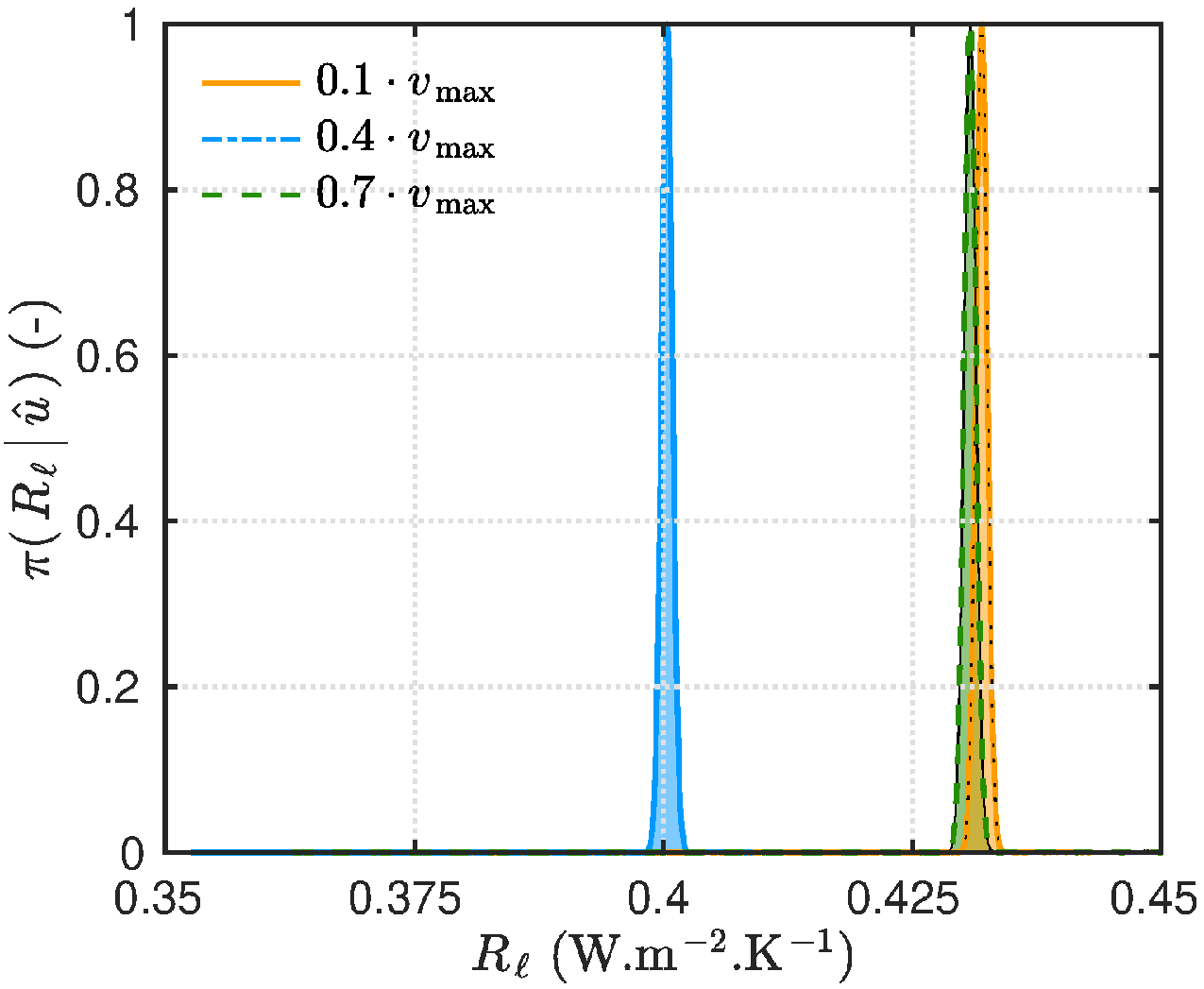}}
\caption{\revision{Posterior distribution of the two estimated surface transfer coefficients $h_{\,t}$ \emph{(a,c)} and $R_{\,\ell}$ \emph{(b,d)}} \revision{ with \emph{(a,b)} and without \emph{(c,d)} AEM.}}
\end{figure}

The marginal posterior distribution of the top and lateral surface transfer coefficients are evaluated and presented in Figures~\ref{fig:pdf_htop_aem} to \ref{fig:pdf_hlat} with and without the AEM. The distributions are \textsc{G}aussian, which means and standard deviations are reported in Table~\ref{tab:parameter_results}. The lateral surface transfer coefficient is almost $2$ orders lower than the top one. In addition, $R_{\,\ell}$ remains in the same range of $0.35$ to $0.4 \ \mathsf{W\,.\,m^{\,-2}\,.\,K^{\,-1}}\,$. It increases slightly according to the climatic chamber ventilator speed. Regarding, the top surface transfer coefficient, the mean values increase according to the speed ventilator. Note that the \emph{a priori} values of coefficients considered for the identifiability is in the range of the estimated ones, validating the preliminary investigations.

A comparison is carried out with results of the parameter estimation problem without considering the AEM in the direct model computations. Some changes are observed between the determined parameters. The coefficient $R_{\,\ell}$ is overestimated in the case without AEM. For the top surface transfer coefficient, the value is underestimated for the case without AEM, except for the low speed ventilator results. Such comparison underlines the importance of considering the AEM model to obtain a reliable estimation of the unknown parameters.

\begin{figure}[h!]
\centering 
\subfigure[\label{fig:Tv10_ft}$10\,\% \cdot v_{\,\max}$]{\includegraphics[width=.45\textwidth]{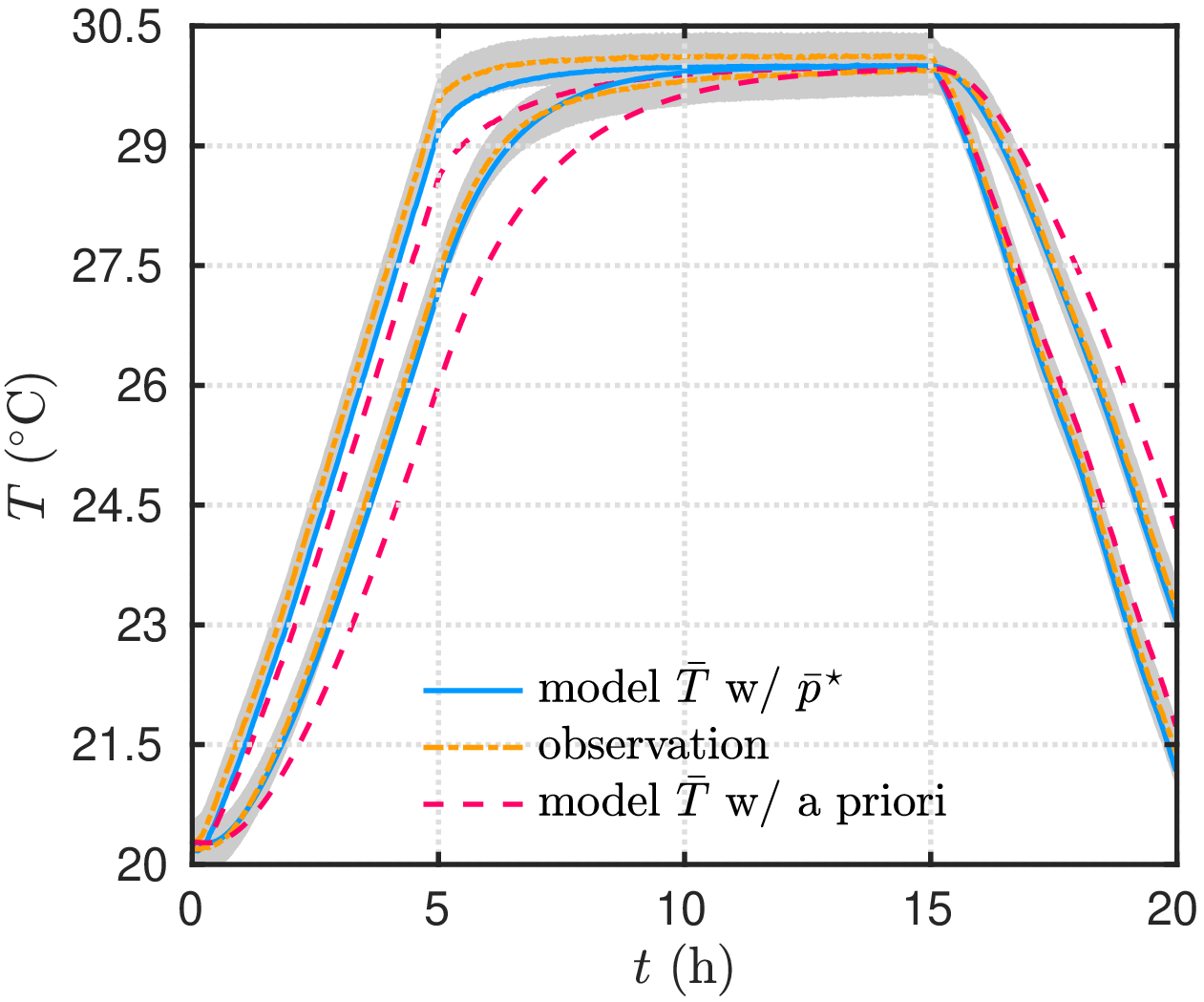}}
\hspace{0.2cm}
\subfigure[\label{fig:Tv40_ft}$40\,\% \cdot v_{\,\max}$]{\includegraphics[width=.45\textwidth]{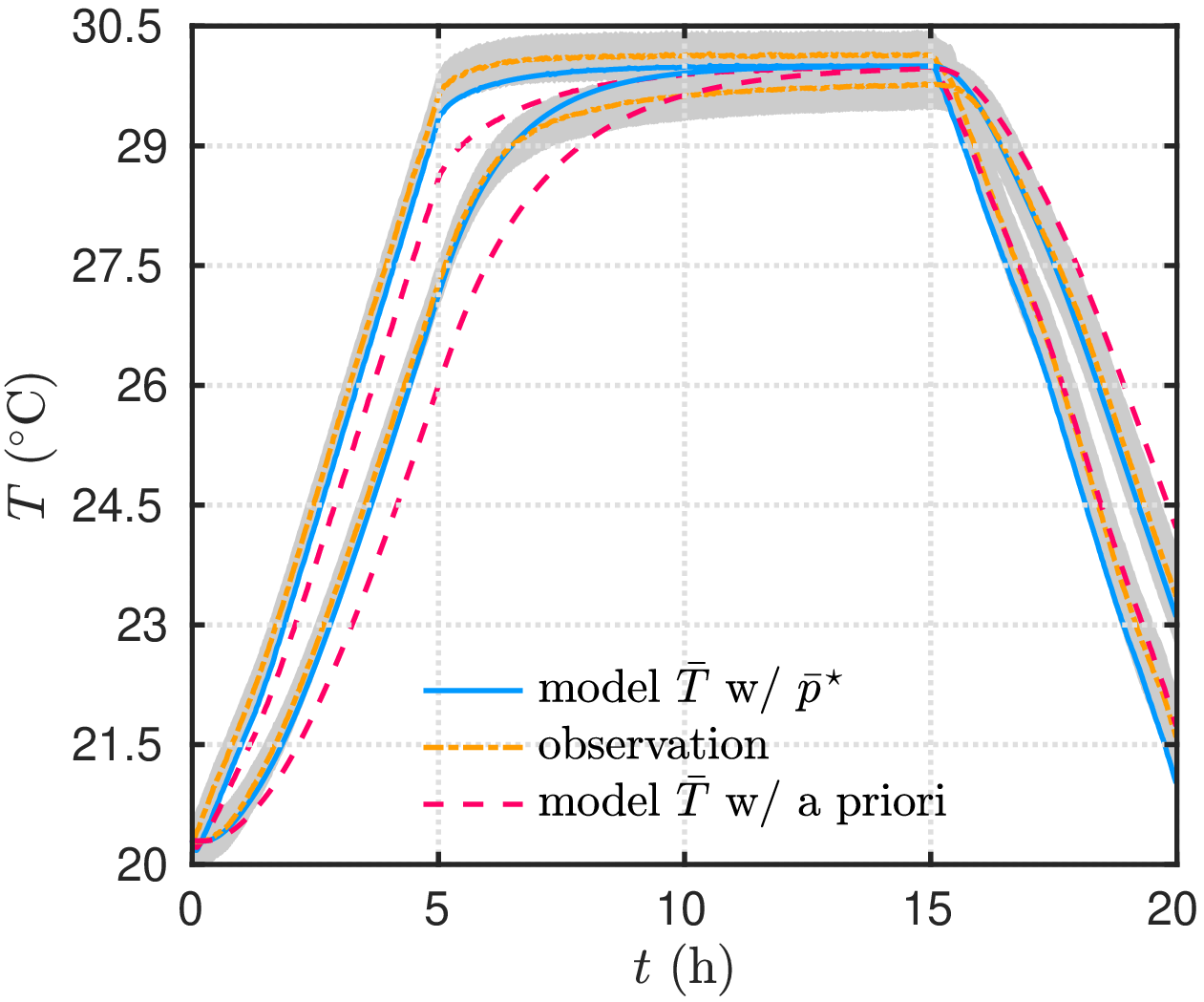}} \\
\subfigure[\label{fig:Tv70_ft}$70\,\% \cdot v_{\,\max}$]{\includegraphics[width=.45\textwidth]{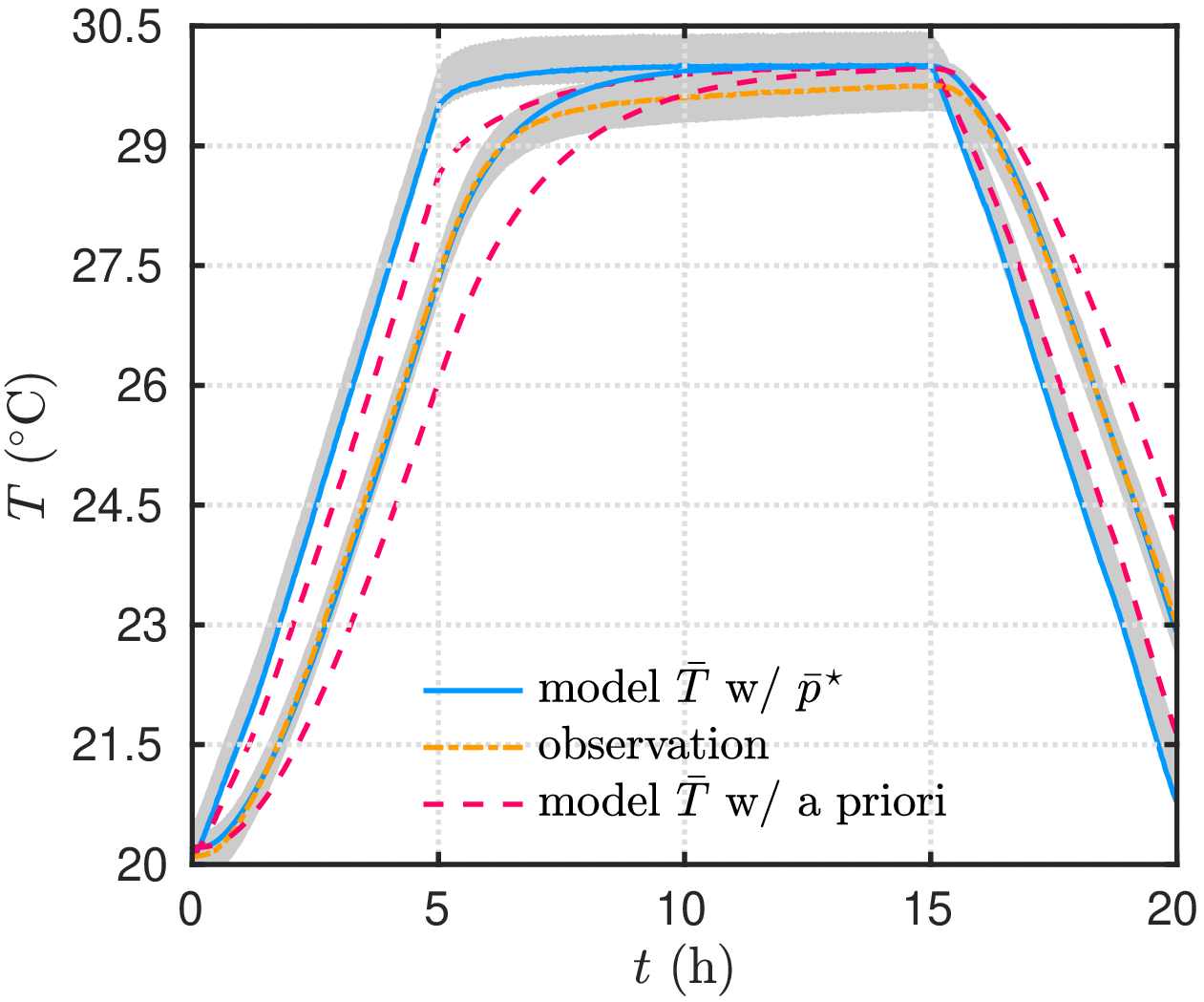}}
\hspace{0.2cm}
\subfigure[\label{fig:Tv100D_ft} $100\,\% \cdot v_{\,\max}$ ]{\includegraphics[width=.45\textwidth]{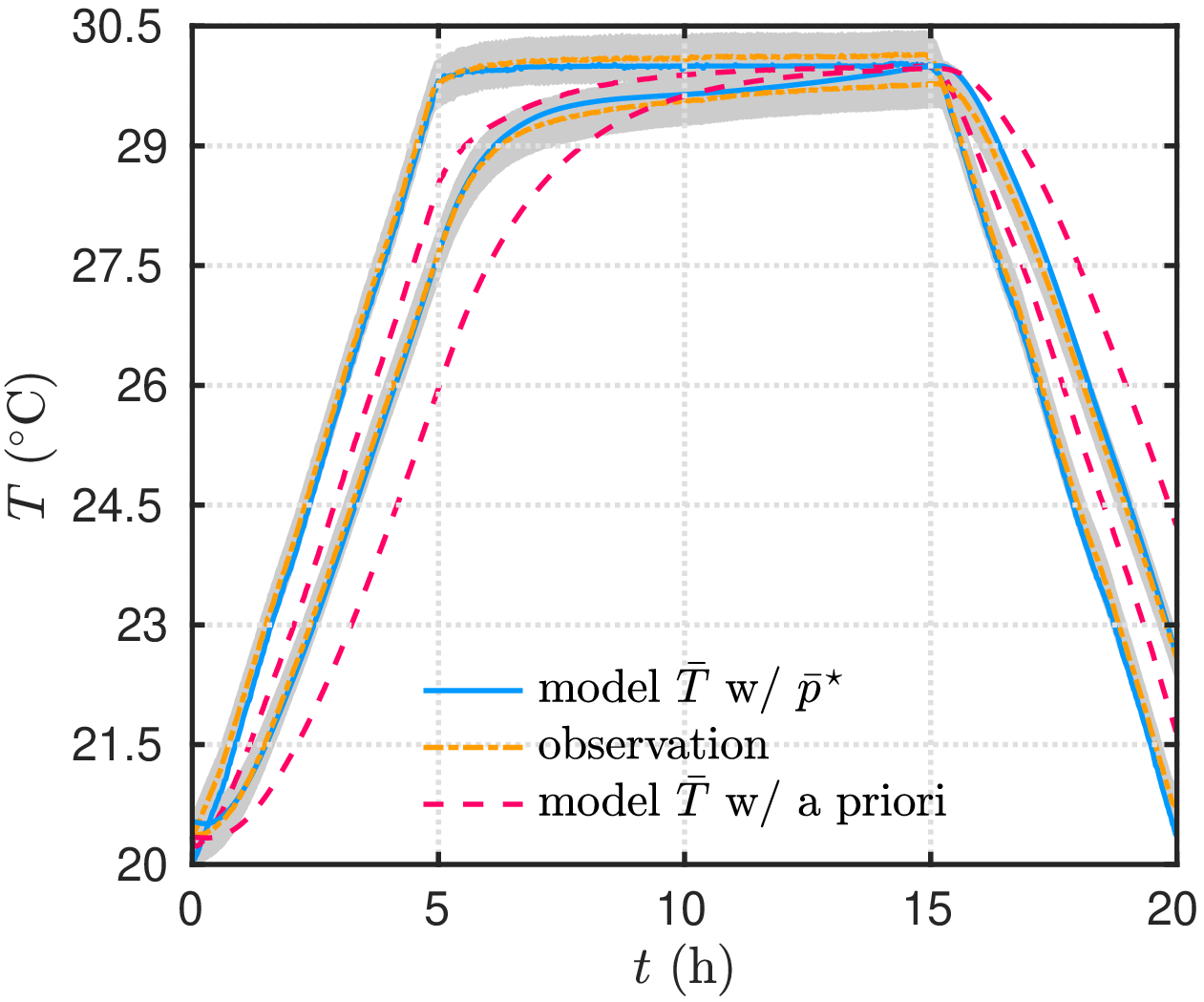}} 
\caption{\revision{Comparison of the experimental observation with the model prediction using \emph{a priori} and estimated parameters.}}
\label{fig:T_ft}
\end{figure}

\begin{figure}[h!]
\centering 
\subfigure[$x_{\,1} \egal 0$ \label{fig:resx1_ft}]{\includegraphics[width=.45\textwidth]{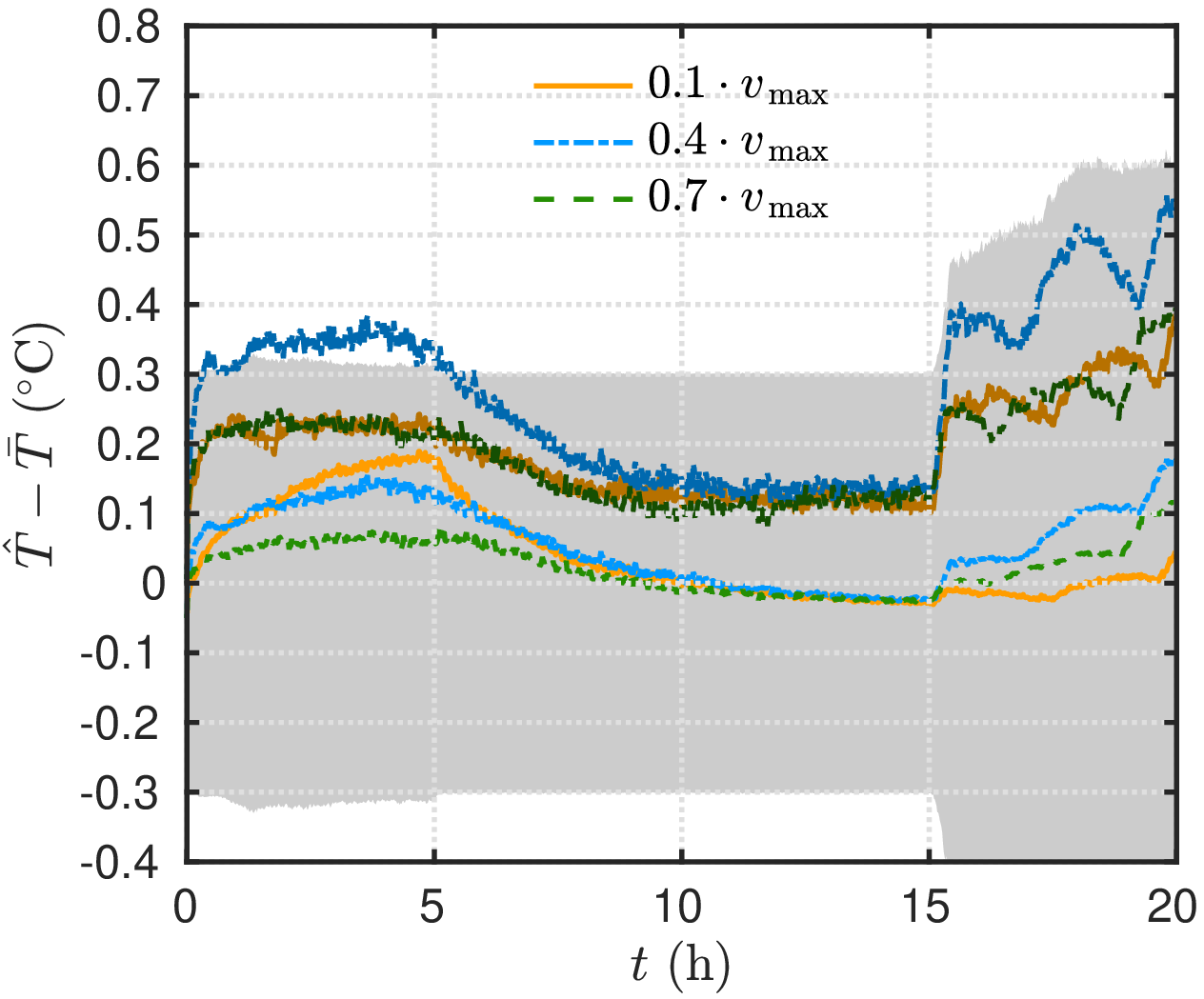}}
\hspace{0.2cm}
\subfigure[$x_{\,2} \egal 0.5 \cdot L_{\,0\,,\,x}$ \label{fig:resx2_ft}]{\includegraphics[width=.45\textwidth]{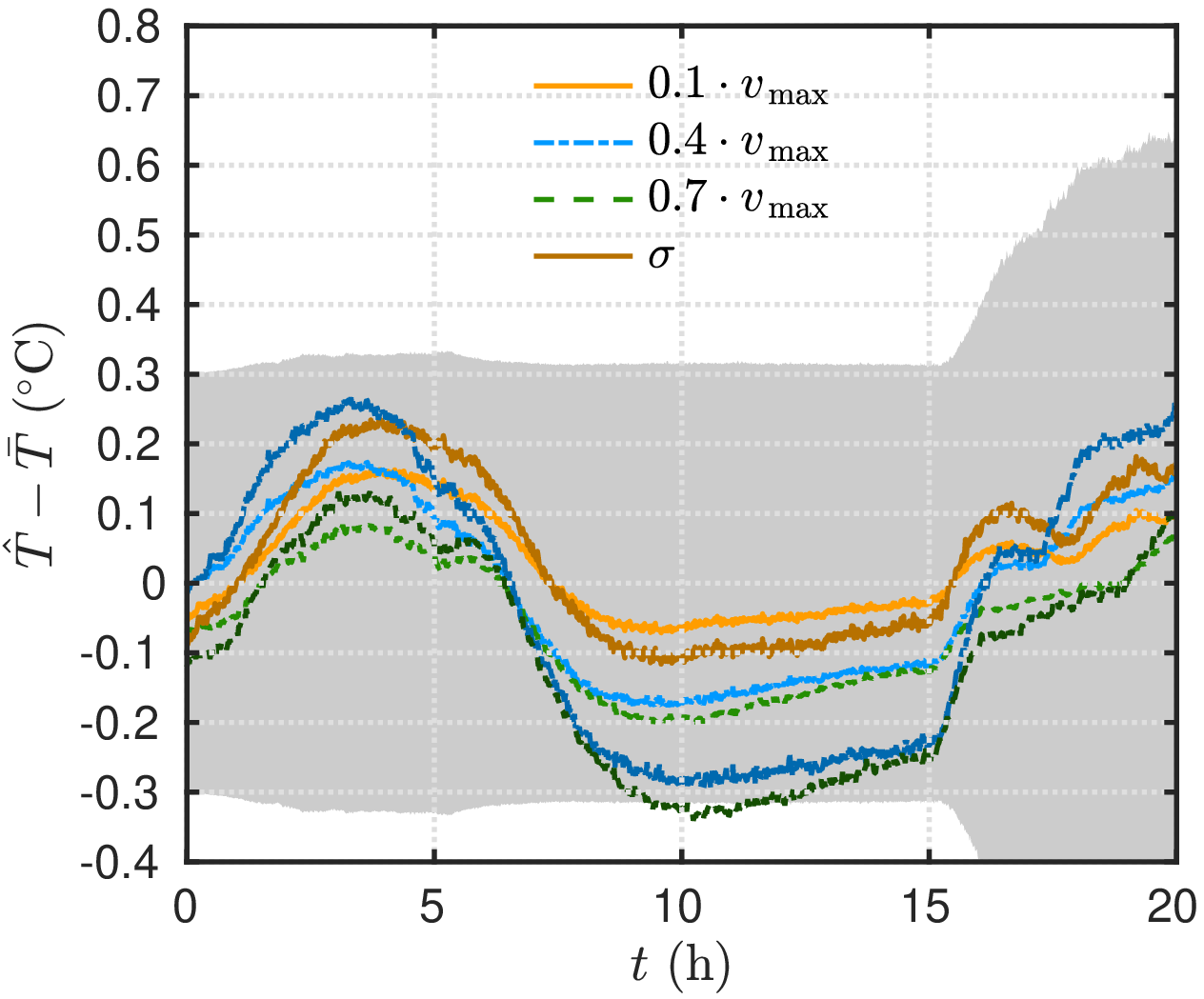}}
\caption{\revision{Time variation of the residual between the model prediction with the estimated parameters} \revision{ and the experimental observations at the two sensor positions. Lighter and darker colors} \revision{ corresponds to the case with and without AEM, respectively.}}
\label{fig:res_ft}
\end{figure}

The $1$D lumped model predictions are computed using the mean of the estimated parameters (with AEM). Results are displayed in Figures~\ref{fig:T_ft} and \ref{fig:res_ft}. The predictions are in good agreement with the experimental observations and remain within the measurement uncertainty. The model is more reliable with the estimated parameters than the \emph{a priori} ones. Moreover, the residuals have a mean close to zeros and no strong correlation \revision{for the case considering the AEM. When not considering the error in the direct model, the} \revision{ residual are higher with higher correlations, indicating an error in the model formulation.} The three experiments show satisfactory results. 

\begin{table}[h!]
\centering
\caption{\revision{Mean and standard deviation of the estimated parameters.}}
\label{tab:parameter_results}
\setlength{\extrarowheight}{.5em}
\begin{tabular}[l]{@{} cc cc cc}
\hline
\hline
& \textit{Fan velocity}
& $\mu \bigl(\, h_{\,t}\,\bigr)$ & $\upsilon \bigl(\, h_{\,t}\,\bigr)$
& $\mu \bigl(\, \, R_{\,\ell}\,\bigr)$ & $\upsilon \bigl(\, R_{\,\ell}\,\bigr)$ \\
& $\unit{\% \cdot v_{\,\max}}$
& \multicolumn{2}{c}{$\unit{W\,.\,m^{\,-2}\,.\,K^{\,-1}}$} 
& \multicolumn{2}{c}{$\unit{W\,.\,m^{\,-2}\,.\,K^{\,-1}}$} \\[4pt] \hline
& $10$ & $11.5$ & $0.29$ & $0.38$ & $0.001$  \\
& $40$ & $16.15$ & $0.23$ & $0.37$ & $0.001$  \\
& $70$ & $21.4$ & $0.74$ & $0.40$ & $0.002$  \\ \hline
Config. \emph{(i)}: $\Omega_{\,h_{\,t}} \egal \bigl[\, 1 \,,\, 40 \,\bigr]\, \mathsf{W\,.\,m^{\,-2}\,.\,K^{\,-1}}$ & $100$ & $94.7$ & $3.21$ & $0.535$ & $0.008$  \\
Config. \emph{(ii)}: $h_{\,t} \, \rightarrow \, \infty\,$ & $100$ & $-$ & $-$ & $0.515$ & $0.012$  \\
\hline
\hline
\end{tabular}
\end{table}

\subsubsection{Case $100\%$ of $v_{\,\max}$}

Estimation results are now presented for the case of maximal speed of the chamber fan ventilator. The \textsc{B}ayesian estimation is performed for two configurations. \emph{(i)} The first one considers the \emph{a priori} domain of variation $\bigl[\, 0.125 \,,\, 5 \,\bigr] \cdot h_{\,t}^{\,\apr} \egal \bigl[\, 0.1 \,,\, 40 \,\bigr] \, \mathsf{W\,.\,m^{\,-2}\,.\,K^{\,-1}}\,$ for $h_{\,t}$ as used in Section~\nameref{sec:practical_identifiability}. \emph{(ii)} The second configuration assumes a model with \textsc{D}irichlet boundary condition at the top surface $h_{\,t} \, \rightarrow \, \infty\,$. In such case, only the lateral surface transfer coefficient $R_{\,\ell}$ needs to be retrieved. Note that in all configurations, the domain of variation of $R_{\,\ell}$ is maintained to $\bigl[\, 0.01 \,,\, 1 \,\bigr] \, \mathsf{W\,.\,m^{\,-2}\,.\,K^{\,-1}}\,$.

\begin{figure}[h!]
\centering 
\subfigure[\label{fig:htop_v100_fn}]{\includegraphics[width=.45\textwidth]{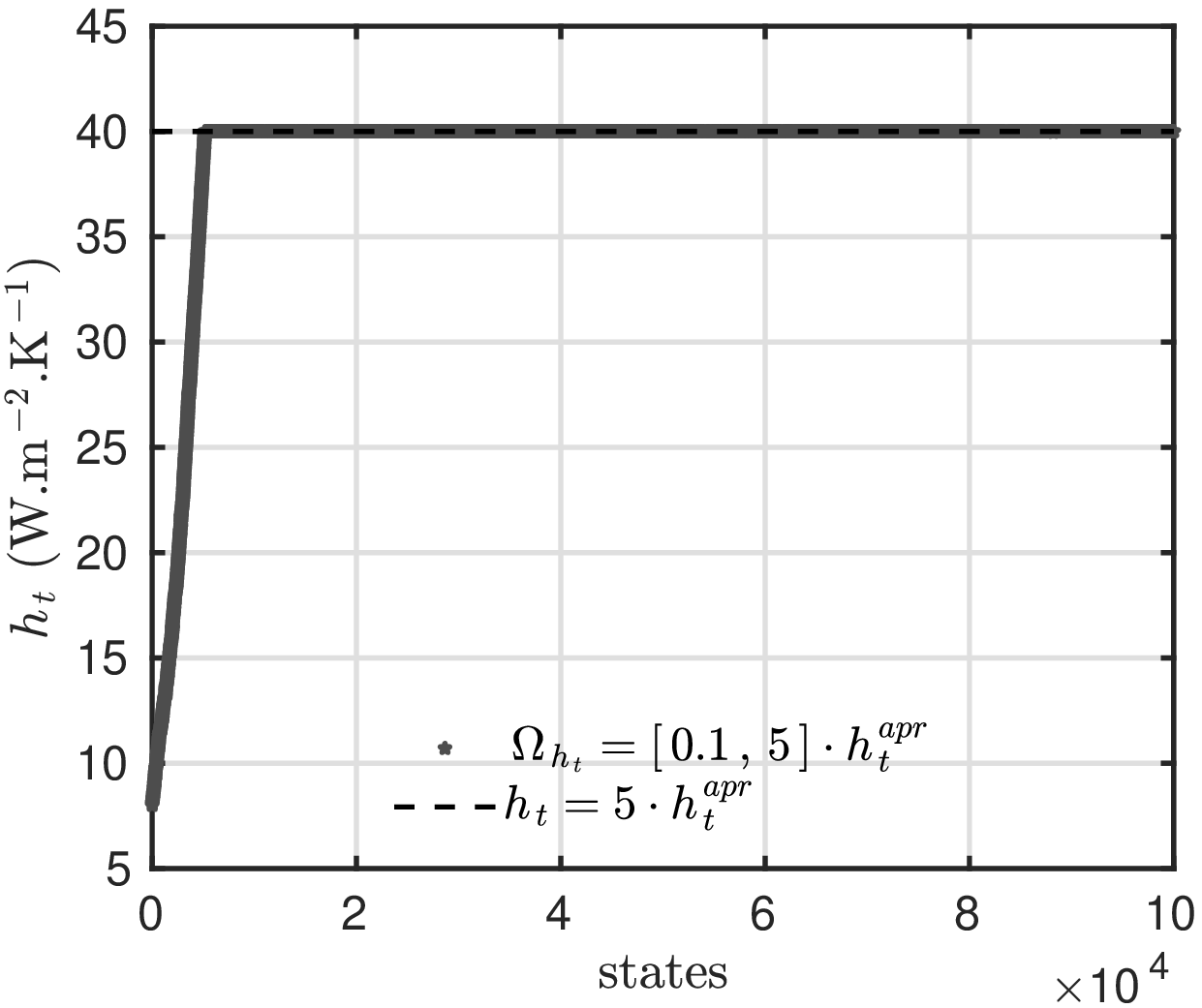}}
\hspace{0.2cm}
\subfigure[\label{fig:hlat_v100_fn}]{\includegraphics[width=.45\textwidth]{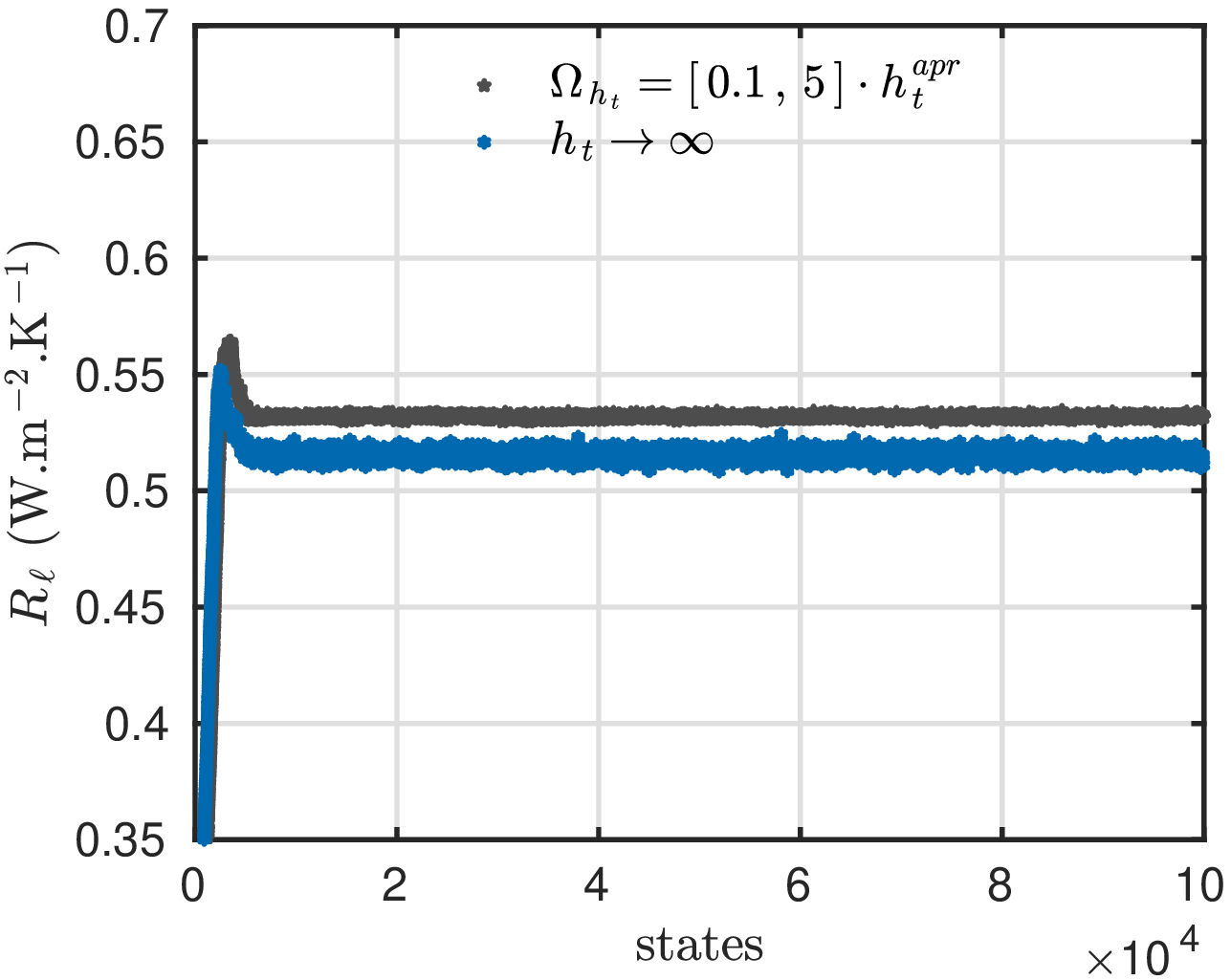}}
\caption{\revision{Variation of the two estimated surface transfer coefficients $h_{\,t}$ \emph{(a)} and $R_{\,\ell}$ \emph{(b)} according to} \revision{ the number of states for the case $100\,\% \cdot v_{\,\max}$.}}
\label{fig:p_v100_fn}
\end{figure}

\begin{figure}[h!]
\centering 
\includegraphics[width=.45\textwidth]{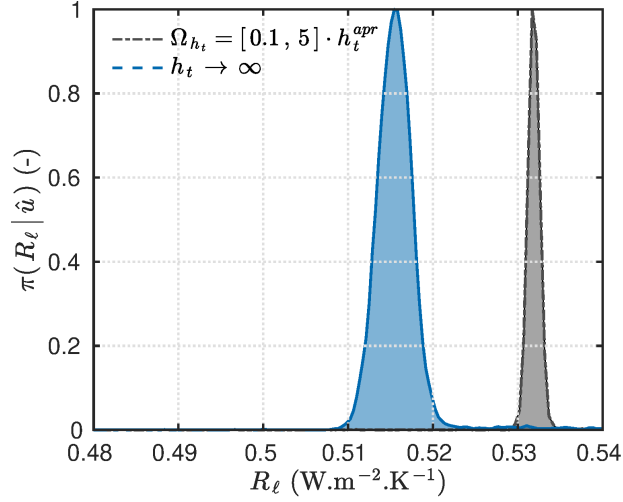}
\caption{\revision{Posterior distribution of the estimated coefficients $R_{\,\ell}$ for the case $100\,\% \cdot v_{\,\max}$.}}
\label{fig:pdf_hlat_v100}
\end{figure}

For the two configurations, the burn-in period is around $1000$ states with similar acceptance rates. The evolution of the likelihood according to the number of state is given in Figure~\ref{fig:likelihood_fn} for the first configuration. Figure~\ref{fig:p_v100_fn} shows the variation of the parameter according to the number of states. For the parameter $R_{\,\ell}$, the chain converges in a similar area. As remarked in Figure~\ref{fig:pdf_hlat_v100}, the probability distribution of $R_{\,\ell}$ are in the same range for both configurations, around $0.52\, \mathsf{W\,.\,m^{\,-2}\,.\,K^{\,-1}}\,$. The mean and standard deviation are synthesized in Table~\ref{tab:parameter_results}. 

\begin{figure}[h!]
\centering 
\subfigure[$x_{\,1} \egal 0$ \label{fig:resx1_ft_v100}]{\includegraphics[width=.45\textwidth]{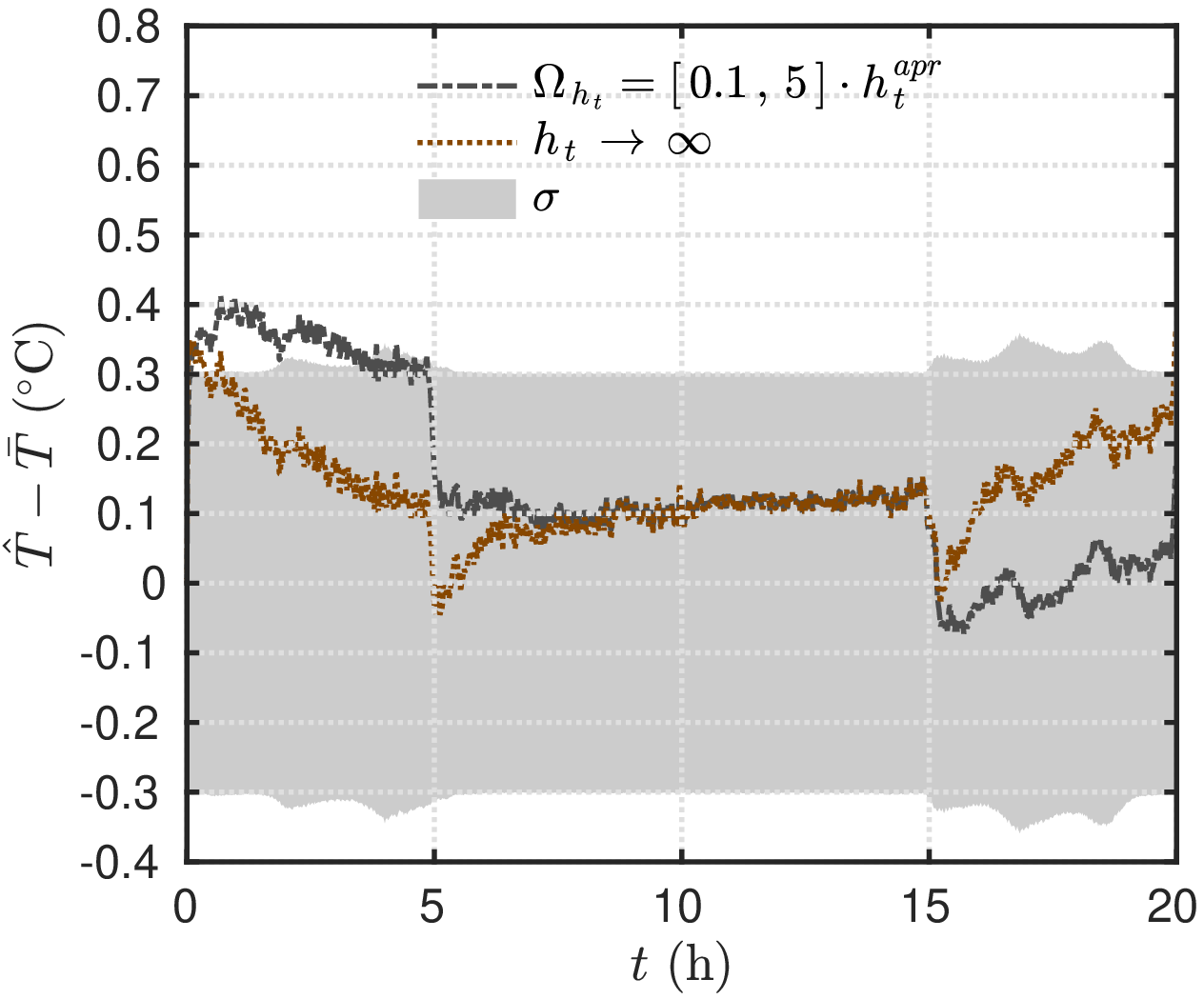}}
\hspace{0.2cm}
\subfigure[$x_{\,2} \egal 0.5 \cdot L_{\,0\,,\,x}$ \label{fig:resx2_ft_v100}]{\includegraphics[width=.45\textwidth]{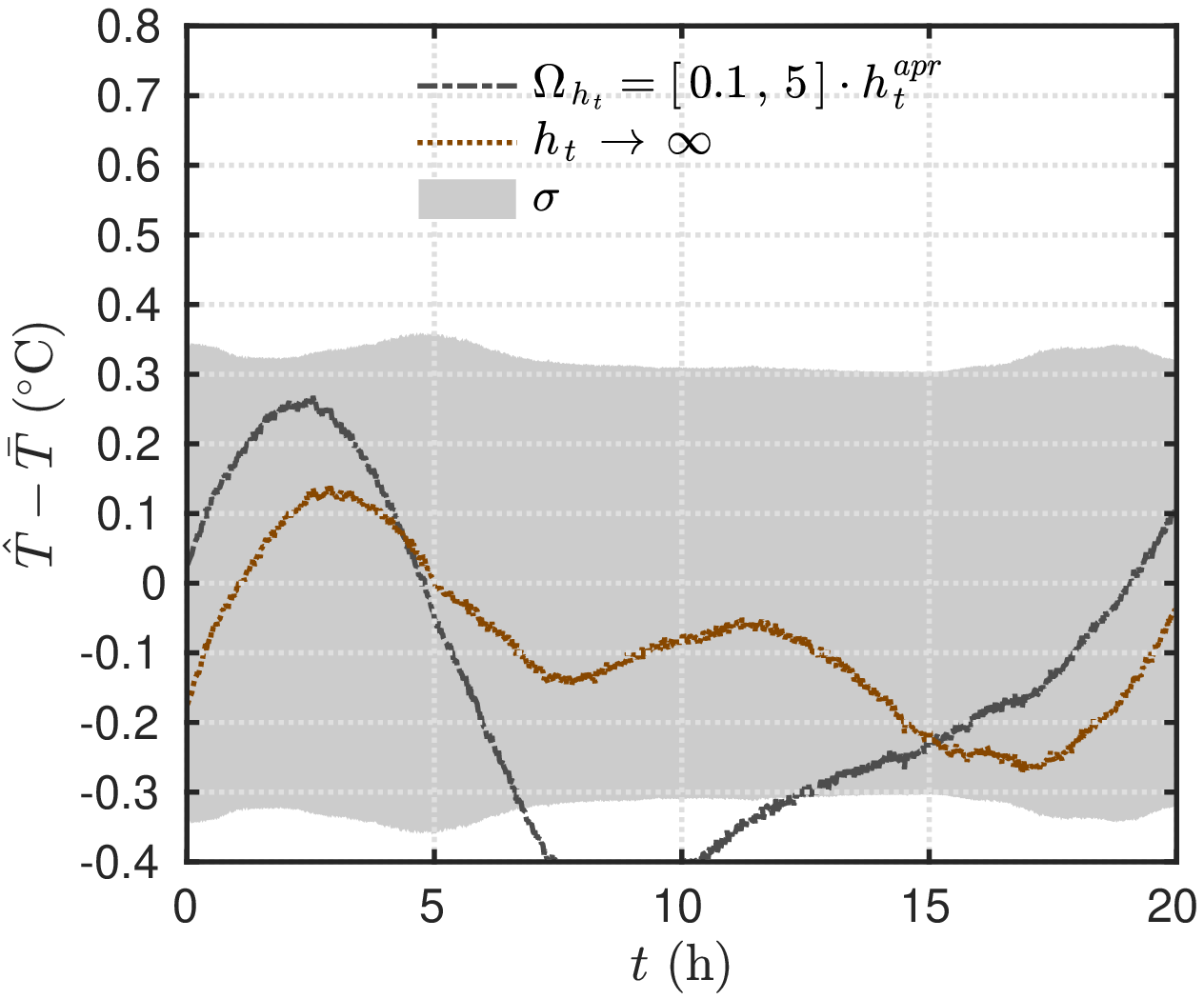}}
\caption{\revision{Time variation of the residual between the model prediction with the estimated parameters} \revision{ and the experimental observations at the two sensor positions.}}
\label{fig:res_ft_v100}
\end{figure}

However, for parameter $h_{\,t}$ the chain is very unstable as remarked in Figure~\ref{fig:htop_v100_fn}. For the first configuration, the mean estimated parameter is $40\, \mathsf{W\,.\,m^{\,-2}\,.\,K^{\,-1}}$ with a very small standard deviation. In other words, the estimates reach the upper bound of the parameter domain of variation. Additional tests have been carried by increasing the upper bound of the domain; still leading to the same conclusions. To continue the investigations, a comparison of the residuals between the measurement and the predictions of the model using the estimated parameter is provided in Figure~\ref{fig:res_ft_v100}. Higher differences are remarked between the two configurations. n the case of \textsc{D}irichlet boundary conditions, the residual is lower for the first step of increase of temperature $t \, \in \, \bigl[\, 0 \,,\, 5 \,\bigr] \ \mathsf{h}\,$. However, for the third part $t \, \in \, \bigl[\, 15 \,,\, 20 \,\bigr] \ \mathsf{h}\,$, when the temperature decreases from $30$ to $20 \ \mathsf{\degC}\,$, the residual is a bit higher. Still, the residual remains centered to zero and in the range of the measurement accuracy for both configurations. Last, Figure~\ref{fig:Tv100D_ft} compares the model prediction of configuration \emph{(ii)} with the experimental results. A very satisfactory agreement is observed. Therefore, given those results, it seems for the case $100\%$ of $v_{\,\max}\,$, the top surface transfer coefficient has a high magnitude so that it corresponds almost to \textsc{D}irichlet boundary conditions.

\subsubsection{\revision{Reliability of the lumped model}}

A lumped one-dimensional model has been proposed and used for the \textsc{B}ayesian estimation to reduce it computational cost. 

\begin{figure}[h!]
\centering 
\subfigure[\label{fig:y2D}]{\includegraphics[width=.45\textwidth]{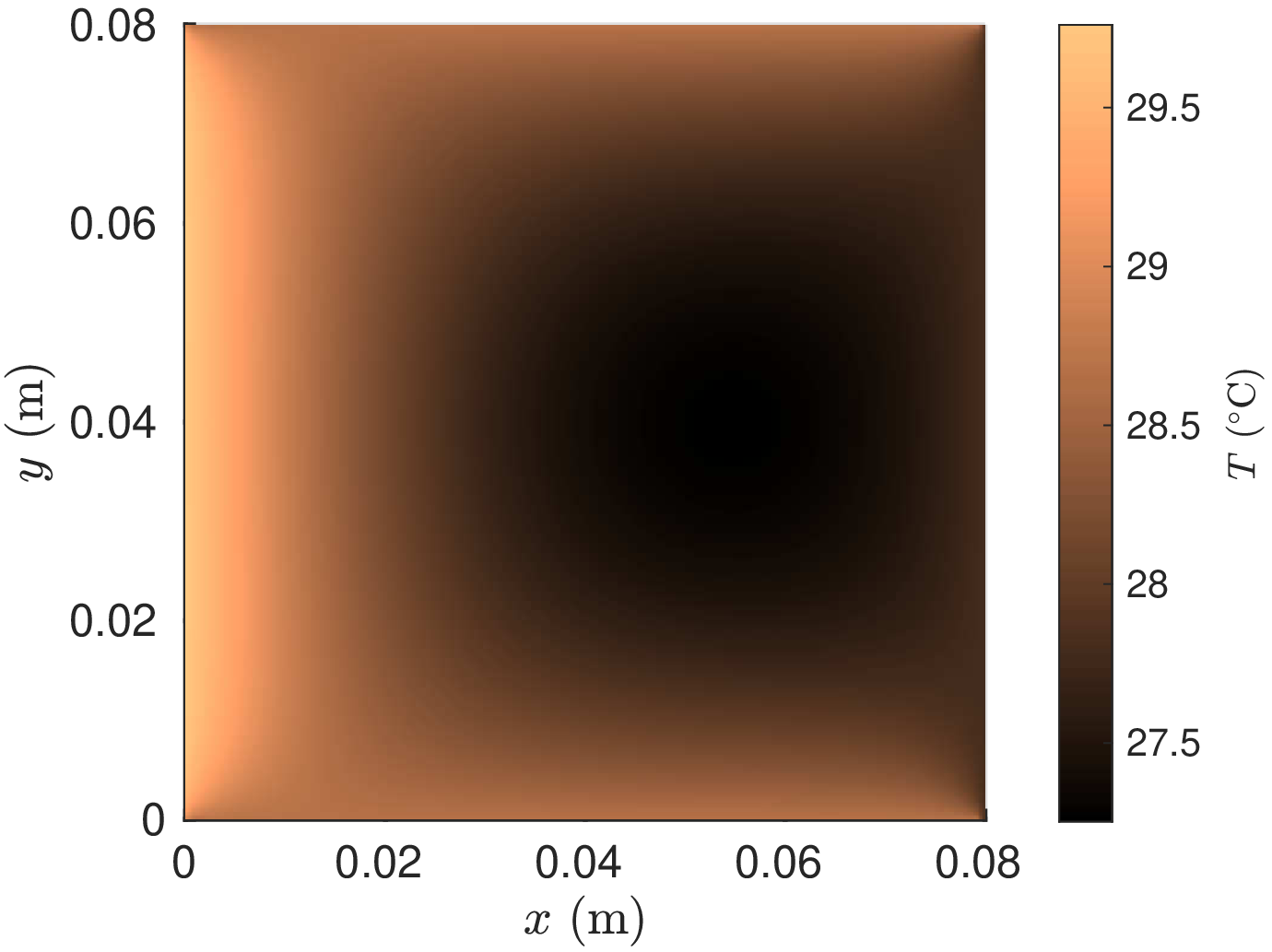}}
\hspace{0.2cm}
\subfigure[\label{fig:Rjq_ft}]{\includegraphics[width=.45\textwidth]{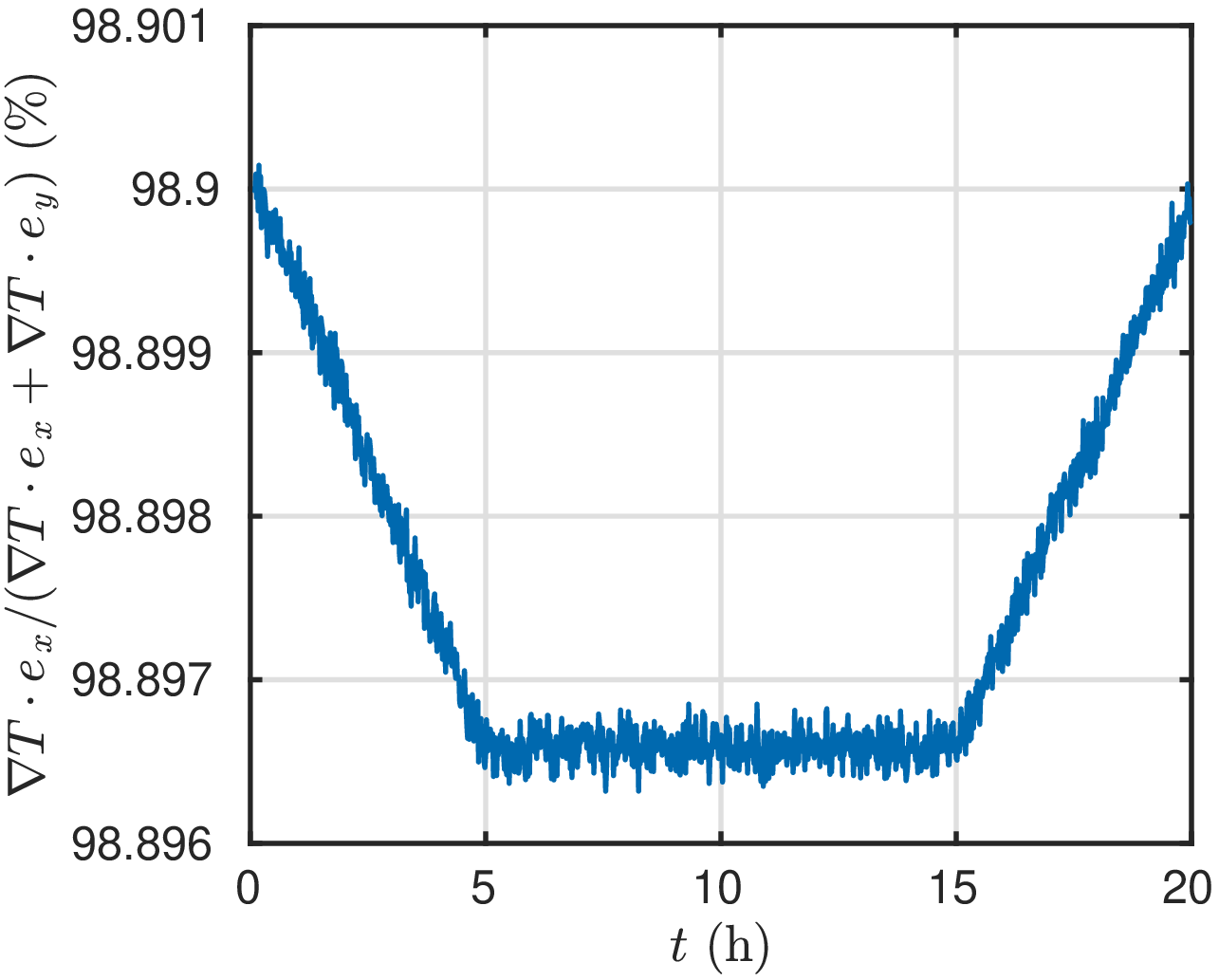}} 
\caption{\revision{Variation of the temperature distribution of the complete model at $t \egal 5 \ \mathsf{h}$ \emph{(a)} and time} \revision{ variation of ratio between the flux in the $x$ direction among the total one \emph{(b)}.}}
\end{figure}

Complementary investigations are carried out to justify that the transfer occurs mainly in the $x$ direction. The predictions of the complete model, considering 2D heat transfer in the wood fiber, aluminum and insulator domains have been performed.  The $2$D model is solved using the \DF ~numerical method introduced in \cite{berger_efficient_2021}. Figure~\ref{fig:y2D} shows a surface distribution of the temperature. It can be noted that at the sensor location $x_{\,1} \egal 0$ and $x_{\,2} \egal 0.04 \ \mathsf{m}\,$, the transfer occurs mainly in the $x$ direction. Furthermore, the time variation of the ratio between the heat flux in the $x$ direction among the total heat flux is given in Figure~\ref{fig:Rjq_ft}. It is observed that $98\%$ of the flux concerns the $x$ direction. The magnitude varies really little according to the boundary condition evolution.

\begin{figure}[h!]
\centering 
\subfigure[\label{fig:jins_ft}]{\includegraphics[width=.45\textwidth]{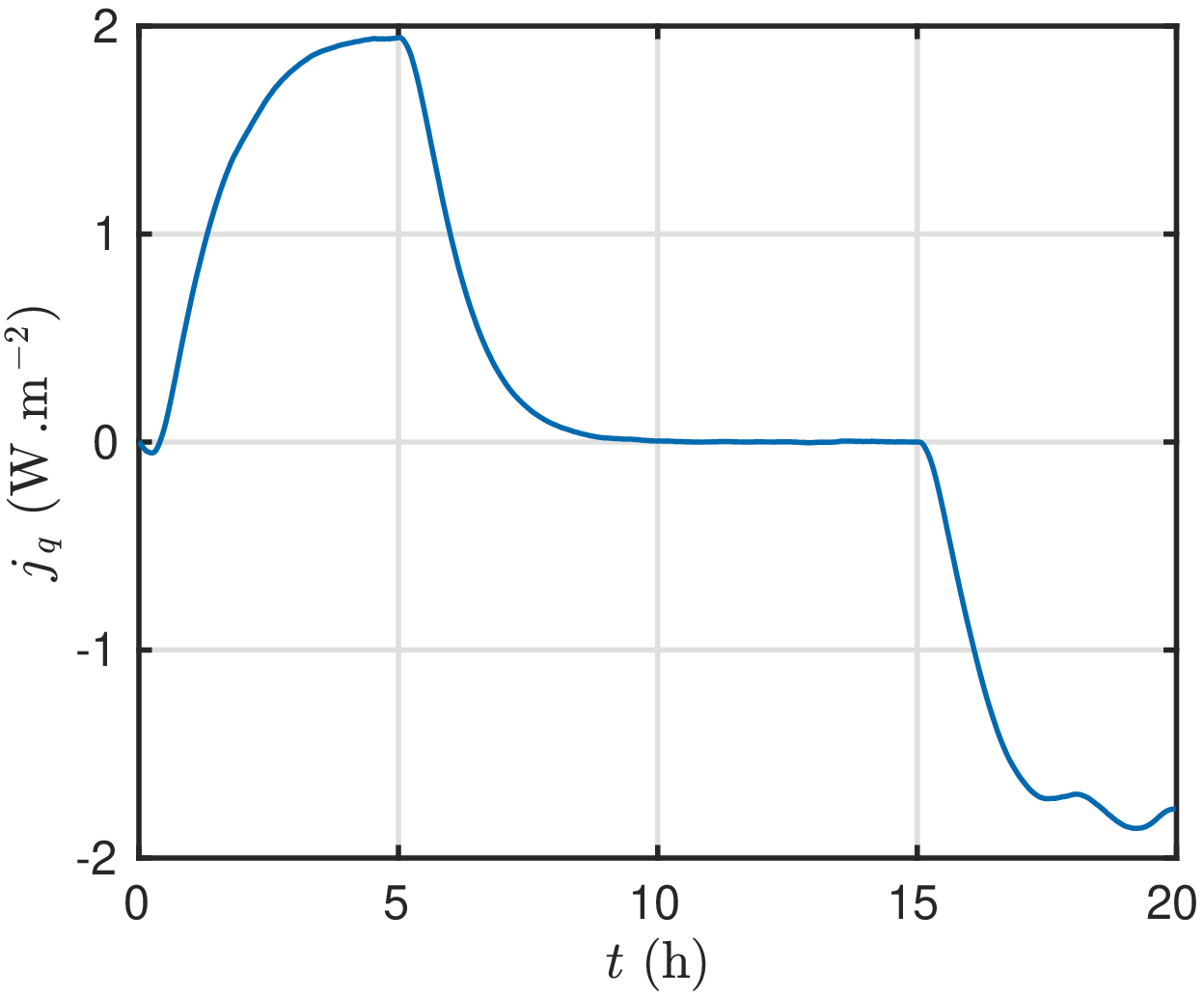}} 
\hspace{0.2cm}
\subfigure[\label{fig:Talu_ft}]{\includegraphics[width=.45\textwidth]{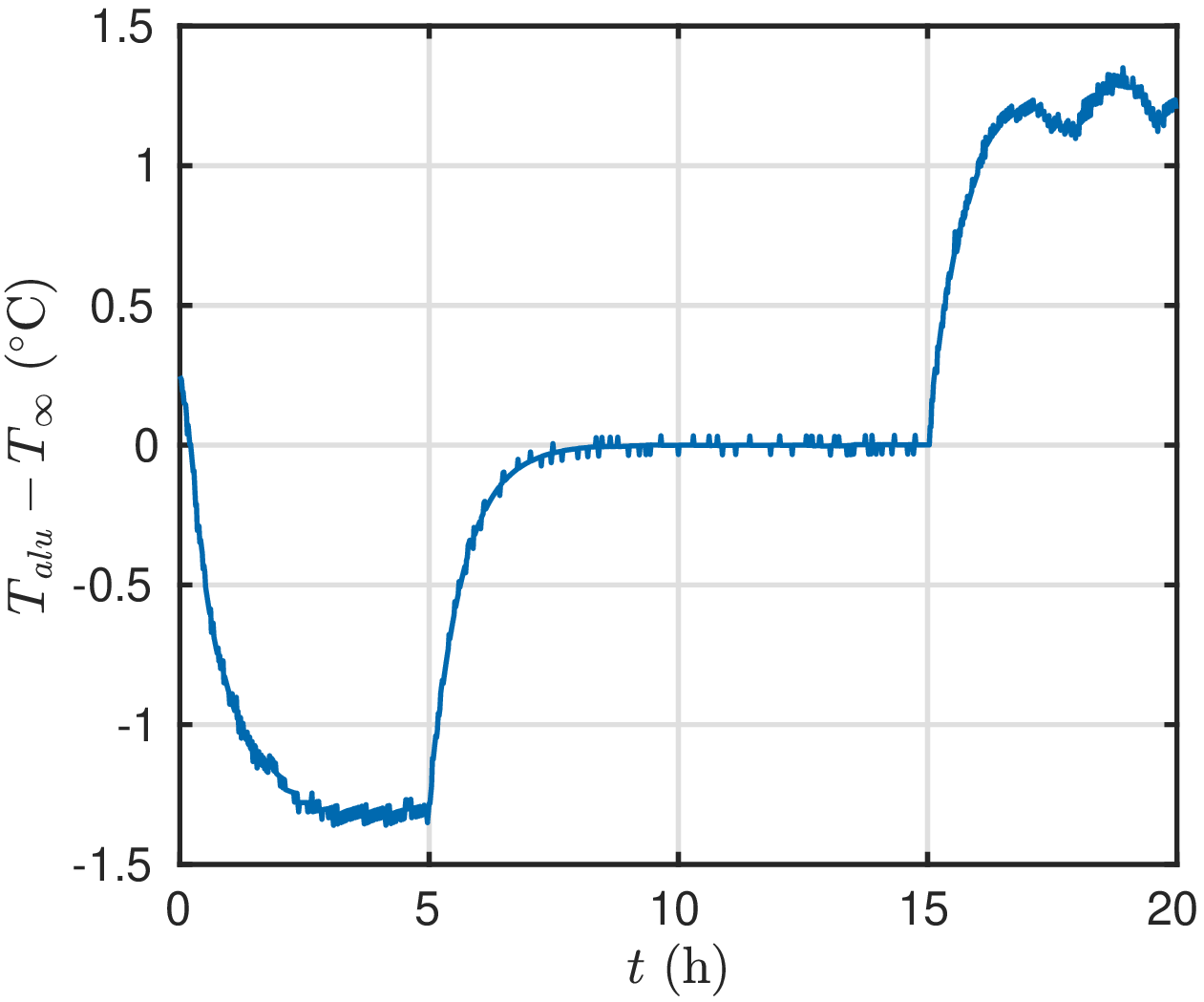}}
\caption{\revision{Time variation of the heat flux from the complete model at the interface between the wood} \revision{ fiber and the insulator \emph{(a)} and of the temperature difference between the aluminum and the} \revision{ climatic chamber  \emph{(b)}.}}
\end{figure}

Figure~\ref{fig:jins_ft} presents the heat flux obtained with the complete 2D model at the interface between the wood fiber and the insulator. The flux is not null reaching an absolute maximum of $2 \ \mathsf{W\,.\,m^{\,-2}}\,$. Thus, even with a $8 \ \mathsf{cm}$ insulator at the bottom of the wood fiber, it does not ensure an adiabatic hypothesis at this interface.

The procedure of sealing the wood fiber lateral faces with aluminum tape may affect the surface temperature. Indeed, Figure~\ref{fig:Talu_ft} compares the temperature in the aluminum computed with the 2D formulation and with the temperature in the climatic chamber. The difference can reach $\pm \, 1.5 \ \mathsf{\degC}\,$. If the lumped model does not consider the transfer through the aluminum, this model approximation have been included by using the AEM in the parameter estimation procedure. The AEM introduces an error between lumped and complete formulations as a \textsc{G}aussian variable preliminary computed so that the 1D model imperfections are taken into account.

\begin{figure}[h!]
\centering 
\subfigure[$0.1 \cdot v_{\,\max}$\label{fig:T_tay_v10_ft}]{\includegraphics[width=.45\textwidth]{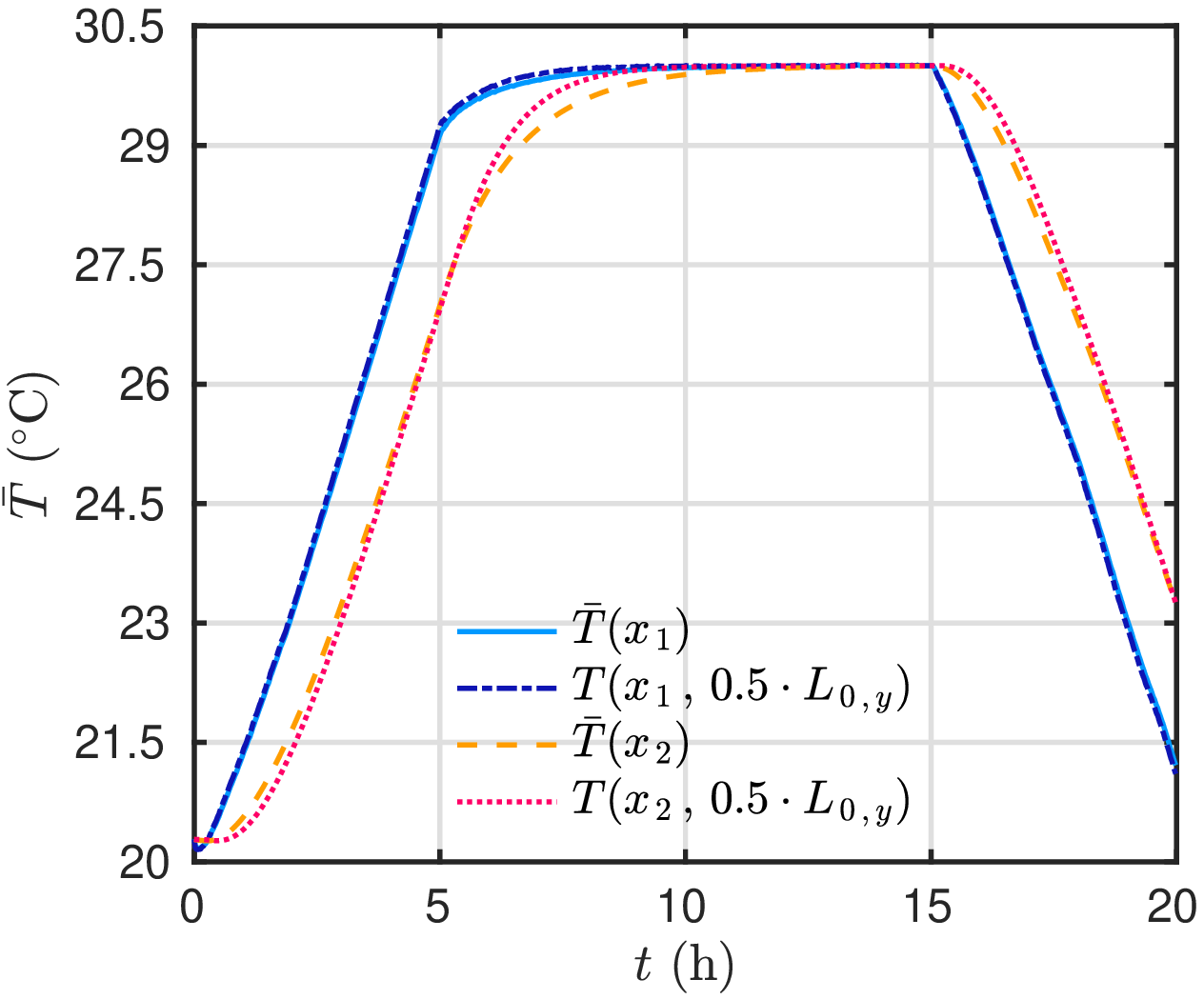}}
\hspace{0.2cm}
\subfigure[$0.7 \cdot v_{\,\max}$\label{fig:T_tay_v70_ft}]{\includegraphics[width=.45\textwidth]{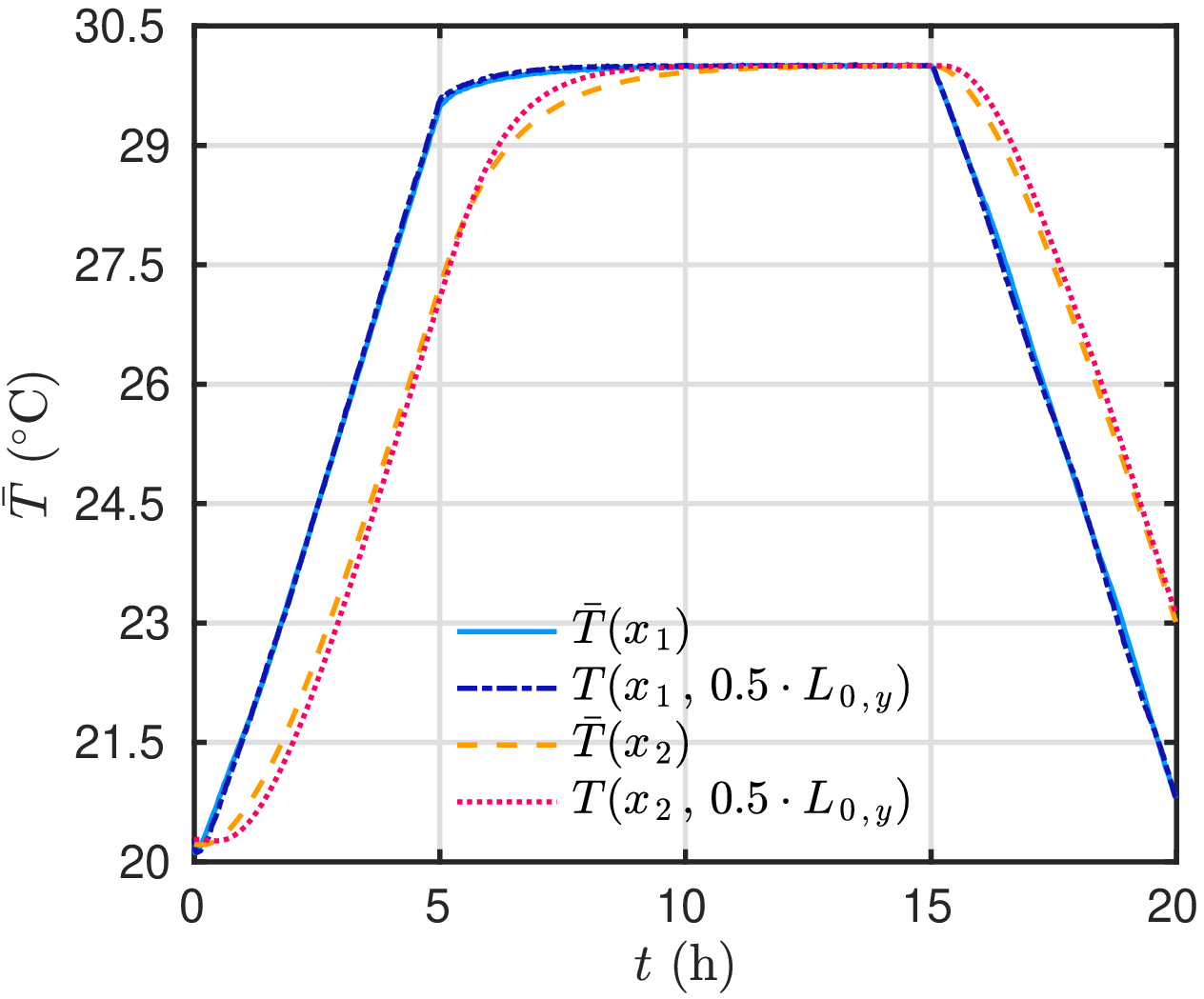}}
\caption{\revision{Model predictions considering the estimated parameters for two chamber fan velocity. } \revision{Results are given for the $1$D lumped model and the $2$D one.}}
\end{figure}

\begin{figure}[h!]
\centering 
\subfigure[$x_{\,1} \egal 0$\label{fig:res2Dx1_ft}]{\includegraphics[width=.45\textwidth]{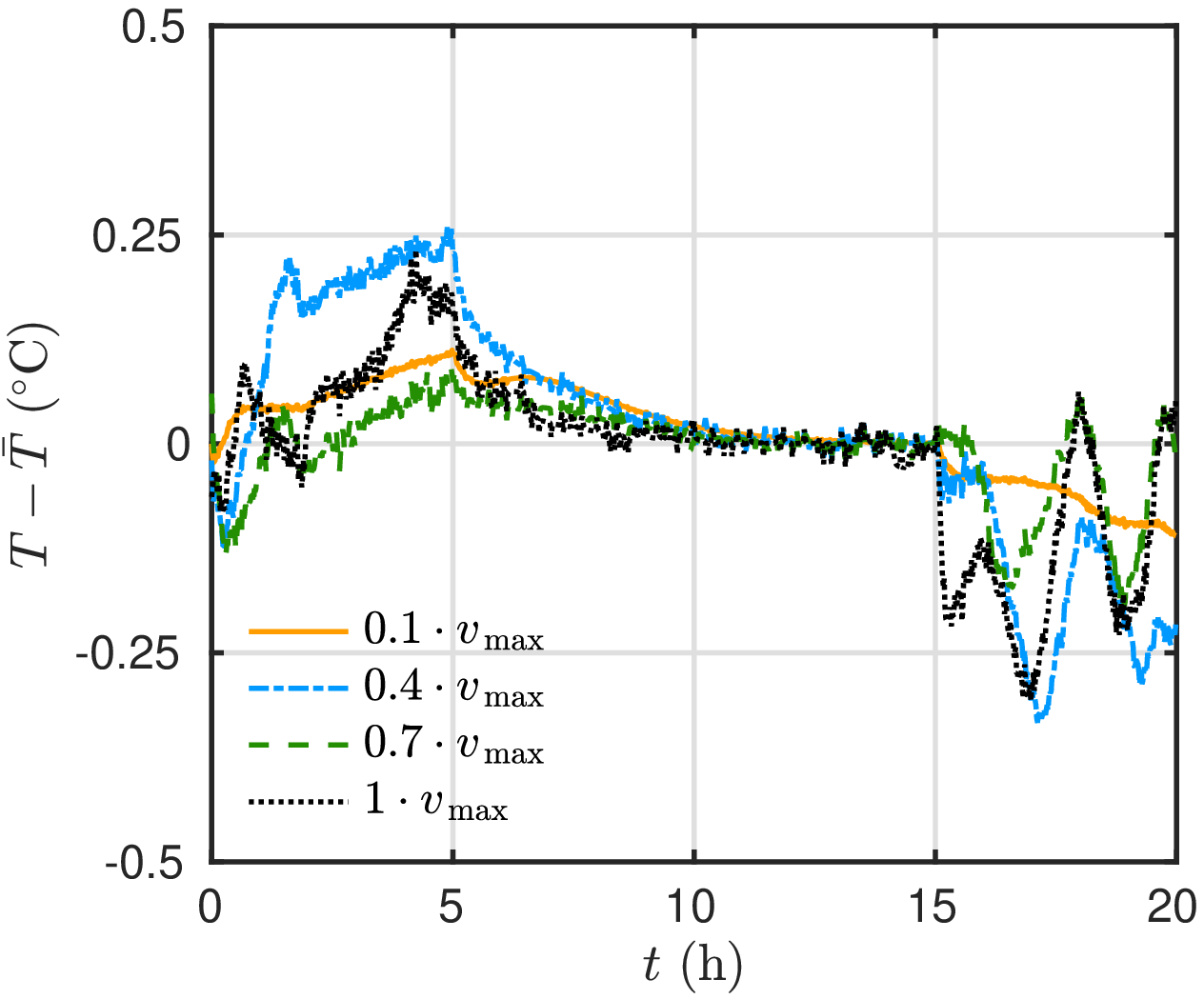}}
\hspace{0.2cm}
\subfigure[$x_{\,2} \egal 0.5 \cdot L_{\,0\,,\,x}$\label{fig:res2Dx2_ft}]{\includegraphics[width=.45\textwidth]{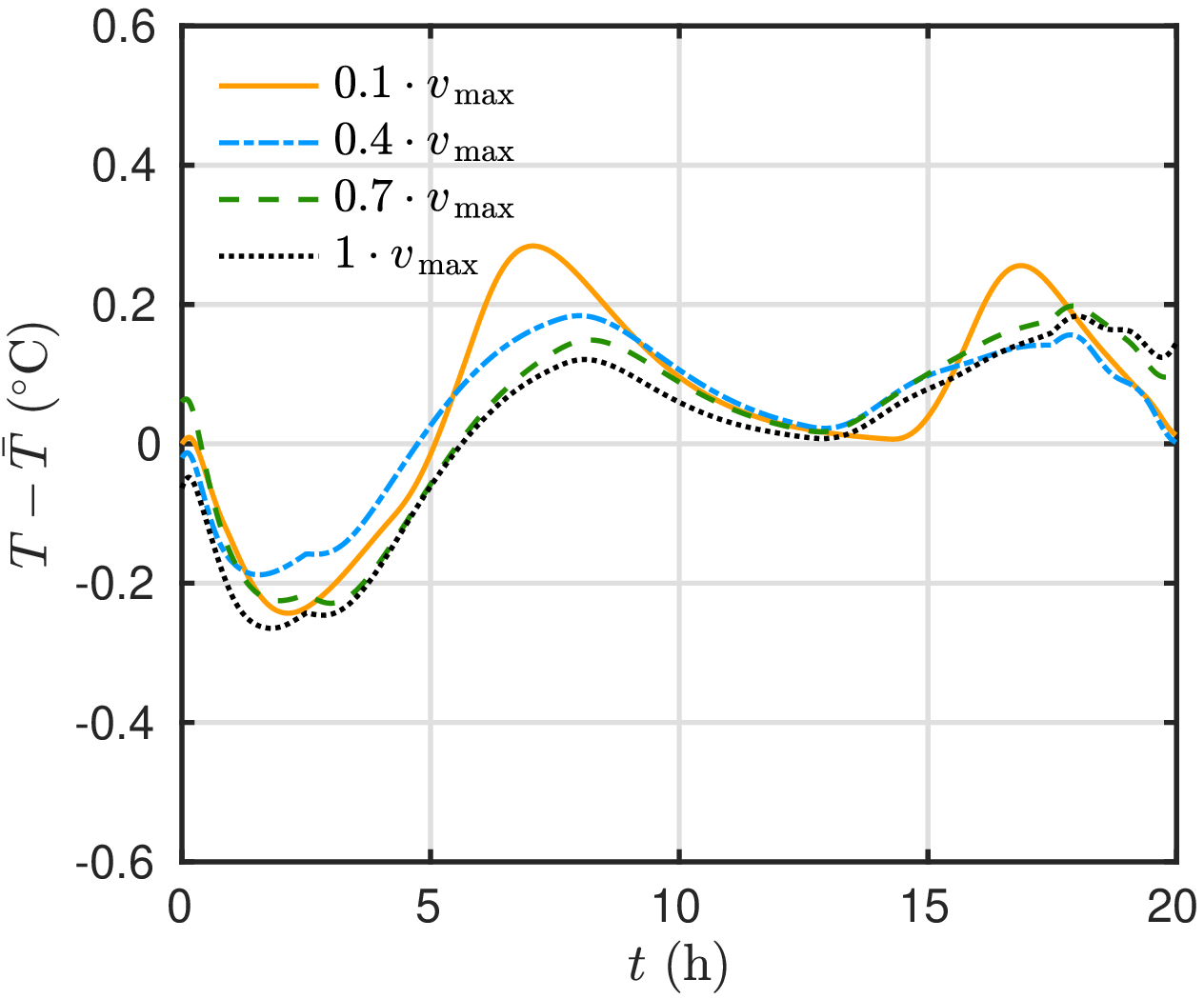}}
\caption{\revision{Time variation of the residual between the model prediction of the $1$D lumped model and the} \revision{ $2$D model, both computed with the estimated parameters at the two sensor positions.}}
\end{figure}

A comparison between the prediction of the lumped model and with a complete two-dimensional model is performed considering the estimated parameters in Figures~\ref{fig:T_tay_v10_ft} and \ref{fig:T_tay_v70_ft}. The difference between the computed temperature is presented for the two sensor positions in Figures~\ref{fig:res2Dx1_ft} and \ref{fig:res2Dx2_ft}. The maximal discrepancy scales with $0.25 \ \mathsf{\degC}$ for the sensor at the top surface. All configurations have a similar behavior. For the sensor inside the material, the error is smaller varying below $0.2 \ \mathsf{\degC}\,$. The error decreases with the ventilator speed. The difference in the model predictions are acceptable compared to the computational burden of the \textsc{B}ayesian estimation using the $2$D model. Table~\ref{tab:cpu_pep} presents the computational cost to solve the inverse problem using the lumped and $2$D models. For the latter, the cost is interpolated considering the computational time of the model to solve one direct problem and an hypothetical number of states required for the MCMC algorithm. The computational cost would be $53$ times higher for the $2$D model than for the lumped one. Note that the \DF ~model is an innovative numerical approach that enables to relax the stability conditions. The computational burden would be exponential using standard numerical methods such as \textsc{E}uler explicit or implicit approaches.

\begin{table}[h!]
\centering
\caption{Computational cost of the \textsc{B}ayesian parameter estimation, measured for the $1$D lumped model and extrapolated for the $2$D model $\bigl(\,t_{\,\cpu\,,\,0} \egal 7.13 \ \mathsf{h}\,\bigr)$.}
\label{tab:cpu_pep}
\setlength{\extrarowheight}{.5em}
\begin{tabular}[l]{@{} ccc}
\hline
\hline
& \multicolumn{2}{c}{\textit{Computational time}} \\
\textit{Number of MCMC states}
& $1$D \textit{lumped model}
& $2$D \textit{model} \\
 $10^{\,5}$
& $t_{\,\cpu\,,\,0}$
& $53 \cdot t_{\,\cpu\,,\,0}$ \\
\hline
\hline
\end{tabular}
\end{table}

\section{Conclusions}

The objective of this work is to characterize the surface heat transfer coefficients of a climatic chamber. The unknown parameters are retrieved through the solution of an inverse two-dimensional nonlinear heat conduction problem. The \textsc{B}ayesian framework with the MCMC algorithm are used to determine the posterior distribution of the two unknown parameters, namely the top and lateral surface transfer coefficients. To decrease the computational cost of the method, a lumped one-dimensional model is introduced. It is based on the assumption that the transfer occurs mainly in one direction. Thus, the problem can be formulated as a $1$D heat diffusion equation with a convective source term. The lumped model approximations are considered within the parameter estimation procedure thanks to the Approximation Error Model.

The experiments are carried for a wood fiber material. Four configurations are investigated according to the variation of the chamber ventilator speeds. For each configuration, a set of three experimental data are generated to decrease the random measurement uncertainty. In addition, a complete propagation of the measurement uncertainties including the sensor position and sensor measurement is realized. To retrieve the unknown parameters, two sensors are placed at the interface of the material with the ambient air and in the middle of the sample. 

The results of the parameter estimation problem enables to determine accurately the \textsc{G}ausian type posterior distribution of the surface transfer coefficients. For velocity in the range of $10\%$ to $70\%$ of the ventilator maximum speed, the mean of the top and lateral coefficients scale between $19.2$ to $24.5\, \mathsf{W\,.\,m^{\,-2}\,.\,K^{\,-1}}$ and $0.18$ to $0.21\, \mathsf{W\,.\,m^{\,-2}\,.\,K^{\,-1}}\,$, respectively. The standard deviation is small and the residual between measurement and calibrated model prediction are very satisfactory. For the maximal velocity configuration, the results highlight that the top surface transfer coefficient is large and the boundary conditions can be modeled as \textsc{D}irichlet type. The lateral surface coefficient remains of the same order as for the other configurations. Complementary investigations are carried to show the reliability of the proposed lumped model compared to the complete two-dimensional one. The error is acceptable and the computational gains in the MCMC algorithm is very important, about $50$ times faster. 

This methodology can be employed to improve the reliability of the models to accurately predict heat transfer in building walls. In \cite{berger_bayesian_2016} it is applied
for real \emph{in-situ} estimation where the thermophysical properties and the time varying surface heat transfer coefficient of a wall are estimated using \textsc{B}ayesian inferences. The AEM is also used to include the error between the heat transfer model and the coupled heat and mass one.

With those results, future investigations should focus on proposing experimental data-set for coupled heat and mass transfer given the important impact of latent effects on building energy efficiency. Future investigations could analyze further the variation of $R_{\,\ell}$ with space. It demands several developments since it requires an accurate parametrization of such space function. A good compromise has to be found between a refined parametrization enabling an accurate representation; inducing a very high computational cost of the parameter estimation problem. The construction of a reduced order model is needed to answer this issue.

\section*{Nomenclature}

\begin{tabular}{lcccl}
AEM &&&& Approximation Error Model   \\
$\Bi$ &&&& \textsc{B}iot number, $\unitt{-}$\\
$c\,$, $c_{\,0}\,$, $c_{\,1}$  &&&& volumetric heat capacity, $\unitt{W\,.\,m^{\,-3}\,.\,K^{\,-1}}$ \\
$e$ &&&& model temperature error, $\unitt{K}$ \\
$f$ &&&& model, $\unitt{-}$ \\
%
$\mathcal{F}$ &&&& \textsc{Fisher} information matrix, $\unitt{-}$ \\
$\Fo$ &&&& \textsc{F}ourier number, $\unitt{-}$ \\
$h$ &&&& heat transfer coefficient, $\unitt{W\,.\,m^{\,-2}\,.\,K^{\,-1}}$ \\
$j_{\,q}$ &&&& conductive heat flux,  $\unitt{W\,.\,m^{\,-2}}$ \\
$\mathrm{k}$ &&&& state indicator, $\unitt{-}$ \\ 
$k\,$, $k_{\,0}\,$, $k_{\,1}$ &&&& thermal conductivity, $\unitt{W\,.\,m^{\,-1}\,.\,K^{\,-1}}$ \\
$L$ &&&& length, $\unitt{m}$ \\
MCMC &&&& \textsc{M}arkov Chain \textsc{M}onte \textsc{C}arlo \\
$\mathcal{M}at\,\bigl(\, \mathbb{R}_{\,>\,0} \,,\, 2 \,,\, 2 \,\bigr)$ &&&& matrices of size $2 \times 2$ with positive real coefficients \\
$n$ &&&& time index, $\unitt{-}$ \\
$\mathcal{N}$ &&&& \textsc{G}aussian distribution, $\unitt{-}$ \\
$N_{\,b}$ &&&& number of states of burn-in period in MCMC algorithm, $\unitt{-}$\\
$N_{\,e}$ &&&& number experimental campaign, $\unitt{-}$\\
$N_{\,t}$ &&&& number of time measurements, $\unitt{-}$ \\
$N_{\,s}$ &&&& number of states in MCMC algorithm, $\unitt{-}$\\
$N_{\,x}$ &&&& number of sensor positions, $\unitt{-}$ \\
$\mathcal{P}$ &&&& probability distribution, $\unitt{-}$ \\
$p\,$, $\p$ &&&& unknown parameter, $\unitt{-}$ \\
$R$ &&&& thermal resistance, $\unitt{W\,.\,m^{\,-2}\,.\,K^{\,-1}}$ \\
$r$ &&&& length ratio, $\unitt{-}$ \\
SGI &&&& Structurally Globally Identifiable \\
$T\,$, $\overline{T}\,$, $\widehat{T}$ &&&& temperature, $\unitt{K}$ \\
$t$ &&&& time, $\unitt{s}$ \\
$U$ &&&& random value sampled, $\unitt{-}$ \\
$\mathcal{U}$ &&&& uniform distribution, $\unitt{-}$ \\ 
$u$ &&&& dimensionless temperature, $\unitt{-}$ \\
$v$ &&&& air velocity, $\unitt{m\,.\,s^{\,-1}}$ \\
$ \boldsymbol{w}$ &&&& parameters walk, $\unitt{-}$ \\
$x\,,\,y\,,\,z\,$ &&&& coordinates, $\unitt{m}$ \\
$x_{\,1}\,,\,x_{\,2}$ &&&& sensor position, $\unitt{m}$ \\
\end{tabular}

\begin{tabular}{lcccl}
\multicolumn{5}{l}{\textbf{Greek symbols}} \\
$\beta$ &&&& acceptance factor, $\unitt{-}$\\
$\delta$ &&&& sensor position uncertainty, $\unitt{m}$\\
$\eta$ &&&& parameter error estimator, $\unitt{-}$\\
$\epsilon$ &&&& small quantity, $\unitt{-}$\\
$\theta$ &&&& dimensionless sensitivity coefficient to $\Bi_{\,t}$, $\unitt{-}$ \\
$\kappa$ &&&& dimensionless conductivity, $\unitt{-}$\\
$\zeta$ &&&& dimensionless capacity, $\unitt{-}$\\
$\mu$ &&&& estimated parameter mean, $\unitt{-}$\\
$\nu$ &&&& estimated parameter standard deviation, $\unitt{-}$\\
$\pi$ &&&& probability distribution, $\unitt{-}$\\
$\sigma$ &&&& measurement uncertainty, $\unitt{K}$\\
$\tau$ &&&& dimensionless time, $\unitt{-}$\\
$\upsilon$ &&&& estimated parameter standard deviation, $\unitt{-}$\\
$\varphi$ &&&& dimensionless sensitivity coefficient to $\Fo$, $\unitt{-}$ \\
$\phi$ &&&& relative humidity, $\unitt{-}$
\end{tabular}

\begin{tabular}{lcccl}
$\chi$ &&&& dimensionless space, $\unitt{-}$ \\
$\psi$ &&&& dimensionless sensitivity coefficient to $\Bi_{\,\ell}$, $\unitt{-}$\\
$\Omega$ &&&& domain, $\unitt{-}$\\
$\omega$ &&&& material moisture content (dry basis weight), $\unitt{-}$\\
\multicolumn{5}{l}{\textbf{Subscripts}} \\
$\mathrm{alu}$ &&&& aluminum/wood fiber  \\
$\cpu$ &&&& computational run time \\
$\mathrm{ins}$ &&&& insulator/wood fiber  \\
$\ini$ &&&& initial condition  \\
$\max$ &&&& maximum value \\
$i$ &&&& material indicator ($i \egal 1$ wood fiber, $i \egal 2$ insulator,  $i \egal 3$ aluminum) \\
$j$ &&&& sensor position indicator  \\
$\ell$ &&&& lateral surface  \\
$\obs$ &&&& observation  \\
$s $ &&&& sensor measurement  \\
$t$ &&&& top surface  \\
$x$ &&&& relative to height  \\
$y$ &&&& relative to width  \\
$z$ &&&& relative to depth  \\
$0$ &&&& reference state  \\
$\infty$ &&&& ambient air  \\
$\sim $ &&&& random measurement  \\
$\chi$ &&&& sensor position  
\end{tabular}

\begin{tabular}{lcccl}
\multicolumn{5}{l}{\textbf{Superscripts}}  \\
$\mathrm{k}$ &&&& state indicator  \\
$n$ &&&& time measurement indicator  \\
$\widehat{ }$ &&&& measured quantity  \\
$\widetilde{ }$ &&&& other parameter \\
$\bar{ }$ &&&& $1$D lumped model quantity\\
$\star$ &&&& candidate parameter in MCMC algorithm  \\
$\apr$ &&&& \emph{a priori}\\
\end{tabular}

\section*{Acknowledgments}

The authors acknowledge the grants from the French Agency for Ecological Transition (ADEME).

\section*{Conflict of interests}

The authors declare that they have no known competing financial interests or personal relationships that could have appeared to influence the work reported in this paper.

\bibliographystyle{unsrt}  

\bibliography{export_bib}

\end{document}